\documentclass[12pt,twoside]{report}

\newcommand{\nc}{\newcommand}
\newcommand{\beq}{\begin{equation}}
\newcommand{\eeq}{\end{equation}}
\nc{\nn}{\nonumber}
\nc{\tr}{{\rm tr}}
\nc{\Tr}{{\rm Tr}}
\def\lb{\left( }
\def\href#1#2{#2}
\def\rb{\right) }

\batchmode
  \font\bbbfont=msbm10
\errorstopmode
\newif\ifamsf\amsftrue
\ifx\bbbfont\nullfont
  \amsffalse
\fi
\ifamsf
\def\IR{\hbox{\bbbfont R}}

\def\IZ{\hbox{\bbbfont Z}}
\def\IF{\hbox{\bbbfont F}}
\def\IP{\hbox{\bbbfont P}}
\else
\def\IR{\relax{\rm I\kern-.18em R}}
\def\IZ{\relax\ifmmode\hbox{Z\kern-.4em Z}\else{Z\kern-.4em Z}\fi}
\def\IF{\relax{\rm I\kern-.18em F}}
\def\IP{\relax{\rm I\kern-.18em P}}
\fi

\newdimen\tableauside\tableauside=1.0ex
\newdimen\tableaurule\tableaurule=0.4pt
\newdimen\tableaustep
\def\phantomhrule#1{\hbox{\vbox to0pt{\hrule height\tableaurule width#1\vss}}}
\def\half{\frac{1}{2}}
\def\drawbox#1#2{\hrule height#2pt
        \hbox{\vrule width#2pt height#1pt \kern#1pt
              \vrule width#2pt}
              \hrule height#2pt}

\def\phantomvrule#1{\vbox{\hbox to0pt{\vrule width\tableaurule height#1\hss}}}
\def\ZZ{\relax{\sf Z\kern-.4em Z}}
\def\sqr{\vbox{%
  \phantomhrule\tableaustep
  \hbox{\phantomvrule\tableaustep\kern\tableaustep\phantomvrule\tableaustep}%
  \hbox{\vbox{\phantomhrule\tableauside}\kern-\tableaurule}}}
\def\squares#1{\hbox{\count0=#1\noindent\loop\sqr
  \advance\count0 by-1 \ifnum\count0>0\repeat}}
\def\tableau#1{\vcenter{\offinterlineskip
  \tableaustep=\tableauside\advance\tableaustep by-\tableaurule
  \kern\normallineskip\hbox
    {\kern\normallineskip\vbox
      {\gettableau#1 0 }%
     \kern\normallineskip\kern\tableaurule}%
  \kern\normallineskip\kern\tableaurule}}
\def\gettableau#1 {\ifnum#1=0\let\next=\null\else
  \squares{#1}\let\next=\gettableau\fi\next}

\def\Fund#1#2{\vcenter{\vbox{\drawbox{#1}{#2}}}}
\def\Asym#1#2{\vcenter{\vbox{\drawbox{#1}{#2}
              \kern-#2pt       
              \drawbox{#1}{#2}}}}
\def\fund{\Fund{6.5}{0.4}}
\def\asym{\Asym{6.5}{0.4}}

\input epsf
\newcount\figno
\def\fig#1#2#3{
\par\begingroup\parindent=0pt\leftskip=1cm\rightskip=1cm\parindent=0pt
\baselineskip=11pt
\global\advance\figno by 1
\epsfxsize=#3
\centerline{\epsfbox{#2}}
\vskip 12pt
{\bf Figure \the\figno:} #1\par
\endgroup\par
}
\def\figlabel#1{\xdef#1{\the\figno 
\mbox{ }}}
\def\encadremath#1{\vbox{\hrule\hbox{\vrule\kern8pt\vbox{\kern8pt
\hbox{$\displaystyle #1$}\kern8pt}
\kern8pt\vrule}\hrule}}

\begin{document}
\baselineskip=1.1\baselineskip
\thispagestyle{empty}
\begin{titlepage}

\begin{center}
\today
\hfill hep-th/9812072
\vspace{2cm}

\centerline{\LARGE \bf Field Theory Dynamics} 
\vskip 0.5cm
\centerline{\LARGE \bf from Branes in String Theory}
\vskip 1.5cm

{\large Andreas Karch }

\vspace{1cm}

{\it Institut f\"ur Physik,\\
Humboldt Universit\"at Berlin\\
Invalidenstr. 110\\
10115 Berlin, Germany}
\end{center}

\vspace{1.5cm}

\noindent {\bf Abstract:} I review certain aspects of Hanany-Witten setups and 
other approaches used to embed (and solve) gauge theories
in string theory. Applications covered include dualities in
4 and 3 dimensions, fixed points in 6 dimensions, phase
transitions between different geometric backgrounds and dualities
between branes and geometry. 

\end{titlepage}

\vskip 20cm
\eject
\newpage
\tableofcontents
\chapter{Motivation and Introduction}

\section{Motivation}

It is the main goal of physics to explain all ratios between 
measurable quantities. The hope is that in the end a very simple,
beautiful and unique mathematical structure emerges, the
theory of everything.

Modern physics is based on two major building blocks: general relativity
and quantum physics. The former expresses all phenomena of
the macroscopic world and especially gravity in terms
of simple geometric concepts. The latter is powerful in explaining
the microscopic world by replacing the notion of particles which
occupy a certain position in space and carry a certain
amount of momentum by the more abstract formalism
of states in a Hilbert space and observables as operators acting
on them. Since the distinction between microscopic and macroscopic
world seems to be rather arbitrary, these two building blocks
should be unified in an underlying theory. More than
that, as they stand, the two concepts are even inconsistent. Straight
forward quantization of
general relativity
leads to infinities in physical processes that can not be tolerated.
A larger theory unifying gravity and quantum theory is hence not
only desired from an aestethic point of view but indeed required for
consistency.

After many years of extensive search for this unifying theory, a single
candidate has emerged: string theory. String theory replaces
the fundamental point like objects of particle physics by 1d
strings
thereby removing the infinities encountered in quantizing general
relativity. General relativity reemerges as a low energy limit
at large distances, where it was tested experimentally. However
at small distances stringy physics takes over and even our concepts
of space and time break down. Similarly ordinary particle physics
as described by the standard model can reemerge in the limit
where gravitational interactions between the particles can be
neglected. The energy scale at which both gravity
and quantum effects become important is set by the Planck scale
and is roughly $10^{19} GeV$. Since this scale
is so huge, it is impossible to just create
the fundamental degrees of quantum gravity in an accelerator and then
look what they are.

Even though string theory has all the ingredients required by modern
physics it is difficult to make contact with physics as we know it.
The major obstacle is that string theory allows for a variety of different
vacua, each of which leads to different physics. No precise
predictions about the low-energy physics (like answers to
the why questions left by the standard model) can
be made without finding a process that determines the right vacuum.
However the most important concepts appearing in string theory,
namely gauge theory, gravity and supersymmetry, are indeed known
or believed to dominate the real world. While the former two are
the bases for all physics described by the standard model and general
relativity, the latter is believed to be of similar importance for
next generation collider physics. Experimental verification
of supersymmetry in the real world would be even more support
that string theory is not just the only known consistent quantum
theory of gravity, but indeed the fundamental theory realized in
our world.

Of special interest are also objects which probe the regime of quantum
gravity, that is they are small enough for quantum effects to be
important and heavy enough to require gravity. Examples of such
objects are black holes close to their singularity or our universe
in very early times, close to big bang. Treatment of
these important issues as well as the vacuum selection problem
requires non-perturbative information about string theory. So
far the perturbative expansion of string theory in terms of 
worldsurfaces was the only definition we had of string theory.
Only very recently tools have emerged that allow us control
certain aspects of the non-perturbative physics behind
string theory, raising the hope that these fundamental
issues can finally be addressed.
\section{Introduction}

During the so called ``second string revolution'' it has become possible to
gain control over aspects of string theory
\cite{gsw1,gsw2,luest} that were not contained
in its perturbation expansion in terms of
worldsurfaces. The major achievements were the discovery
of D-branes \cite{polchinski,rpolchinski}
as one of the non-perturbative objects in string theory
and the realization of the role of duality symmetries in string
theory \cite{various}, relating
two seemingly different theoretical descriptions to one and the same
physical situation.

Dualities have been of similar importance in gauge theories.
Using dualities it has been possible to solve the IR behaviour
of certain
quantum field theories exactly \cite{SW} and get a lot of non-perturbative
information even in situations with less restrictive symmetries. However
in all those setups supersymmetry is a vital ingredient and it is not
clear yet how these methods can be generalized to non-supersymmetric
situations.

Non-perturbative string theory and Super Yang Mills (SYM) gauge
theories are indeed deeply related. The dynamics of the D-branes
which are so important for our understanding of non-perturbative
string-theory is basically governed by SYM.
This connection can be used in a twofold way: dualities and other
obscure aspects of field theory only discovered recently, like
non-trivial fixed points of the renormalization group with mutual
non-local objects becoming massless, find their natural place in string
theory, where they can be easily visualized.

On the other hand some problems which seem intractable in string theory can
be mapped to questions in gauge theory, which are under much better
control. Indeed it has been proposed that all non-perturbative aspects
of string theory can be encoded in large $N$ SYM theory \cite{bfss}.

It is the purpose of this work to highlight some of the insights gained
with the help of this deep connection between gauge theory and string
theory.

In Chapter 2 I will review the construction of D-branes and explain
how gauge theory determines their dynamics. I will also comment
on the important role D-branes played in understanding string
theory dualities, since these dualities lead to the discovery
of an 11-dimensional theory called M-theory that basically
summarizes all the non-perturbative insights we gained
about string theory and is the natural arena for visualizing
aspects of SYM which are hard to understand 
from the field theory point of view.

In Chapter 3 I will introduce a setup first used by Hanany and Witten
to study 3d gauge theories embedded in string theory. In
Chapter 4 I will use
this setup to study certain aspects of 6-dimensional and 4-dimensional
physics. The interplay will allow us to understand certain
non-trivial fixed points and transitions which are
obscure from the field theory point of view, to say the least.
But we can also learn about string theory from the
correspondence.
Among many other things it will allow us to show that in string theory
chiral vacua can be smoothly deformed into non-chiral vacua,
perhaps taking a small step towards a more detailed understanding of the 
vacuum selection problem.

In Chapter 5 I will show how the other aspects of the
gauge-theory / string-theory connection are related to
the Hanany-Witten setups by a series of string-theory dualities.
We will see that the different techniques used to explore the correspondence
might be more or less powerful in various situations, but that
in the end we are guaranteed to obtain the same results, no matter
how we chose to embed our gauge theory under consideration
into string theory.

In Chapter 6 I will discuss what open problems remain and where
we can go from here.

\chapter{D-branes and non-perturbative effects in string theory}

\section{The breakdown of perturbation theory}

String theory as we used to know it was only defined via
its perturbation series. That is a given scattering process
receives contributions from worldsheets of various topology.
Higher genus surfaces are weighted with higher powers of $g_s$,
the string coupling (which is the expectation value
of a dynamical field, the dilaton: $g_s \sim e^{<\Phi>}$)

\vspace{1cm}
\fig{Propagation of a string from its perturbative
definition}
{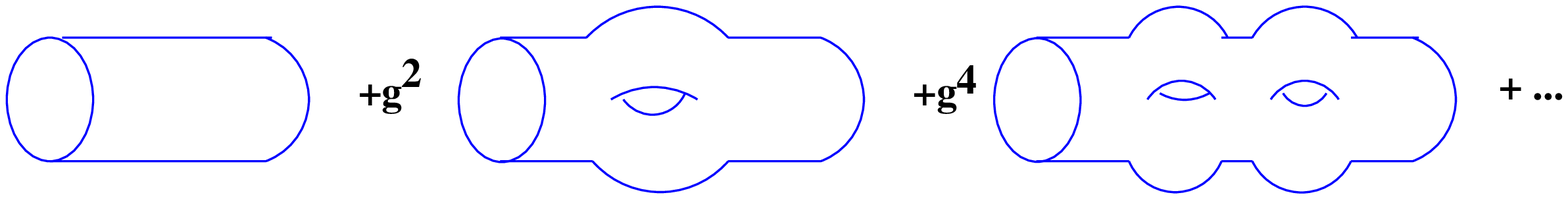}{10truecm}
\figlabel{\perturb}
\vspace{1cm}

To calculate the contribution from a single diagram in the 
perturbation series depicted in Fig.\perturb, we have to solve
a conformal field theory on the worldvolume of the given topology
and than integrate over all possible deformations (moduli). This
is often possible. 2d conformal field theories are very
constraint due to the high amount of symmetry and many
calculational tools are available. At weak coupling only
diagrams of low genus contribute and we can actually calculate
the amplitudes.

However this perturbative definition clearly fails when we are
at strong coupling. Here we really have to calculate
an infinite number of diagrams, since higher topologies are no longer
suppressed. Worse than that, a generic scattering process
may also receive contributions that are not even visible at all
in the perturbation series, even if we would be able to sum up the
infinite diagrams. Such contributions are suppressed as $e^{-1/g_s}$
(or as $e^{-1/g_s^2}$). These contributions have a series expansion
around strong coupling ($g_s \rightarrow \infty$) 
$$
e^{-1/g_s}=1-\frac{1}{g_s} + \frac{1}{2 g_s^2}- \frac{1}{6 g_s^3} + \ldots
$$
None of these terms has the right power to match any of those appearing
in the ordinary perturbation series which only contains
positive powers of $g_s$. These are purely non-perturbative
effects.

From field theory it is well known that there are indeed phenomena
giving rise to such non-perturbative contributions. The most famous
example are instantons. Instantons are stable solutions to the
Yang-Mills equations of motion that are centered in space and time.
Their existence is due to the fact that Yang-Mills can have topologically
distinct vacua. Instanton solutions interpolate between
different vacua and their stabelness is therefore guaranteed
by topology. Arguing on the bases of the cluster
decomposition principle one can show \cite{weinbergbook}
that in order to define a consistent quantum field
theory we indeed have to sum over all possible instanton
backgrounds when performing the path integral, so these
configurations do contribute to scattering processes.
Calculating the classical action corresponding
to an instanton configuration one finds that it is indeed suppressed
as $e^{-1/g_{YM}^2}$. This example also gives us an intuitive feeling
why such things will never appear in the perturbative expansion:
while perturbation theory expands around a given vacuum, non-perturbative
contributions arise from tunnelling processes and interpolation between
different vacua. But this is also why it is so crucial to
understand non-perturbative states in string theory: solving
the vacuum problem, that is what is the right string theory ground state
and how did nature pick it, requires detailed understanding
of precisely these processes.

Similar effects are due to solitonic objects like
monopoles or domain walls. They are again stable solutions
to the equations of motion centered in space. This enables
us to interpret them as particles (or higher dimensional
objects) in our theory. They have masses which go like
$1/g^2_{YM}$. At weak coupling they are very heavy and can be
neglected. However at strong coupling they should be included. 
Virtual monopoles running in loops will be suppressed by
$\sim e^{-1/g^2_{YM}}$ due to the $e^{-\mbox{action}}$ factor
in the path integral, signalling a non-perturbative contribution.

In the same spirit we can try to identify solitonic objects
with mass $1/g^2$ in string theory in order to identify non-perturbative
string states. By studying supergravity (SUGRA),
the low-energy field theory limit of
string theory, one indeed finds a whole zoo of such objects, 
generically called p-branes.
Among more exotic objects there exists the magnetic dual
of the string, the NS5 brane with tension $1/g_s^2$
and the so called Dirichlet (D) branes, whose tension at weak
coupling only grows as $1/g_s$. Understanding those objects
should enable us to learn about non-perturbative effects in
string theory. They will be the topic of the rest of this work.

\section{A string theoretic description of D-branes}

To understand the contribution to the amplitude by a given brane configuration,
one can study perturbative string theory in the background
of the branes at weak string coupling. 
For the D-branes this is straight forward, once one realizes
that D-branes are space-time defects at which open strings can end.

\vspace{1cm}
\fig{String in the background of a D-brane}
{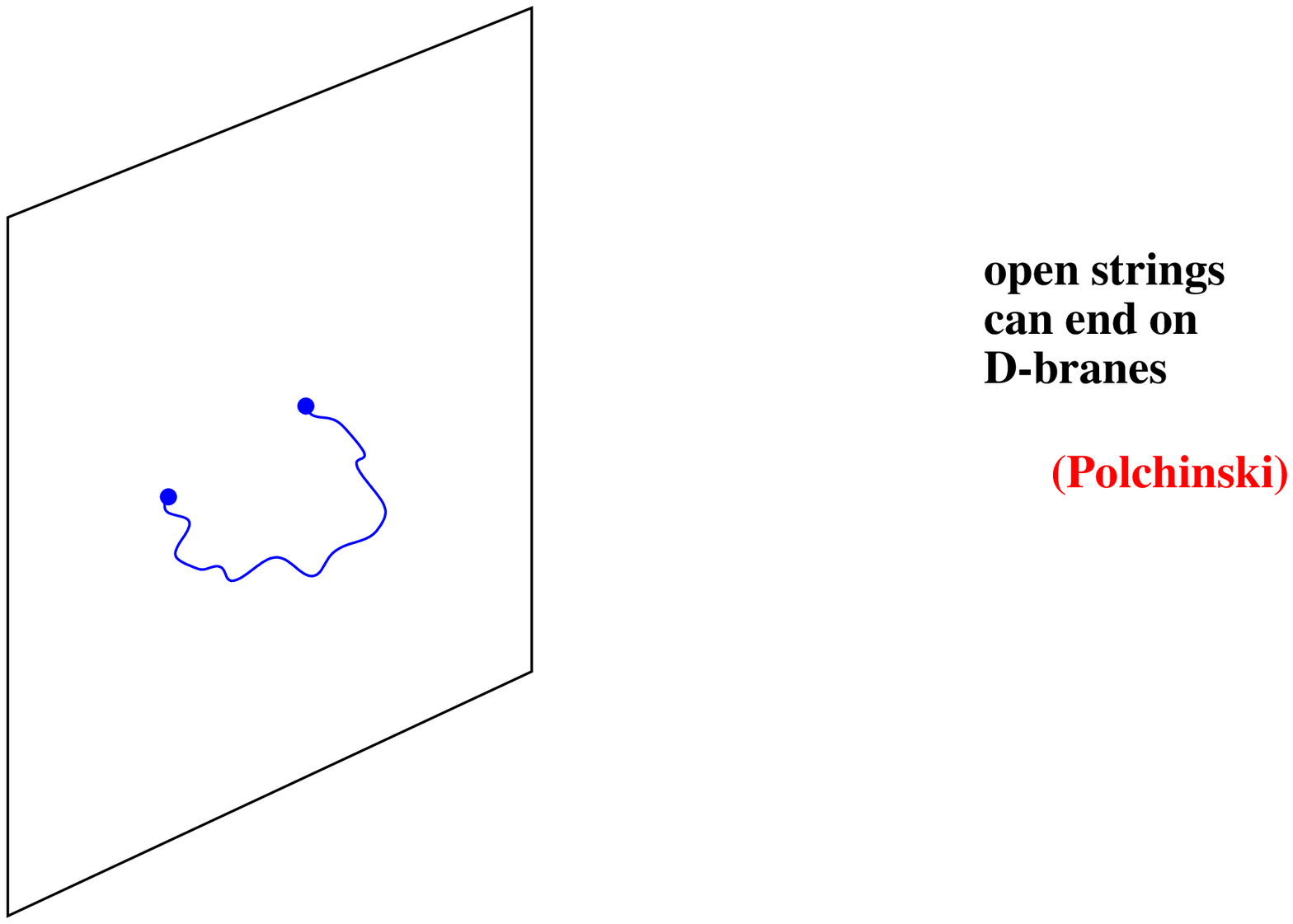}{10truecm}
\figlabel{\dbrane}
\vspace{1cm}

Figure \dbrane shows this basic concept of D-brane physics. It was known
since the early days of string theory that in open string theory
one can as well impose Dirichlet boundary conditions (the end of the
string is at a fixed position) as the usual Neumann boundary
conditions (the end is free to move, no momentum is allowed to
flow of the end). One usually neglected this possibility, since
it introduced hyperplanes (the planes on which the endpoints are forced
to stick) which break Lorentz invariance. It was the achievement
of Polchinski to show \cite{polchinski,rpolchinski} 
via an explicit 1-loop open string calculation
that these space-time defects carry charge under the RR gauge fields
of string theory and calculating their tension to be
\begin{equation} 
\label{tension}
T_{Dp}= \frac{1}{l_s^{p+1} g_s}
\end{equation}
where $l_s$ and $g_s$ denote the string coupling and length respectively.
These properties allow us to identify them with the stringy version
of the solitonic solutions
of SUGRA which I already called D-branes before.

Now it is straight forward to do everything we are used to from perturbative
string theory in the background of the D-branes. Quantizing the
oscillator modes of the string theory in the presence of the
modified boundary conditions one finds that the massless
spectrum of the open strings ending on the D-brane are given by
a SYM multiplet living on the brane worldvolume. That is,
for the D9 brane we find the usual ${\cal N}=1$ SYM multiplet consisting
of the vector gauge field and the gauginos in 10d. All other
branes yield dimensional reductions of those to the 
appropriate worldvolume dimension.

Demanding conformal invariance on the string worldsheet yields equations
of motion for the space-time fields by setting the $\beta$ function
of the 2d conformal theory on the worldsheet to zero, order by order
in the string tension (which plays the role of the coupling
constant in the 2d theory), just like in the well known case
of Neumann boundary conditions. Writing down an action that yields
these equations one obtains as an effective action for the D-brane
theory a supersymmetric Dirac-Born-Infeld action with Wess-Zumino couplings
to the bulk fields \cite{leighd}

\begin{equation}
\label{DBI}
S=S_{DBI} + S_{WZ} = - \int d^{p+1} \xi \; e^{- \Phi} \sqrt{-\mbox{det}
(g_{ij} - {\cal F}_{ij}}) + 
\int d^{p+1} \xi \; C e^{\cal F}
\end{equation}
where ${\cal F} = F -B$, $C= \sum_{r=0}^{10} C^{(r)}$ is
a formal sum over all the form fields present in the IIA/B
supergravity and the integral always picks out the right form
to go with the right power of ${\cal F}$ from the exponential.
The fields should be understood as pullbacks from superspace
to the worldvolume. 

The low-energy approximation of this action, that is the expansion
to lowest order
in $l_s^2$  (the string length), yields SYM on the
worldvolume. This is in accordance with the analysis
of the massless spectrum. The gauge coupling can be read off
from (\ref{DBI}) to be
\begin{equation}
\label{gYM}
g^2_{YM}=l_s^{p-3} g_s. 
\end{equation}

As in the case of fundamental string theory the scalars on the worldvolume
define the position of the brane in the transverse space. Via the DBI
action these are coupled to the worldvolume gauge fields. This is an important
property of the D-brane action which we will explain
in more detail in the following. Basically a brane can absorb the flux
of a charged particle by bending in transverse space, balancing
the force from the gauge fields with its tension, that is with
the worldvolume scalars. A flat D-brane breaks half of the supersymmetries
(since the open string spectrum only has half of the supersymmetries
of the closed spectrum). The supersymmetries
preserved have to satisfy \footnote{which can e.g. be seen
by analyzing the Killing spinor equations in the background
of the D-brane soliton solution}
\begin{equation}
\label{SUSYD}
\epsilon_L = \Gamma_0 \cdot \ldots \cdot \Gamma_p \epsilon_R
\end{equation}
where the preserved supercharge is $\epsilon_L Q^L + \epsilon_R Q^R$
and $Q^L$, $Q^R$ are the supercharges generated by left- and right-
moving degrees of freedom in the surrounding type II string theory.
Choosing a non-trivial embedding
generically breaks all the supersymmetries. If the embedding
geometry allows for some Killing spinors, lower fractions of supersymmetry
may be preserved.

If we try to repeat this analysis for NS5 branes we run into
trouble. The NS5 brane metric looks like a tube, the dilaton
blows up if we move towards the core and any conformal field theory
description breaks down. Only asymptotically, the NS5 brane can be
described by a well known conformal field theory, a WZW model \cite{chs}.
The NS5 brane worldvolume theory is not accessible by purely
perturbative string techniques. However we will see later
that we can deduce its properties by string dualities.

\section{D-branes and gauge theory}

By now we have gained some insights in the dynamics governing
D-branes. We have learned that there is
a very deep connection between D-branes and gauge
theories. We will analyse how some of the
most interesting aspects of D-branes are captured
by simple field theoretic phenomena.
This discussion will pave the way for the
discussion in the following chapters, where I will
exploit the D-brane / gauge theory correspondence to learn about string
theory as well as about gauge theory.
The general philosophy is that we consider certain limits of string theory,
in which the gravity and heavy string modes (the bulk modes) decouple, 
leaving us just
with the open string sector described by SYM.
The basic quantities that control this limit are the Planck scale
$M_{pl}$ and the string scale $M_s$ (the inverse of the string length).
They satisfy
\begin{equation}
\label{pls}
M_{pl}^4 g_s = M_s^4
\end{equation}
which just shows the relation between string frame and Einstein frame.
Sending $M_{pl}$ to infinity is the same as sending Newton's constant
to zero, so gravity is decoupled. Taking $M_s$ to infinity sends
all excited string states to infinite mass effectively decoupling them, too.
This can be done at finite string coupling, keeping an interacting SYM theory.

\subsection{Gauge theory on the worldvolume}

As we have seen, the effective theory on the worldvolume
is given by a DBI action. We want to analyse
this world volume theory in the limit, where the bulk physics 
decouples, that is we get rid of gravity and other closed
string modes.
We only keep the
degrees of freedom on the brane. Expanding
the DBI action in $l_s^2$ (which explicitely shows up together
with every $F$) it is
easy to see, that in the $l_s \rightarrow 0$ limit the theory on the
worldvolume of the D$p$-brane reduces to $U(1)$ SYM in p+1 dimensions.
The amount of supersymmetry preserved by a given brane
is determined by its embedding in space-time,
as discussed above. 
A flat brane always preserves half of the 32 supercharges of
type II theory, leading to maximally supersymmetric Yang-Mills on
the worldvolume.
Now let us consider what happens if $N$ D-branes coincide.
This situation was analysed by Witten \cite{boundstates}.

A single D-brane supports on its worldvolume a single $U(1)$ multiplet.
These massless states arise from a string starting and ending on the same
brane. The mass of a state is given by the length of the string times
the string tension. The massless vector hence arises from a zero
length string
starting and ending at the same point. Each of the ends of the strings
carries a Chan-Paton label of the gauge group, that is an
index in the fundamental representation, so the vector 
multiplet
is correctly left with a fundamental and an antifundamental index, 
an adjoint field. Clearly  nothing 
happens to these states if many
D-branes coincide. However, whenever two branes are close,
there are new 
states that become important.
Strings stretching from one brane to the other yield states whose
mass is determined by the distance between the branes. They carry
a fundamental Chan Paton index of the U(1) of the brane
they start and end on respectively. It is natural to identify
those as W-bosons of a broken U(2) gauge group. The distance
between the branes determines the Higgs expectation value.
When $N$ branes coincide all the W-bosons become massless
and the full $U(N)$ gauge symmetry becomes visible.

In order to obtain different gauge groups one can consider D-branes
coinciding on top of space-time singularities. An example of
such a singularity which is under control from perturbative
string theory is an orientifold plane, the fixed plane of a $Z_2$
orbifold action, that combines worldsheet parity with a
space time reflection in $r$ coordinates. The resulting
$p+1=10-r$ space-time fixed plane is called an O$p$ orientifold plane.
A similar calculation like that of Polchinski's determination
of the D-brane charge shows that the orientifold is
also charged under the same RR field as the D$p$, where
the relative value of the charge is given by
\footnote{Here and in what follows I will always consider the
D-brane and its $Z_2$ mirror as different objects, each carrying
charge $q_D$. If one wants to consider only physical D-branes one
should assign them charge $2 q_D$ in these conventions.}
\begin{equation}
q_O= \pm 2^{p-4} q_D.
\end{equation}
The sign is determined by a discrete choice of the precise way
one performs the projection. When $N$ D-branes coincide on-top
of the orientifold (and hence also coincide with their $N$ mirrors),
only oriented strings stretching between the branes will yield
new massless gauge bosons, leading to an $SO(2N)$ ($USp(2N)$) gauge
theory on their worldvolume for an orientifold of negative (positive)
charge. The best known example is the type I string. If we mod
out IIB just by world-sheet parity we basically produce an O9. Since
this is a space-filling brane we have to cancel the RR charge, forcing
us to use the negatively charged orientifold with 32 D-branes on
top of it, yielding an $SO(32)$ gauge theory, as expected.

\subsection{Compactifications and D-branes}

There is a seemingly different way that D-branes can be described by
gauge theories. If we consider compactifications of string theory,
we will have non-perturbative states in the resulting lower-dimensional
theory from D-branes wrapping cycles of the compactification
manifold. The mass of these states is just given by the tension
of the brane times the volume of the cycle (and therefore has the
$1/g_s$ dependence signalling a non-perturbative state). At certain
points in the moduli space of compactifications 
some of these cycles may shrink to zero 
size, leading to new massless states in the low-energy theory. Some of
these states are usually massless vectors, giving rise to non-perturbative
gauge groups.

\vspace{1cm}
\fig{Non-perturbative states from D-branes on shrinking cycles}
{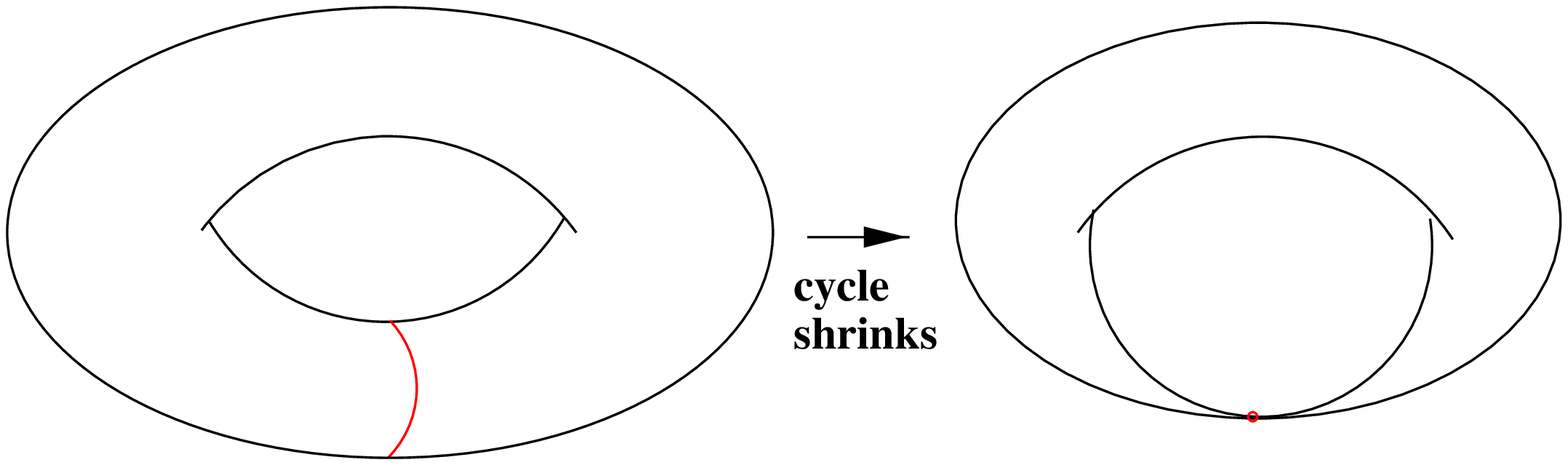}{10truecm}
\figlabel{\dbrane}
\vspace{1cm}

\section{Engineering Gauge theories}
With the two mechanisms at hand we can try to engineer gauge theories,
that is we make up a string theory geometry with branes that realize
a certain gauge theory we want to study. Combining
the two basic mechanisms discussed above 
in various ways there are several possibilities
to do so. Basically all these different approaches described in
the literature can be separated in three classes. As I will
discuss in the last chapter they are actually
equivalent. There I will also give a more technical discussion
for the specific case of ${\cal N} =2$ theories in 4d.

\subsection{Geometric Engineering}

A geometric engineer tries to cook up a string background
that
captures all aspects of the gauge theory she wants to study in
the geometry of the compactification manifold.
In 
order to focus on the gauge theory modes, one has to decouple all stringy
modes and all bulk modes. That is one has to send $M_s$ and $M_{pl}$
to infinity. Let me for simplicity of notation discuss the case
of a K3 compactification ( see \cite{aspinwall} and
references therein), engineering an ${\cal N}=(1,1)$
or ${\cal N}=(1,0)$ supersymmetric gauge
theory in 6d for type IIA or the heterotic string respectively.
It should be clear that these principles work the same way
in other examples.
The relevant scale is the 6d Planck scale given according to a KK
ansatz by
\begin{equation}
M_{Pl,6}^4 = V_{K3} M_{Pl,10}^8.
\end{equation}
Decoupling gravity therefore effectively amounts to decompactifying
the K3. Since the 6d gauge coupling of the perturbative gauge groups
already present in 10d are also given via the KK ansatz as
\begin{equation}
g^{-2}_{YM,6} = V_{K3} g^{-2}_{YM,10}
\end{equation}
they decouple in the same decompactification limit.
The only gauge groups that survive are the non-perturbative gauge
groups that arise via wrapping branes around vanishing cycles in the 
manifold. All information about the gauge theory is therefore encoded
in the local singularity structure of the K3.

The basic example is IIA on an ALE space, that is a non-compact version
of K3. An ALE is a blow-up of an $R^4/\Gamma$ orbifold, where $\Gamma$
is a discrete 
subgroup of $SU(2)$. Since spinors transform as a $(2,1) + (1,2)$
under the $SO(4) = SU(2) \times SU(2)$ spacetime rotations, embedding
the orbifold in just one of the $SU(2)$ factors leaves half of the
spinors invariant and hence also half of the supersymmetries unbroken.
Since K3 can 
be written as an orbifold of $T^4$, these orbifold singularities
can arise locally in the geometry of K3. The statement that $\Gamma$ should
be a subgroup of $SU(2)$ is equivalent to demanding that the holonomies of
K3 only fill up $SU(2)$ and not the full $SO(4)$ of a generic 4d manifold.
In order to obtain gauge dynamics, the local description in terms
of the ALE is all we need. We expect new gauge dynamics when
we move to the singular point, the orbifold itself.

The ALE space has topological non-trivial cycles.
\vspace{1cm}
\fig{Non-trivial 1-cycle on $R^2/Z_6$.}
{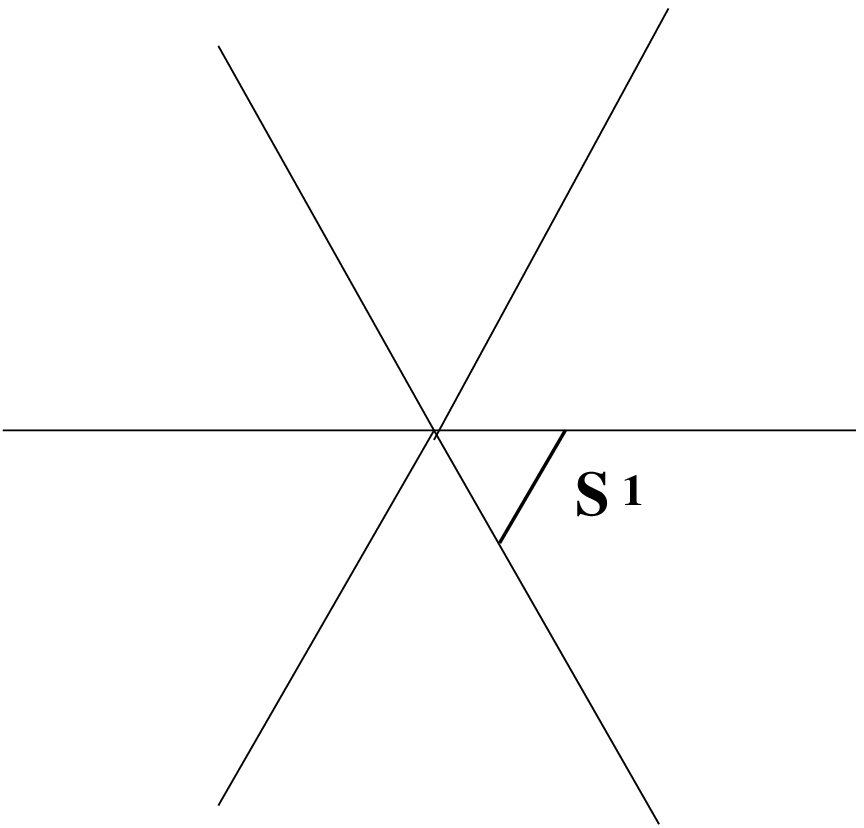}{6truecm}
\figlabel{\cycles}
\vspace{1cm}
Figure \cycles illustrates non-trivial $S^1$s arising from an
$R^2/ Z_6$ orbifold. Similarly we get 2-spheres on the
ALE. These 2-spheres shrink to zero size at the orbifold point. New
massless states arise from D2 branes wrapping these cycles.
The intersection pattern of the 2-cycles will determine the gauge group.
Luckily all discrete subgroups of $SU(2)$ can be classified by
an ADE pattern, where the corresponding Dynkin diagram gives us
precisely the information about the intersection numbers of the
vanishing spheres. The resulting gauge theory has a non-abelian
ADE gauge group.

\subsection{Branes as Probes}

``Branes as Probes'' is the most natural way if we want to learn something
about string theory from Yang-Mills theory. The idea is that in order
to study what happens to a given string background once
one takes into account all the quantum effects, one probes the background
with a D-brane \footnote{In this language one
could view string theory as we used to know it as probing space-time with
a fundamental string.}. On the worldvolume of the D-brane we will as usual find
a gauge theory. The background geometry will be encoded in this gauge
theory via the matter content, the amount of unbroken supersymmetry and
the interaction potentials. Solving the quantum gauge theory will
teach us about the quantum behaviour of the background. This
technique has been very successfully used for probing
D$p+4$ branes and O$p+4$ planes with D$p$ branes \cite{
Probes,Seiberg3,seiberg5d,seiberg6d,senf}, as well as
probing orbifold singularities with D$p$ branes \cite{douglasmoore, int}.

In both cases the matter content and classical superpotential of the gauge
theory can be analyzed by perturbative string theory. For
the higher p branes, we will find new states on the D$p$ worldvolume
corresponding to the zero modes of strings stretching between
D$p$ and D$p+4$ branes in addition to the gauge multiplet already
present from the D$p$-D$p$ strings.

In the case of the orbifold we first include all the twisted sectors
required in string theory for consistency by including all the mirror
D-branes and strings stretching in between them. Then we project
onto states invariant under the orbifold group and this way
obtain the corresponding spectrum.

\subsection{Hanany-Witten setups}

Hanany and Witten (HW) introduced a setup of intersecting branes realizing
$d=3$ ${\cal N}=4$ gauge theories. The gauge theory again lives on
the worldvolume of D-branes. The other branes make
the gauge theory interesting by breaking SUSY and introducing new matter.
Since we are now only dealing with flat branes in flat space,
many things become very intuitive. Moduli and parameters just
correspond to moving the branes around and are very easy to visualize.
As advertised above I will show in the end, that all the 3 approaches
are actually equivalent, so by studying the intuitive HW setups
we can get non-trivial results about quantum string backgrounds
by considering the ``dual'' branes as probes setup.
The next chapter is devoted to an extensive review of the
HW idea, so I won't go into any details at this point.

\section{D-branes and dualities}
\subsection{String Dualities and M-theory}
Probably the most important application of D-branes so far
is the idea of string-dualities, the statement
that one and the same physical system has
two dual descriptions. The concept of duality
was already discussed long ago in the context of field theories,
as I will explain in more detail in the
Chapter 3. In string theory duality was first detected in the
form of T-duality \cite{tduality}. Studying the spectrum of bosonic string
theory on a circle of radius $R$

\begin{eqnarray}
H=\frac{1}{2} p_R^2 + p_L^2 +\mbox{oscillators} \\
\nonumber p_R= \frac{1}{\sqrt{2}} ( \frac{l_s}{R} n - \frac{R}
{l_s} m) \\
\nonumber p_L= \frac{1}{\sqrt{2}} ( \frac{l_s}{R} n + \frac{R}
{l_s} m)
\end{eqnarray}

One sees that due to the presence of winding modes characterized
by the integer $m$ as well
as momentum modes $n$
around the circle, the states are invariant under an exchange of the two
if one simultaneously takes $R$ into $l_s^2/R$. This invariance under
 $R \rightarrow
1/R$ exchange can be shown to be a symmetry of amplitudes to all
orders in perturbation theory and is believed to be valid
even non-perturbatively. The two compactifications
are T-dual to each other. For the superstring this T-duality
works almost the same. For example type IIA on $R$ is dual
to IIB on $l_s^2/R$.
In this case the $p+1$ form fields from the RR sector
T-dualize into $p+2$ and $p$ form fields, depending on whether
we take the components along or transverse to the compact direction.
Since the D$p$ branes couple to these fields, T-duality transverse to
the worldvolume produces a $Dp+1$ brane while T-duality along
a worldvolume direction leaves us with a $Dp-1$ brane.

More interesting are dualities relating one string theory
at weak coupling to another string theory at strong coupling.
Many dualities of this type have been  discovered over the
recent years. However non of them can be proven by a direct calculation.
Since by definition we compare a strongly coupled with a weakly coupled
theory, only one side is accessible to calculations. Duality then
amounts to a prediction for the strong coupling behaviour of the other
theory. The reason why most string theorists nevertheless believe
in the validity of these dualities is that they can be checked in
several ways. The most important check is the matching of objects
which are BPS. They preserve some fraction of the supersymmetry
and are therefore protected by the superalgebra from
any renormalization.
We have already encountered some of these objects: D-branes.
This way certain properties of these non-perturbative
states which dominate the strong coupling theory can be calculated
and they can be matched onto the perturbative states at weak coupling.

One of the examples I am going to consider several times in this work is
the selfduality of type IIB string theory. Type IIB with
coupling $g_s$ is dual to type IIB with $1/g_s$.
Including the axion $a$ we can build a complex coupling 
$\tau=\frac{a}{2 \pi}+\frac{i}{g_s}$. Combining the $g_s$ to $1/g_s$ duality
with the invariance of the axion under shifts of $2 \pi$, a
whole $SL(2,Z)$ of dual theories can be constructed. The NS
2-form field combines with the RR 2-form into an $SL(2,Z)$ doublet.
The objects coupling to them, the fundamental F1 and the D1 string are
exchanged under the strong-weak coupling duality. More general
$SL(2,Z)$ transformations take the F1 into a $(p,q)$ bound state
of $p$ fundamental and $q$ D-strings. Similarly their magnetic
duals, the NS5 and the D5 brane form an $SL(2,Z)$ doublet.
Since there is only one 4-form field, it has to be a singlet under
$SL(2,Z)$ and hence the D3 brane stays invariant under
all duality transformations. Since the low-energy
effective actions of the dual theories are supposed to agree,
the Planck scale has to remain invariant, therefore
using (\ref{pls})
the dual string scale has to be $M_s g_s^{1/2}$.

Basically all string dualities can be summarized as the existence
of an conjectural 11d theory, called M-theory, which contains all
the string theories as well as 11d SUGRA as perturbative expansions
in certain limits.

\vspace{1cm}
\fig{All known string theories as well as 11d SUGRA are just
different perturbative expansions of an overarching 11d M-theory.}
{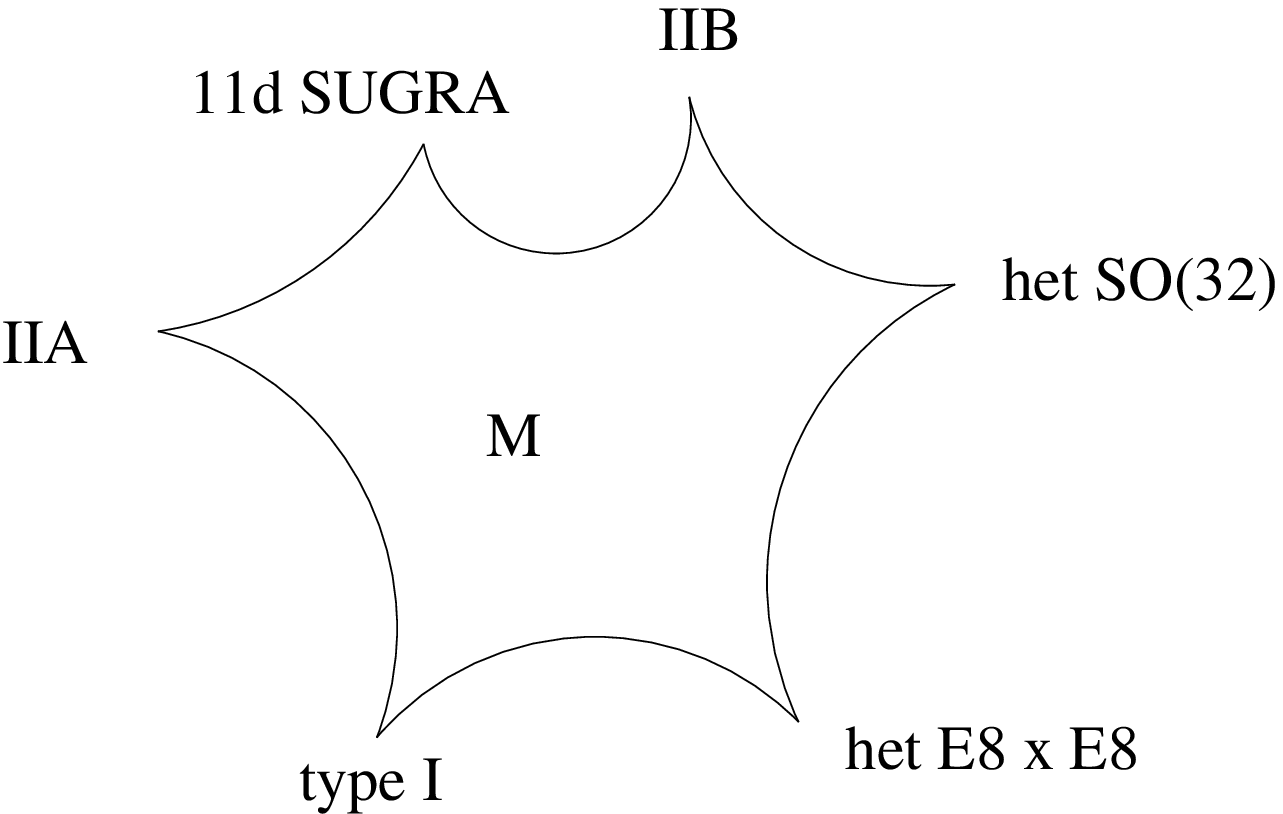}{10truecm}
\figlabel{\mtheorie}
\vspace{1cm}

Since I am going to use this M-theory picture in what follows, let
me briefly present as a defining duality of M-theory the duality
between 11d SUGRA and type IIA, which originally led to the discovery
of M-theory \cite{various,townsendm}.
According to this proposal M-theory on a circle
is type IIA string theory with the IIA coupling and string length
given in terms of the 11d Planck length and the radius of the 11th
dimension $R$ as
\begin{equation}
\label{11d}
g_s^2 = \frac{R^3}{l_{pl}^3} \; \; \; l_s^2 = \frac{l_{pl}^3}{R}
\end{equation}
These relations can be obtained by comparing the low-energy effective actions.
The relation really constitutes a strong-weak coupling duality: at
very large $R$ IIA becomes strongly coupled and we lose all control.
However in the 11d picture as $R$ becomes bigger the curvature becomes
smaller and SUGRA becomes a good approximation. Similar at very small $R$
the curvatures are Planckian in 11d, so SUGRA fails to capture the physics,
however perturbative string theory is a good description. To
describe couplings of order 1, we need the yet unknown full fledged M-theory.

The appearance of the 11th dimension can be seen from studying D-branes.
D0 branes are non-perturbative states, whose mass goes to zero in
the strong coupling limit. $N$ D0 branes are believed to form
a unique threshold bound state (that is with zero binding energy)
\cite{boundstates,sethistern}.
They therefore led to a tower of states with mass $\frac{N}{g_s l_s}$.
It is natural to identify these as momentum modes around the 11th dimension
of radius $R=\frac{1}{g_s l_s}$.
M-theory also provides us with a nice organization principle for
all the other branes. From 11d SUGRA we learn that M-theory has two
extended objects, the M2 and the M5 brane. Together with three more
complicated solutions that only arise upon compactification of at least one
more direction, the wave (momentum mode around the circle) with
mass $1/R$, its magnetic
dual, the KK monopole 6 brane with tension $R^2/l_{pl}^9$ and an M9 brane
with tension $R^3/l_{pl}^{12}$, they give rise to all brane solutions
in the perturbative limits of M-theory.

\subsection{Matrix Theory}

Having said the above, it would clearly be desirable to find a microscopic
definition of M-theory. The only candidate that has emerged so far
is matrix theory \cite{bfss}. The idea behind this approach is to
quantize the theory in a special frame, called the infinite momentum frame,
where only a very limited amount of the original degrees of freedom are
visible. What we do is boost ourselves as observers infinitely along
a compact direction, so that of all modes with momentum $N/R$ only
those with positive $N$ survive. This has to be considered as
$N$ and $R$ both go to infinity with $N/R$ also going to infinity.

Since we want to work at finite $N$ to do any realistic computation,
one would like to study a reference frame that is described by finite
$N$ matrix theory and reduces to the IMF in the $N \rightarrow 
\infty$ limit. Such a frame exists, the discrete light cone frame.
Therefore we want to study discrete lightcone quantization (DLCQ) of
M-theory. This was conjectured to be described by the finite $N$
matrix model in \cite{finite}. DLCQ formally can be thought of as
quantizing the theory on a compact lightlike circle. 
This notion seems to be rather counterintuitive. Indeed it was
shown in \cite{seisen} that the best way to think about the DLCQ is to
consider it as a limit of a compactification on an almost lightlike
circle, that is
\begin{equation}
{x \choose t} \sim {x \choose t} +{ \sqrt{ {R^2\over 2} +
R_s^2} \choose - {R\over \sqrt{2}} } \approx  {x \choose t} +{ {R
\over \sqrt{2}}  + {R_s^2 \over \sqrt{2} R} \choose - {R \over
\sqrt{2}}}
\end{equation}
where $R_s << R$. For $R_s \rightarrow 0$ this reduces to a lightlike
compactification with radius $R$.
Now this compactification is just a Lorentz boost transform
of an ordinary spacelike compactification on $R_s$ with
boost parameter
\begin{equation}
\beta = {R \over \sqrt{R^2+2 R_s^2}}\approx 1-
{R_s^2 \over R^2}.
\end{equation}
Therefore Seiberg's final result can be stated as follows: DLCQ of any system
is Lorentz equivalent to the $R_s \rightarrow 0$ limit of a
spacelike compactification on $R_s$. But we know what M-theory
on a vanishing circle is: it is just weakly coupled IIA!
Since this is a rather
familiar theory, it is very easy to identify the relevant degrees of freedom.

So matrix theory is just the weak coupling limit of type IIA string theory.
The relevant degrees of freedom surviving are the carriers of positive
momentum around the (vanishing) circle in the 11th dimension. But we
already identified them as D0 branes. Studying matrix theory
of compactified M-theory, we find that the limit on the IIA side
shrinks all the radii to zero, forcing us to perform a T-duality.
This way the D0 branes turn into different branes. 
But after all one finds that matrix theory is defined via the worldvolume
theory of a certain brane. Some of these are rather exotic. E.g.
matrix theory on the $T^4$ is described by a D4 brane at very strong
coupling, so that it turns into an M5 brane
\cite{rozali,brs}. Similar the $T^5$ compactification
is described by D5 branes at strong coupling and hence IIB NS5 branes
at weak coupling \cite{newstrings}.
The $T^6$ is described by D6 branes at strong coupling
\cite{mand,mhan}.
However in this case one automatically keeps some of the bulk
modes, so that the interacting world volume theory is not
decoupled from gravity \footnote{Existence of limits
decoupling the bulk while keeping an interacting theory on the brane
are required for the ``brane proof'' of the existence of certain
fixed points in higher dimensions. I will discuss the existence
and non-existence of these limits in the following chapters.}.
Higher dimensional compactifications are plagued by similar problems. Even 
though we can still define matrix theory as the worldvolume theory
of some brane, this description is not any more useful than saying
M-theory is a consistent 11d theory of gravity, since the
corresponding worldvolume theories do not decouple from the
bulk gravitons in the matrix limit
\footnote{Some doubts have been voiced, whether it is legitimate
to neglect all the effects of the modes with zero momentum around the
compact circle, which became infinitely heavy and were integrated out
by keeping only the positive momentum modes. Doing field theory,
these zero modes carry all the information about the non-trivial
vacuum structure. In DLCQ the vacuum is trivial. 
The description we presented so far
might have to be modified due to the effects
of the zero modes.}.

The matrix conjecture this way elevates the SYM / non-perturbative
string theory correspondence to a principle: not only does
the worldvolume 
SYM capture important aspects of non-perturbative string theory,
it is used as a definition of ``all of M-theory''. 
So learning something about the worldvolume theory of branes
will always automatically bring us a step further towards the
goal of understanding the fundamental theory of everything.

By analyzing the size of D0 bound states some evidence can be found that 
matrix theory is holographic, that is information in a given
space-time volume grows like the area surrounding the volume, not
like the volume itself
\cite{holo1,holo2}. This is supposed to be a genuine property of
quantum gravity. The idea is that the best you can do is to fill
up your volume with a big black hole and the entropy of the
black hole grows with its horizon area. Until recently it has
been totally unclear how such a principle could be implemented
in string theory. Maldacena's proposal
\cite{Maldacena} led to a beautiful
realization of holography in spaces with negative cosmological
constant. So far matrix theory is our only candidate for a
holographic description of Minkowski space.

\subsection{Worldvolume theory of the NS5 brane}

As we have seen it is easy to understand the worldvolume theory of the
D-branes from analyzing the modes of strings with Dirichlet boundary
conditions. The NS5 and M5 branes are a little bit more elusive, but
using dualities we can say something about them, too.
First consider the IIB NS5 brane. Type IIB is selfdual, that is
type IIB with coupling $g_s$ and string scale $M_s$ is dual to type IIB with
coupling $1/g_s$ and string scale $M_s g_s^{1/2}$, holding $M^4_{pl}=
M_s^4 g_s$ fixed. The dual theory is the theory of IIB D1 strings, which
play the role of fundamental strings at strong coupling. This
duality takes NS5 into D5 brane. From this we learn that as the D5 the
NS5 will be governed by 6d SYM with gauge coupling
\begin{equation}
g^2_{NS5}= \frac{1}{M_s^2}.
\end{equation}
The supersymmetries preserved by an flat NS5 brane living in 012345 space
will satisfy
\begin{equation}
\label{SUSYNS}
\epsilon_L=\Gamma_0 \Gamma_1 \Gamma_2 \Gamma_3
\Gamma_4 \Gamma_5 \epsilon_L \; \; \; \;
\epsilon_R= \pm \Gamma_0 \Gamma_1 \Gamma_2 \Gamma_3
\Gamma_4 \Gamma_5 \epsilon_R 
\end{equation}
where the $\pm$ depends on whether we consider IIA or IIB.

The M5 brane can be best understood in the limit where M-theory
is well described by 11d SUGRA. Here $N$ coinciding M5 
branes can just be thought
of as a soliton given by the following SUGRA solution, 
with $F$ being the field strength
of the 3-form vector potential \cite{horr}:
\begin{eqnarray}
ds^2&=&f^{-1/3} dx^2_{||} + f^{2/3} \left ( dr^2 + r^2 d \Omega^2_4 \right
) \\ \nn
F_{\alpha_1 \ldots \alpha_4} &=&\frac{1}{2} \epsilon_{\alpha_1 \ldots
\alpha_4} \partial_{\alpha_5} H \\ \nn
f&=&1+\frac{\pi N l_p^3}{r^3} .
\end{eqnarray}
Analyzing the zero modes of this soliton one can calculate that the
worldvolume supports a 6d ${\cal N}=(2,0)$ supersymmetric
tensor multiplet. From 11d SUGRA - IIA duality we immediately
learn that the IIA NS5 brane hence also supports a (2,0)
tensor multiplet. However this time one of the 5 scalars in
the tensor multiplet lives on a circle (the one parametrizing
the position of the 5brane in the 11th dimension). Using
a normalization in which the scalar has mass dimension 2
\footnote{This is the natural normalization since the scalars sit
in the same multiplet as the 2-form $B_{\mu \nu}$ which must
have mass dimension 2, so that the $B$ Wilson line is dimensionless}
the radius of this circle is $M_s^2$.

\chapter{Exploiting the SYM D-brane correspondence}

\section{Classical Hanany-Witten setups}

After we have learned in principle
how to use the SYM D-brane correspondence
in order to engineer certain gauge theories in a stringy setup,
we now want to exploit this correspondence and study
possible applications. One of the
most prominent and intuitive setups used to learn about gauge
theories from string theory is the Hanany-Witten (HW) setup
\cite{HW}. Let
me briefly review the basic ideas. A very exhaustive review
of these setups and their applications can be found in
\cite{givkut}.

\subsection{Branegineering}

The idea behind Hanany-Witten setups is to study branes in flat space
and get interesting gauge theories by having many flat branes intersecting
each other. The matter content can be determined by simple, intuitive rules.
Similarly deformations and moduli become easily visible. In the last
chapter of this work I will establish a dictionary mapping HW setups
to the other approaches, where the interesting dynamics is hidden in 
the background geometry. This way HW setups can
be used to encode complicated looking information about deformations and
phase
transition in harmless looking brane moves. In several cases aspects of
string theory in the background of the intersecting branes can be solved,
leading to interesting results about the quantum gauge theory on the brane.

We start with the maximally supersymmetric (16 supercharges) SYM on
the worldvolume of a D$p$ brane. For definiteness let me discuss the case $p=4$.
\footnote{As compared to $p=3$ in the original work of \cite{HW}.}
The idea will be the same in the other dimensions. 

In order to go to more interesting physics with lower supersymmetry,
we have to project out some of the degrees of freedom. This
can be achieved by letting our D4 branes, which I henceforth will call
color branes ($N_c$ of those will give rise to SUSY QCD with $N_c$ colors),
end on another brane. The boundary conditions will do the job
of projecting out certain states.

The concept of a brane ending on branes is a straight forward generalization
of the defining property of D-branes as an object on which strings can
end. Indeed all of the ``brane ends on brane'' configurations
used in this work can be obtained from this defining setup via
duality:
\vspace{1cm}
\fig{Configurations dual to a fundamental string ending on a
D-brane.
}
{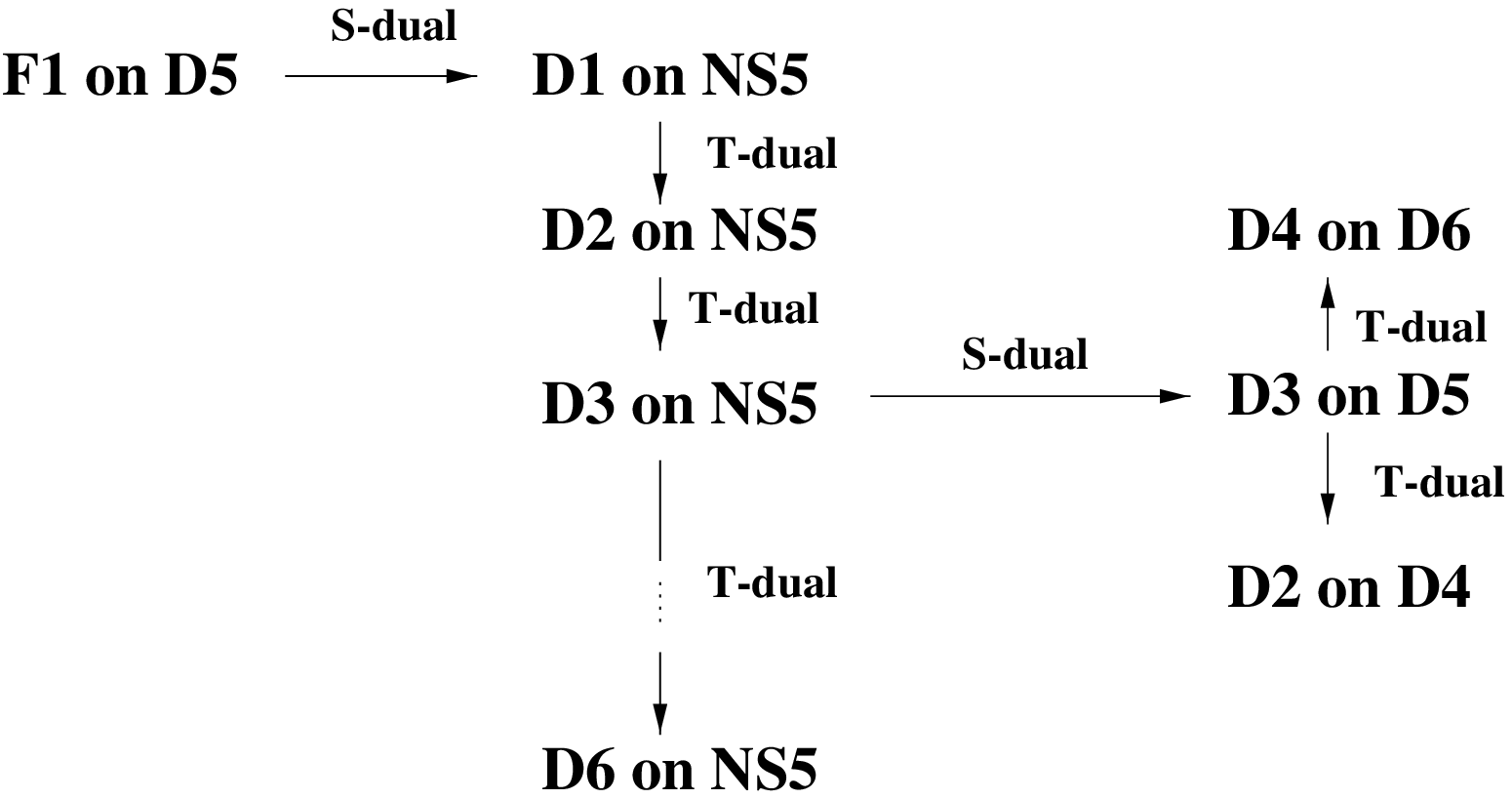}{10truecm}
\figlabel{\who}
\vspace{1cm}
Figure \who illustrates this chain of dualities.
Another way to understand the who-ends-on-whom rules is to study
the worldvolume theories. The end of a brane is a charged object.
In order for a given brane to be allowed to end on another brane,
there should better be a field on the worldvolume that can carry
away that charge. In the case of the fundamental string ending
on the D-brane, the end of the string is charged electrically
under the worldvolume gauge field. Similarly we can explain the other
setups of figure \who. For example the
D2 brane ending on a IIA NS5 brane is
the string like object
charged under the 2-index tensor gauge field on the worldvolume,
the D3 brane ending on the NS5 or D5 is the magnetic monopole on the
worldvolume (which is a 2-brane in 6 dimensions).

In addition the non-vanishing field strength induced on the brane
couples via the DBI (\ref{DBI}) action to the scalars which describe the
embedding of the brane and the brane has to bend in order to compensate
the force exerted by the gauge field. Indeed it was shown in
\cite{callmald} that the DBI action allows for stable soliton
solutions which can be interpreted as a string ending on a D-brane.
Much can be understood
about bending just on the base of symmetry arguments.
As we will see in the following bending is really a quantum effect in 
the field theory. So for now I will neglect bending and only
discuss the ``classical'' setup.

We choose to let our D4 branes end on two NS5 branes. This way the D4
brane stretches only over a finite interval bounded by the NS5 branes.
Usually one chooses this finite interval to be in the 6 direction.
This will do several things for us: for one certain degrees of freedom
will be projected out by the boundary conditions, as desired. In addition,
since our gauge theory now lives on a compact interval, the
low energy theory will be governed by the KK-reduction on the interval.
Thus effectively we will be dealing with a $p$ dimensional theory.
Last but not least the addition of the NS5 brane will break some
more supersymmetry. We will remain with 8 supercharges, that is
in the $p=4$ case with ${\cal N}=2$, as can be checked
explicitly using (\ref{SUSYD}) and (\ref{SUSYNS}).
The following table displays
the worldvolume directions of these branes. The additional
D6 brane will be explained soon.

\begin{center}
\vspace{.6cm}
\begin{tabular}{|c||c|c|c|c|c|c|c|c|c|c|}
\hline
&$x^0$&$x^1$&$x^2$&$x^3$&$x^4$&$x^5$&$x^6$&$x^7$&$x^8$&$x^9$\\
\hline
NS 5&x&x&x&x&x&x&o&o&o&o\\
\hline
D 4&x&x&x&x&o&o&x&o&o&o\\
\hline
D 6&x&x&x&x&o&o&o&x&x&x\\
\hline
\end{tabular}
\vspace{.6cm}
\end{center}

The presence of these branes breaks the Lorentz symmetry group
$SO(9,1)$ down to $SO(3,1) \times SO(2) \times SO(3)$. While
the first part is our obvious Lorentz group in 4d, the rest
should better have an interpretation as the R symmetry of
the corresponding superalgebra. It has to be an R symmetry
due to the fact that we interpreted the worldvolume scalars
as positions of the branes in this ``internal'' part of spacetime,
therefore the scalars and fermions on the worldvolume
naturally transform like vectors and spinors under this internal
Lorentz group. A symmetry group acting differently on the
fermions and scalars in a supermultiplet is an R symmetry.
Indeed the R symmetry of the ${\cal N}=2$ SUSY algebra
is $U(2) = SU(2) \times U(1) = SO(3) \times SO(2)$.

To determine from perturbative string theory
which degrees of freedom are projected out would require an analysis
in the background of NS5 branes. Luckily there is an easier way
to figure out what is going on, by remembering once again
that the scalars describe
the position of the brane. Without the NS5 branes, we had 5 scalars
in the 5d ${\cal N}=2$ vectormultiplet, describing the position of
the D4 brane in 45789 space. In addition we will get a 6th scalar after
KK reduction from the 6-component of the vector itself. Under 
the 4d ${\cal N}=2$ SUSY 4 of these
scalars constitute the bosonic part of a hypermultiplet (HM),
while the vector
together with the 2 other scalars forms the bosonic part of the 
vectormultiplet (VM). Now requiring the D4 brane to end on the NS5 brane fixes
its position in 789 space. The only scalars surviving are the 45 motion.
From supersymmetry it than follows, that the surviving multiplet is
the VM and the HM is projected out. The same will be true for any D brane
ending on NS5 branes. So indeed in order to branegineer gauge
theories a D brane ending on NS5 branes will be the starting building block.
The gauge coupling of the gauge theory is also encoded in this
simple setup. Due to the KK reduction the inverse gauge
coupling is proportional
to the distance $L$ between the 2 NS5 branes. Using (\ref{gYM})
the precise value for the gauge coupling
of the $p$ dimensional gauge theory from
a D$p$ brane suspended between two NS5 branes is
\begin{equation}
\label{gHW}
g^2_{YM}=\frac{l_s^{p-3} g_s}{L}
\end{equation}
Moving the brane along the 45 direction turns
on vevs for these scalars. This corresponds to moving
on the Coulomb branch of the gauge theory. At generic points
of the Coulomb branch all the D4 branes will be at a different
45 position and we are left with an unbroken $U(1)^{\mbox{rank}}$
gauge group, justifying the name.

As already indicated in the brane table above, using (\ref{SUSYD})
and (\ref{SUSYNS})
there is one more brane
one can add without spoiling any further SUSY, a D6 brane.
Analyzing the possible brane motions one finds, that a D4
between two D6 branes supports a hypermultiplet while
the vector is this time projected out. A D4 suspended
from NS5 to D6 is stuck and hence supports no scalars
and since we have ${\cal N}=2$ this means no degrees of freedom at all.
So in order to have a gauge theory we will keep having the
D4 end on the two NS5 branes. Is there any new multiplet that arise from
the presence of $N_f$ ``flavor'' D6 branes? The D4 D6 system only
contains D-branes, so it is easy to deal with. 
In addition to the strings
ending only on the D4 brane, we will have strings ending on the
D4 and on the D6. Analyzing their massless sector one finds
$N_f$ hypermultiplets in the fundamental representation of $SU(N_c)$.
Naively one can understand this from the fact that such strings will
have an $N_c$ Chan Paton factor on their one end and an $N_f$
Chan Paton factor on their other end. 

The dynamics of this system in the IR will only be determined
by the lightest branes. In the case of HW setups these will
be the lowest dimensional objects around, that is the D4 branes.
The $SU(N_f)$ symmetry is really a global symmetry, the 
gauge bosons from D6 D6 strings decouple.
This general philosophy that only the smallest brane contributes
to the dynamics carries over to HW setups in other dimensions,
where we suspend D$p$ color branes between NS branes with $Dp+2$
branes taking over the role of flavor branes. Motions of
light branes will correspond to moduli, while motions
of heavy branes are parameters of the theory.

Let us consider the limits involved in more detail.
Let me discuss the case of D3 branes between NS5 branes
and flavor D5 branes as in the original work of \cite{HW}.
Since this is in type IIB the worldvolume of the NS5 brane
supports just SYM and is easier to discuss. 

In order to decouple
gravity and higher string modes we send $M_{pl}$ and $M_s$ to infinity.
We hold $g^2_{YM}=g_s/L$ fixed. This sets the scale for all the gauge
theory modes. In order to decouple the Kaluza Klein modes from the
interval we want this to be much less than $1/L$, that is we
have to consider weak string coupling. Indeed in this limit the gauge
coupling
on the NS5 and D5 branes, which is $1/M_s^2$ and
$g_s/M_s^2$ goes to zero, justifying in a quantitative manner
our assertion that at low energies the only surviving
modes are those of the lowest dimensional brane, leading
to a 3d gauge theory. 

To summarize, we have shown that in the decoupling limit
\begin{equation}
\label{decouple}
l_s \rightarrow 0, \; \; l_{pl} \rightarrow 0, \; \;
L \rightarrow 0, \; \; g^2_{YM} \mbox{ fixed}
\end{equation}
the HW setup with $N_c$ D$p$ color and $N_f$ D$p+2$ flavor branes
describes an interacting p dimensional theory
with $SU(N_c)$ gauge group and $N_f$ flavors.

\subsection{There's so much more one can do}

So far we allowed ourselves to freely jump between dimensions
in order to relate the classical setups. This is indeed possible by
applying a simple T-duality to the original 3d setup. As long as
we act inside the worldvolume of the NS5 brane it will stay an
NS5 brane, while the color and flavor branes loose or gain
a dimension whether we T-dualize a worldvolume or a transverse direction.
The maximal dimension we can achieve this way is 6 from D6 branes
between NS5 branes. These will be the main focus of this work.
Gauge theories in more than 6 dimensions have a minimal of 16
supercharges. This is a classical constraint just following
from the size of the spinor representation of $SO(6,1)$. 
Our branes know all this. In any dimension the 8 supercharge
HW setup looks as follows:

\begin{center}
\vspace{.6cm}
\begin{tabular}{|c||cccccc|c|c|c|c|}
\hline
&$x^0$&$x^1$&$x^2$&$x^3$&$x^4$&$x^5$&$x^6$&$x^7$&$x^8$&$x^9$\\
\hline
NS 5&x&x&x&x&x&x&o&o&o&o\\
\hline
D$p$&x&$\ldots$&up& to&$x^{p-1}$&$\ldots$&x&o&o&o\\
\hline
D$p+2$&x&$\ldots$&up& to&$x^{p-1}$&$\ldots$&o&x&x&x\\
\hline
\end{tabular}
\vspace{.6cm}
\end{center}

In order to go down to 4 supercharges we need yet another kind of
brane. One thing we can do is to rotate one of the players already
present \cite{egk,rotate}.
Checking the unbroken supersymmetries according to
(\ref{SUSYD}) and (\ref{SUSYNS}) we see that for example
an NS5' brane living in 012389 space will do the job. It will
break 1/2 of the supersymmetries presented in the setup so far.
In addition the $SO(3)=SU(2)$ part of the R-symmetry corresponding
to rotations in the 789 plane is broken to $SO(2)=U(1)$ as required.
Analyzing the massless modes on a D4 brane suspended between
NS and NS' we now find that all the scalars are locked. Only
the ${\cal N}=1$ vector multiplet survives the projection.
The same amount of SUSY will be preserved
if we choose to rotate the second NS5 brane by any nonzero angle $\theta$
in the 4589 plane. The adjoint chiral multiplet from the decomposition
of the ${\cal N}=2$ vector multiplet under ${\cal N}=1$ will receive
a mass $m= \tan \theta$. For $\theta=0$ one recovers the ${\cal N}=2$ theory,
for $\theta = \pi/2$ the adjoint decouples all together.
In \cite{egk} they also found that when we use $k$ coinciding
NS5' branes a superpotential $W=X^{k+1}$ is created rather
than the mass term $X^2$ which we obtained for $k=1$.
One can argue for a term like this by studying which flat directions
such a term lifts in the classical field theory and than comparing
with the possible brane motions.
The second possibility is rotating the flavor branes \cite{aharony}.
This will leave the matter content untouched. Instead the
superpotential $W= X Q \tilde{Q}$ required for
an ${\cal N}=2$ theory will be turned off continously with
$\theta$. 

All these rotations are 
only possible for HW setups in 4 and lower dimensions, again
reflecting the fact that 4 is the maximal dimension for SYM with
4 supercharges. Introducing even more branes with other rotations we can
as well engineer gauge theories with 2 and 1 supercharges in 3 and 2
dimensions. Of course it is no problem to also write down
configurations that preserve no supersymmetry at all. A generic
setup will do so. 

Since rotating branes as we have seen just corresponds to perturbing
${\cal N}=2$ theories, only a very restricted class of ${\cal N}=1$
theories may be constructed this way. For example (except for
one exotic exception I will introduce later) no chiral gauge theories
can be constructed this way. In order to do so it is
necessary to generalize the idea of suspending a D$p$ brane
on an interval in order to have a p dimensional gauge theory  to having 
$Dp+1$ branes on a rectangle (a brane box) \cite{HanZaff}
\footnote{An equivalent way of viewing this is to view
NS 5 branes at an orbifold as it was suggested in \cite{poppitz}. For
us it seems more natural to either have the advantages of 
branes as probes or the HW setup. However the hybrid 
construction yields the same answers.}. This
way generic models can be constructed, at least on the classic level.
Recently it has been shown \cite{leighrozali} 
that they are also good as a quantum
theory as long as we choose an anomaly free matter content. This
proof is done in the equivalent brane as probe picture. Since the 
relation between the various approaches will be one of my main subjects
I will postpone a discussion of brane boxes to later
chapters.

Let me at this point introduce another
possible construction used to introduce flavors. It will also be
the basic building block for product gauge groups. The basic
question to analyze is the following: let us consider three
NS5 branes at 3 different positions in the 6 direction. We will put
$N_c$ D4 branes on the first and $N_c'$ D4 branes on the second
interval. What is the low energy field theory corresponding to this? We will
definitely get an $SU(N_c) \times SU(N_c')$ gauge theory. Additional
matter will be produced from the D4 D4 strings stretching
from the $N_c$ to the $N_c'$ branes. They have one Chan Paton
factor in each gauge group. So we would expect to obtain bifundamental
matter that is a hypermultiplet fundamental under both gauge
groups. This can again be verified by checking that the allowed brane
motions correspond to the classic moduli space of the gauge theory.
Taking the third NS brane to infinity the second gauge group decouples
and becomes a global symmetry. 
This way one can introduce flavors via semi-infinite D4
branes to the left or right.

Indeed the two ways of including flavors are related if we take into account
the HW effect: whenever a flavor D5 crosses and NS5 a D3 is created
in between them. This process holds in all the T-dual configurations
as well. 
That brane creation must occur can be seen
by looking at the so called linking number of the NS5 and
D5 brane, which is
a topological invariant. In order to have it unchanged in brane crossing
prcesses, a D3 brane has to be created when D5 and NS5 pass.
So
if we move all the flavor branes to the far left or right, we will
create semi-infinite branes. The number of flavors didn't change, just
their realization. There are many equivalent pictures yielding the
same gauge theory. 

There is one more classical parameter we can introduce: the relative
position of the NS5 branes in the 789 space. This corresponds
to turning on an FI term $\kappa$, that is adding
\begin{equation}
\int d^4 x d^4 \theta \kappa D
\end{equation}
to the Lagrangian, where $D$ is the auxiliary field in the
vector multiplet. Once more one can prove this by analyzing the
possible brane motions in the presence of this term and compare with
classical gauge theory.

We have now all the tools in order to branegineer quite arbitrary
classical gauge groups. The space of all possible
brane motions corresponds to the classical moduli space of the gauge
theory. We already identified the 45 motions as the Coulomb branch.
The Higgs branch is seen if we split the color branes along the
flavor branes. As shown earlier such a brane segment between two
color branes supports the 4 scalars in a hypermultiplet, corresponding
to moving the segment off, leaving a broken gauge group. A generic
point on the Higgs branch then would look as follows:

\vspace{1cm}
\fig{A maximally broken situation on the Higgs branch in a
theory with 4 supercharges. The numbers denote the complex scalar degrees
of freedom associated with motions of the given brane piece. The s-rule
has been taken into account.}
{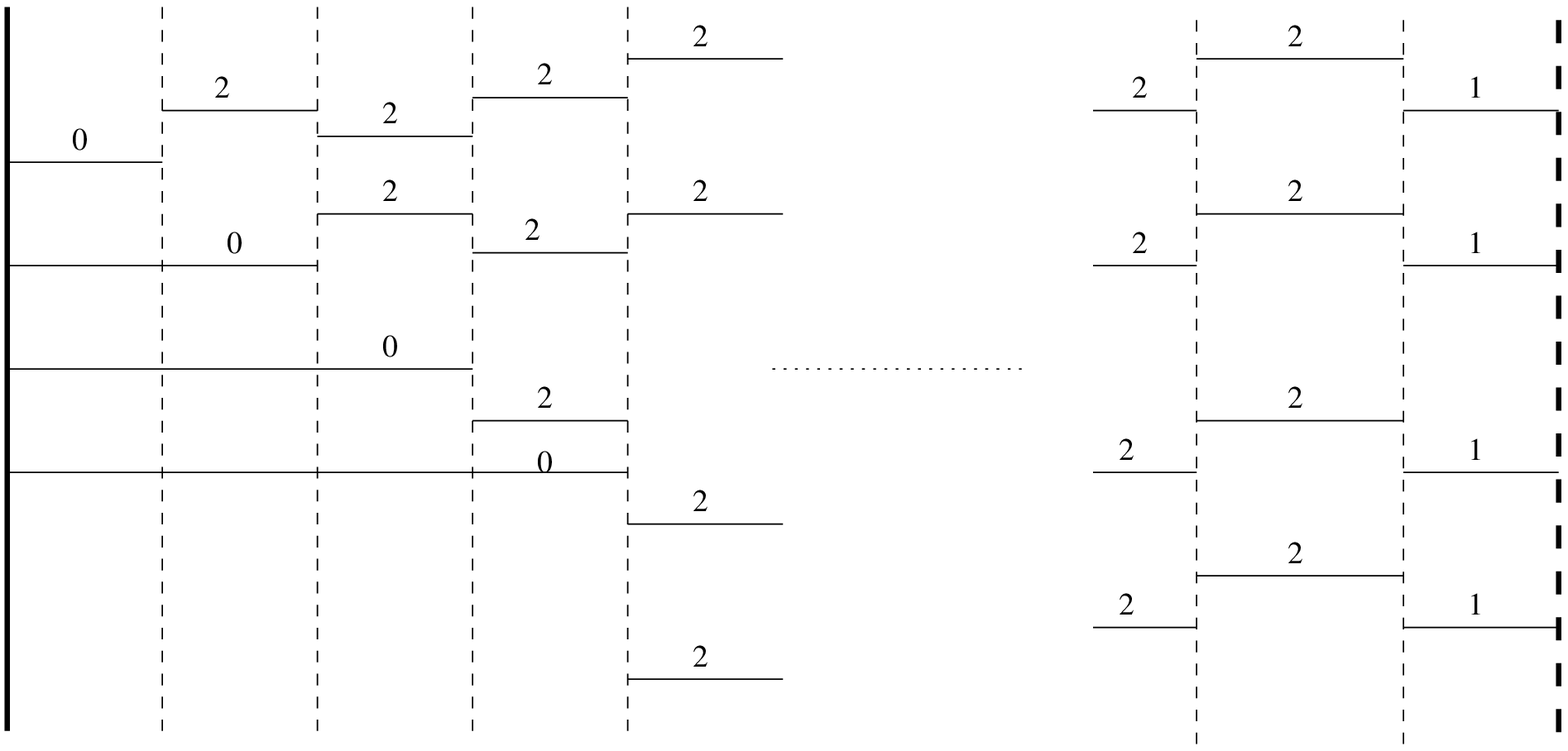}{12truecm}
\figlabel{\higgs}
\vspace{1cm}

In the picture, the fat line is the NS brane, the fat broken
line is the NS' brane, the horizontal lines
denote color and the broken vertical lines denote flavor branes.
For a theory with 8 supercharges,
the NS' has to be replaced by an NS branes and both sides
of the picture look the same. 
The picture does not depend on the dimension (it has to be 4 or
less since we are only dealing with 4 supercharges).
The numbers denote the complex degrees of freedom associated to moving
around the brane pieces. 
In order to reproduce the results from classic gauge theory we have
to take into account the 
so called s-rule \cite{HW}: given a single pair of NS and flavor brane,
only a single color brane is allowed to end on them.
Summing up all the degrees of freedom we find that
the complex dimension of the Higgs branch is (for $N_f \geq N_c$)
\begin{equation}
d_{Higgs}=  (2N_f - N_c) N_c= 2 N_f N_c -N_c^2
\end{equation}
in perfect agreement with the field theory counting, where
we just count the scalar fields that are not eaten by the Higgs
mechanism,
\begin{equation}
d_{Higgs}= \mbox{number of scalars } - \mbox{ number of vectors}
=2 N_f N_c -N_c^2. 
\end{equation}

The Higgs branch
is not visible when we use semi-infinite branes to realize the flavors.
Turning on FI terms (that is separating the NS branes in 789 space
for 8 supercharges or just in 7 space for 4 supercharges) kills
the Coulomb branch of the ${\cal N}=2$ theories, as is expected. Since
the color branes are only allowed to extend in the 6 direction in 
order to preserve SUSY they have to split on the flavor branes, that is we
have to be on the Higgs branch.

\section{Solving the quantum theory}
\subsection{Bending and quantum effects}

Above I argued in a qualitative manner that a brane has to bend 
if another brane ends on it in order to balance the force exerted by
the flux it has to support, since the end of the other brane represents
a charge on its worldvolume. Let me discuss this point in
a bit more of a 
quantitative manner following \cite{wittenn2,callmald}.
Before doing so I'd like to show that bending indeed is
a quantum effect from the point of view of the SYM.

Let me discuss the ${\cal N}=2$ setup in 4d since this was the main
example I discussed on the classical level so far.
The classical value of the Yang-Mills coupling is given
by $g^2_{YM}=\frac{g_s l_s}{L}$ where $L$ again denotes the
length of the interval. In order to focus on the SYM modes
in this setup we want to take the decoupling limit
(\ref{decouple}). The tension of the D4 brane and the NS5 brane
are $\propto \frac{1}{g_s} $ and $\propto \frac{1}{g_s^2} $ respectively.
Since $g_s$ is proportional to $g^2_{YM}$ we see that for
zero coupling the NS5 brane is infinitely heavy and hence is not pulled
or bent by the D4 brane. Turning on the loop corrections in the
gauge theory is reflected in the bending of the branes.

As pointed out earlier, the DBI action couples the scalars and
the vector multiplet, so a non-trivial flux on the brane (as
it is induced by the end of another brane) can be supported
by simultaneously turning on vevs for the scalars, that is
bending the brane since the scalars describe the
position of the brane. We are only interested in the $l_s
\rightarrow 0$ limit, so the SYM approximation of the DBI is sufficient.
We are looking for a stable setup with non-vanishing
flux. One way to construct this is to look at BPS setups.
One can construct them by demanding some preserved supersymmetry.
The no force condition is then guaranteed. This discussion
was carried out in \cite{callmald}.
The same results for the bending as 
found by Callan and Maldacena can be obtained
in a more qualitative fashion following the discussion of Witten
and noting that
\begin{itemize}
\item the bending should be proportional to the net charge,
that is: \newline branes ending on the left - brane ending on the right
\item the scalar terms in the SYM action demands a minimal area
embedding of the brane, therefore the bending should be
a solution of a Laplace equation
\item the bending can only depend on the worldvolume directions
transverse to the end of the other brane and for symmetry reasons
should only depend on the radial distance
\end{itemize}
From this discussion it follows that in order to solve
for the bending of the NS branes in a p dimensional HW setup
we should solve the Laplace equation in the $6-p$ NS5 brane
worldvolume directions transverse to the end of the D$p$ brane,
leading to $r$, log, $1/r$, $\ldots$ bending in 
5,4,3,$\ldots$ dimensions, where $r$ denotes
the radial coordinate along the NS5 brane worldvolume
away from the end of the D$p$ brane. The coefficient in front of the
$r$ dependence will be proportional to the net number of branes.
In 6d the transversal space is zero dimensional. No bending
can occur, no field strength can be supported. The
net number of branes has to be zero.

In theories with 8 supercharges the only perturbative corrections
to the $\beta$ function come from 1-loop. Since the gauge
coupling is encoded in the length of the interval,
the $r^2$, $r$, log, $1/r$, $\ldots$ bending in 6,5,4,3,$\ldots$ dimensions
reflects precisely the known 1-loop running of the gauge coupling
in these dimensions.

There is one peculiar effect due to the bending. As I stated above,
only motions of the light branes are moduli of the setup. Taking
into account the bending we should make this slightly more precise:
only such motions leaving the asymptotic form of the heavy branes untouched
will correspond to moduli. Consider the simplest setup of $N_c$ color
branes on the interval. We argued above that the classic
theory describes $U(N_c)$ gauge group. A generic point on the Coulomb branch
will have an unbroken $U(1)^{N_c}$ corresponding to moving
the $N_c$ branes independently along the NS5 branes. If we
have a $1/r$ or faster fall of in the bending, quantum mechanically
the color branes only created a little dimple on the NS5 and 
the asymptotic of the NS branes will stay unchanged. However
if we have logarithmic or linear bending, changing the center of
mass would correspond to changing the asymptotic behaviour. Therefore
on the quantum level the center of mass $U(1)$ part is frozen
out and we are only left with an $SU(N_c)$ gauge theory.
This reflects the field theory statement that only in 3 and lower
dimensions a $U(1)$ gauge theory can lead to interacting IR physics.

Once we have frozen out the $U(1)$ we have to find a new interpretation
for the 789 position of the branes. An FI term is only possible
for abelian gauge groups (the auxiliary component of the vector
multiplet transforms as an adjoint and is not gauge invariant unless
we are dealing with $U(1)$). On the other hand $SU$ theories
have a new branch: the baryonic branch \footnote{Going
from $SU(N_c)$ to $U(N_c)$ basically corresponds to
gauging baryon number}. Interpreting the 789 position as
a deformation that forces us on this baryonic branch
yields the right moduli spaces when compared with gauge theory
calculations.

In addition to the effects from loops there are also
corrections due to instantons. These instantons can be seen
directly in the brane picture. It is known that a D$p-4$ brane within a D$p$
brane satisfies the 4 dimensional YM instanton equations \cite{BranesWithin}.
So D0 branes are
instantons within D4 branes. To interpret these D0 branes as instantons in
the 0123 spacetime we should consider Euclidean D0 branes whose world-line
stretches along 
the 6 direction so that they are contained within the D4 branes between
the NS5 branes \cite{Barbon1, Barbon2, BrodieD0}.
This way the field theory objects and quantum
effects have been mapped to D-branes and their properties. The problem is now
to solve the
theory after including all these effects.

\subsection{Lifting to M-theory}

So far we have seen how the quantum corrections manifest themselves
in the stringy embedding. Now we are going to actually solve them. This
analysis was performed in the remarkable work of Witten \cite{wittenn2}. To
implement this solution we use the IIA M-theory duality and view our
setup as an 11d setup. For large values of the radius $R$ of the eleventh
dimension 11d SUGRA will be a good approximation. Since
both the NS5 and the D4 lift to an M-theory M5 brane,
our whole setup will be described by just a single
M5 brane in 11d. For this brane to be a
solution of the 11d equations of motion, it will be determined
by the requirement that it lives on a minimal area cycle (this
way Laplace equation sneaks in again). Therefore
the shape of our branes will be solely determined by solving the problem
of soap bubbles: we fix the asymptotics of the branes and they will
arrange themselves to live on a minimal area cycle.

Before I move on and show that indeed all quantum effects, perturbative
as well as non-perturbative, are indeed incorporated in the shape
of the M5 brane in the 11d SUGRA limit, let me first discuss
the validity of this limit. Let me discuss once more the precise decoupling
limit from the IIA point of view. Here we want to decouple gravity,
heavy string modes and KK modes from the finite
interval, that is send $M_s$ and $M_{pl}$ to infinity and $L$ to zero,
while keeping the gauge coupling $g^2_{YM}= \frac{g_s}{M_s L}$ fixed.
Translating into 11d units $g^2_{YM}=\frac{R}{L}$ is supposed to
be fixed as $L$ goes to zero. But this requires us to also take
$R$ to zero. This is the opposite limit of the one where 
we are able to solve!
The only reason why I nevertheless will go on and do some
calculations in what follows is that holomorphic quantities will
be protected and we can calculate them at any value of $R$, even
though they will only correspond to gauge theory quantities in the small
$R$ regime. Also qualitative aspects, like e.g. whether
the theory confines, are supposed to agree.
But it has been shown in \cite{notmatch} that in general unprotected
terms do not agree. This is a pity, since the holomorphic information
is encoded in the SW curve and
was known already from pure field theory considerations \cite{SW}.
The branes only give us a nice organizing principle for analyzing the
holomorphic quantities in complicated situations, where field
theory ``guess and check'' methods like they are usually employed
in order to obtain SW curves do not work anymore, like in situations
with many product groups \cite{wittenn2} or matter content that leads to non
hyperelliptic curves, e.g. two-index symmetric tensors \cite{LL}.

In order to get new information we have to solve the full string theory
in the background of NS and D4 branes, a task that seems too hard with
today's tools. This is a problem that is common to most approaches
of trying to get new gauge information out of string theory, including
the most recent one, the Maldacena large $N$ conjecture, about which
I will make some more comments in what follows. There is always one regime
where we can easily compute and another regime where we want the answer.
However for Maldacena's case to get the full answer we have to do
string theory on $AdS_5 \times S_5$ with some RR flux turned on. This
hasn't been solved so far, but due to the large symmetries of the
background at least there is hope.

Above I have argued that in M-theory we can solve for the exact shape
of the M5 brane by solving the minimal area condition given
a set of boundary conditions, which incorporate the classical
input. Requiring SUSY of the low energy effective action amounts
to restricting to supersymmetric cycles, a special subclass
of minimal area cycles \cite{becker,becker2}.
A cycle is supersymmetric if a brane wrapping it preserves
some amount of supersymmetry. For 2-cycles this
condition directly translates into holomorphicity.
A 2-cycle obtained as the zero-locus of a holomorphic
function of 2 complex coordinates will  preserve 1/2,
as the zero locus of 2 equations in 3 variables 1/4
of the original supersymmetries.

Now let me go ahead and show that indeed all the quantum effects
are incorporated in the shape of the brane. After this I will 
write down the solution as obtained in \cite{wittenn2}.
We already found that the bending of the brane incorporates
the perturbative corrections. In the type IIA setup the
non-perturbative corrections have to be put in by hand. They
are represented by Euclidean D0 branes stretching along
the 6 direction. In order to incorporate all instanton effects
we would have to sum over all setups bound with a given number of D0s.
What becomes of the D0 branes once we lift to 11d? Remember
that the 11d origin of a D0 brane is momentum around
the compact $x^{10}$ circle, that is a bound state with
D0 branes corresponds to adding 10 momentum. The 10 position
of the D4 brane would be a function of time:
\begin{equation}
\dot{x}^{10} \neq 0 \; \; \; \Rightarrow \; \; \; x^{10}=x^{10}(t).
\end{equation}
Our field theory instantons correspond to Euclidean D0 branes,
that is their worldvolume stretches along the finite interval
in the 6 direction instead of stretching in time. According
to the same philosophy they should correspond to ``twist''
in the 11d setup, that is
\begin{equation}
{x'}^{10} \neq 0 \; \; \;\Rightarrow \; \; \; x^{10}=x^{10}(x^6).
\end{equation}
All quantum effects lead to ``bending'' and ``twisting'' in 11d
and are hence incorporated in the shape of the brane.

Now let me move on and present the solution. Let me discuss
the general setup of $n+1$ NS5 branes leading to a product
of $n$ $SU(N_{\alpha})$ gauge groups.
We choose as complex coordinates $t=e^{-s}=e^{-(x^6+ix^{10})/R}$
and $v=x^4 + i x^5$, so that a D4 brane is at $v=const.$ and
an NS5 brane at $t=const.$ This is the only complex structure
in which our two ingredients can be written as holomorphic functions.
Since our 2-cycle asymptotically will look like NS5 or D4,
we have to choose this structure. Taking the exponential
in the definition of $t$ beautifully takes into account
the compactness of $x^{10}$.
Using $x^{10}$ as
$\theta$-parameter one can introduce a complex coupling constant
\begin{equation}
\tau_\alpha={\theta_\alpha\over 2\pi}+i{4\pi\over g^2_\alpha}=
i(s_\alpha-s_{\alpha-1}).
\end{equation}
where the $s_{\alpha}$ denote the positions of the various NS5 branes.
At the one loop level the 1-loop running of the
gauge coupling from the logarithmic bending is simply generalized to
\begin{equation}
s=\sum_i\log (v-a_{i,\alpha})-\sum_j\log(v-b_{j,\alpha}),
\end{equation}
where $a$ and $b$ denote the position of the D4 branes to the
left and right of the NS5 brane under consideration.

In order to incorporate all the non perturbative effects we have to solve
for a
surface $\Sigma$ in the four-manifold ${\bf R}^3\times S^1$ which is
parametrized by the coordinates $s$ and $v$.
As stated above, ${\cal N}=2$ space-time supersymmetry requires that
$s$ varies holomorphically with $v$, such that $\Sigma$ is a Riemann surface
in ${\bf R}^3\times S^1$.
Using $t=\exp(-s)$, $\Sigma$ is defined
by the complex equation
\begin{equation}
F(t,v)=0.\label{sw}
\end{equation}
At a given value of $v$, the roots of $F$ in $t$ are the positions of the
5-branes, i.e. $F$ is a polynomial of degree $n+1$ in $t$. On the other
hand, for fixed $t$, the roots of $F$ in $v$ are the positions of the
IIA 4-branes.
Recall briefly the situation of a model with two 5-branes, i.e. $n=1$.
This model is described by the
curve
\begin{equation}
F(t,v)=A(v)t^2+B(v)t+C(v)=0\label{sw2},
\end{equation}
where $A$, $B$ and $C$ are polynomials in $v$ of degree $k$.
More specifically, the zeroes of $A(v)$ ($C(v)$) correspond to the
positions of the semi-infinite 4-branes ending from  the left (right) on
the 5-branes.
On the other hand, the polynomial $B(v)$ belongs to the $k$ 4-branes
suspended between the two 5-branes.
After suitable rescaling and shifting of $v$ and $t$, one
obtains for
pure $SU(k)$ gauge theory simply  $A=C=1$.
$B(v)$ is then a polynomial of the following form:
\begin{equation}
B(v)=v^k+u_2v^{k-2}+\dots +u_k,
\end{equation}
where the $u_i$ are the order parameters of the theory.
In order to include, for example, $N_f$ flavors of hypermultiplets
from the right, $C(v)$ takes the form
\begin{equation}
C(v)=
f\prod_{j=1}^{N_f}(v-m_j).
\end{equation}
It is important to note that these curves precisely
coincide with the Seiberg-Witten
Riemann surfaces \cite{SW}.
The entire holomorphic
 ${\cal N}=2$ prepotential is encoded
in the curve $F(t,v)=0$.

\chapter{Applications of the brane construction}
As we have seen in the previous chapter there
is a deep connection between gauge theories and
string dynamics. In the following I will show
how this connection can be used to understand
certain aspects of gauge theory and
the phase structure of string theory, especially
the transitions between topologically distinct vacua.

\section{Dualities in the brane picture}
\subsection{Exact S-duality}
\subsubsection{General Idea}
In the previous chapter I discussed that there
are certain symmetries in string theory, relating one
theory at strong coupling to another theory at weak coupling.
More generally speaking, in theories with a free parameter
(e.g. the coupling) we identify theories with different
values of this parameter. The easiest example is the
self-duality of type IIB string theory, where this
free coupling is the string coupling and the duality
symmetry identifies the theory at $g_s$ with the theory
at $1/g_s$. If we also include the axion, $g_s$ is enhanced
to a complex coupling parameter and we have a whole $SL(2,Z)$
symmetry acting on it.
This strong weak coupling duality exchanges fundamental string and D-string.
Taking into account the full $SL(2,Z)$ one finds a whole zoo
of $(p,q)$ strings in type IIB. This 
kind of duality is usually called S-duality.

The idea that such a duality may also exist in field theory is
very old. It is easy to convince oneself that classical Maxwell
theory is invariant under the exchange of
$$E \leftrightarrow -B, \; \; \; j_{el.} \leftrightarrow j_{magn.}. $$
Of course this ``duality'' can only be valid if we introduce
magnetic charges as well as electric  charges. Since in
quantum mechanics consistency requires that
electric charge $e$ and magnetic charge $g$ satisfy the Dirac
quantisation condition
$$ e g = 2 \pi n$$
we see that any implementation of this 
electric magnetic duality symmetry in a
quantum theory would automatically provide an S-duality,
since the charges also play the role of coupling
parameters, and hence strong electric coupling corresponds
to weak magnetic coupling. This idea of realizing 
a quantum version of electric magnetic duality was first
voiced in \cite{montonen} and is hence referred to
as Montonen-Olive duality. The problem is that in a standard
field theory, like QED, the coupling constant is not a good
operator, but runs according to the renormalization group. Therefor
it is expected that S-duality is only realized in finite
theories. One way to guarantee finiteness is to consider
maximally supersymmetric Yang-Mills where supersymmetry
guarantees cancellation of all divergencies.

Indeed by now it is believed that ${\cal N}=4$ SYM realizes
Montonen-Olive duality. Like in IIB string theory 
the $g \leftrightarrow 1/g$ duality is enhanced to a
full $SL(2,Z)$ duality once we include the theta angle and
its invariance under $2 \pi$ shifts. Since at least one of the two theories
we want to identify with each other is at strong coupling, it is again
impossible to directly prove the duality. But several
consistency checks have been performed, the most convincing
being Sen's calculation \cite{senpq} establishing the existence
of all the $(p,q)$ dyon states $SL(2,Z)$ dual to the electron
multiplet.

Another way to establish this Montonen-Olive duality
is to exploit the gauge theory- string theory correspondence.
The idea is that by embedding the gauge theories into string theory,
string dualities directly translate down into dualities
of the field theory. Of course this is not a ``proof'' of the duality,
but one reduces the number of independent assumptions to just
string dualities.

Indeed in the case of S-duality in $d=4$, ${\cal N}=4$ SYM it is
straight forward to realize this idea. We just 
study flat D3 branes in uncurved space. According to (\ref{gYM})
the gauge coupling on the D3 brane is just $g^2_{YM}=g_s$.
$(p,q)$ dyons in the field theory are $(p,q)$ strings ending
on the brane. S-duality of IIB leaves the D3 brane
invariant. Field theory duality is 
just what is 
left of the string theory duality after we decoupled the bulk modes.

\subsubsection{S-duality in ${\cal N}=1,2$}

There are some examples of finite theories with less
than 16 supersymmetries. In ${\cal N}=2$ the $\beta$ function
is solely determined at 1-loop, $\beta \propto 2 N_c - N_f$.
So just by choosing the right matter content the theory
is finite. It is believed that all these theories do posses
an S-duality acting on their coupling constant. This was first
established for the $SU(2)$ case with 4 flavors in \cite{SW2}.

In order to find the S-duality in the brane picture one first
performs the lift to M-theory.
Finiteness translates into a no bending requirement.
The duality group is then the
homotopy group of the resulting Riemann surface
\cite{wittenn2}. Since
this analysis only depends on having the curve and not whether
we obtain it from brane physics or just from field theory this
statement does not shed much new light on ${\cal N}=2$ S-duality.
The story becomes clearer when going to the branes as
probes picture. Finiteness corresponds to cancellation of
tadpoles in the orbifold background, as I will show when discussing
the duality between these two approaches. The gauge theory
can be realized as D3 branes on top of an ADE singularity.
As in the ${\cal N}=4$ case S-duality of the embedding IIB
string theory directly translates down to S-duality of
the gauge theory. In general the duality group will be bigger
than just $SL(2,Z)$ since it will also include transformations that
just relabel the gauge group factors. These are obvious symmetries.
Since they in general do not commute with S-duality a large discrete duality
group is generated.

The orbifold construction carries over straight forwardly to
${\cal N}=1$. Now we have to consider
D3 branes on top of an $C^3/ \Gamma$ singularity
with $\Gamma \in SU(3)$. Here we have to distinguish two kinds of tadpoles
\cite{leighrozali}: tadpoles from twist elements that leave a 2d plane
fixed (so that they look like ${\cal N}=2$) cancel only in
a finite theory, all other tadpoles have to cancel in order for the
theory to be free of anomalies. This can be understood as follows:
a non-vanishing tadpole corresponds to a net charge. In
a compact space this has to vanish. If our orbifold has an r dimensional
fixed plane, we are dealing with a charge in r dimensions. If r is bigger
than 2 there are no problems with this. For r=2, as we have to
deal with in the ${\cal N}=2$ case and for the special tadpoles
in ${\cal N}=1$ the charge will lead to a logarithmic divergence.
I will prove that this divergence is nothing but the running
of the gauge coupling in Chapter 4.
So cancellation of tadpoles of the first kind
($r=0$)
is a necessary requirement, tadpoles of the second kind
($r=2$) vanish
only in theories with no running. Again one can see the
finiteness requirement in the dual brane box picture
as a no-bending requirement \cite{HanFinite}.
S-duality is again established trivially by the embedding via
D3 brane in type IIB.

There are certainly other ${\cal N}=1$ S-dual pairs. The 
list of finite theories, which can be constructed using
the methods of \cite{leighstrassler} or \cite{george}
exhibits many examples which can not be realized in any brany
way so far,
and probably most of them exhibit
some kind of S-duality. One way one might hope to generate such S-dual
pairs is to scan through all finite theories whose Seiberg
dual is known. Before embarking on this discussion let me
explain Seiberg duality and its brane realization.

\subsection{Seiberg duality and Mirror symmetry}
\subsubsection{Universality}

Another interesting aspect of field theory that can be
addressed quite systematically in a brany language is
universality. The phenomenon of universality is due
to the effect that the renormalization group (RG) flow
is irreversible. Evolving a theory towards the IR, certain
information about the theory is lost. Several possible
deformations of the theory are irrelevant and do
not lead to new IR physics. Many different theories hence
flow to the same fixed point. The detailed information
encoded in the UV physics does not matter and all physical
IR properties are just encoded in the structure of the fixed
point theory. Since many different theories flow to the
same fixed point their IR physics is described
by the {\it universal}
properties of the fixed point.
All theories flowing to the same fixed point are usually referred
to as a universality class. This is the closest we can
get to duality in the context of theories with a running coupling.
The dual theories (which are usually still referred to as
electric and magnetic) no longer are identically, but nevertheless
describe the same physics in the IR.

In field theory it is often very difficult to establish whether
two different theories belong to one and the same universality
class. Usually one identifies certain possible deformations
of the theory as irrelevant. This is done analysing their
quantum dimensions.
Once we know the precise dimension (that is after
quantum corrections have been taken into account) of a given operator 
we can read off from the
dimension if the corresponding dimensionful
coupling increases or decreases once we multiply all length
scales involved with a certain scaling factor. The corresponding
operators are called relevant or irrelevant. The former lead
to a different IR description, latter leave the IR
description unchanged. Of special interest are dimensionless
operators. They are usually referred to as marginal. 
Existence of an exactly marginal operator leads to a series of
FPs which can be smoothly deformed into each other by
tuning the marginal coupling, that is a fixed line.

This way one can establish that theories which
have the same Lagrangian up to some irrelevant operators belong
to the same universality class. A much more spectacular example
of such a matching was found in \cite{seibergdual}, where it was established
that even two theories with a completely different gauge group
can belong to the same universality class. Usually this phenomenon is
called Seiberg duality. This nomenclature is due to the fact
that one can view the two different UV descriptions as the
dynamics of electric and magnetic variables respectively. The
name duality might be a little bit misleading, since the
two systems do describe different physics away from the fixed point.
It is however quite impressive that they do describe
the same physics in the IR. 

Most statements about dualities of this kind are still
conjectural, even though a huge amount of evidence has been
accumulated. It would be desirable to gain a better
understanding of this phenomenon of duality
from the branes / SYM correspondence. The basic
strategy is the following: we should identify certain
brane moves as irrelevant. This can for example be done
by matching the resulting gauge theories which are known
to belong to the same universality class.
Of course an intrinsic stringy explanation
of why such an brane move should leave the IR physics unchanged
would be desirable. It would basically complete the proof of 
Seiberg duality. But once we have identified a certain brane move
as irrelevant, we can produce a vast variety of field
theories belonging to common universality classes, by
applying the irrelevant brane move to various
configurations.

One example of such an irrelevant brane move is changing
the 6 position of the various branes involved. Since
the 6 direction is the one along which the color branes
stretch, it seems reasonable to assume that the 
low energy physics shouldn't depend on the 6 positions involved.
Effectively we performed a Kaluza Klein reduction along the
6 direction to obtain the p dimensional physics from
the p+1 dimensional worldvolume of the color branes.
In a Kaluza Klein reduction we throw away everything
but the zero mode (that is the constant mode) along the
compact direction. We do not expect that
our low energy physics is sensitive to any structure
(that is to the brane positions) in the KK reduced direction.
Let me show in the following how one can use this assumption
to obtain Seiberg duality and its 3d cousin, mirror symmetry,
from this brane move. After that I will briefly comment
on the validity of the assumption, that this brane
move is indeed irrelevant.

\subsubsection{Seiberg duality}

Probably the most famous example of an IR duality of this
type is the equivalence of SUSY QCD with $N_f$ flavors
and $N_c$ colors with another SUSY QCD, also with $N_f$ flavors,
but with $N_f-N_c$ colors and additional singlet fields called
mesons, coupling via a superpotential $M q \bar{q}$ with the
quark fields. Let me explain the phase diagram of SUSY QCD
as it was discussed in \cite{seibergdual}.
For other gauge groups the phase structure will
look very similar,
with $r=N_f/N_c$ replaced by the ratio $\mu_{matter}/ \mu_{adj}$
where $\mu$ denotes the quadratic index.

\vspace{1cm}
\fig{Phase diagram for SUSY QCD}
{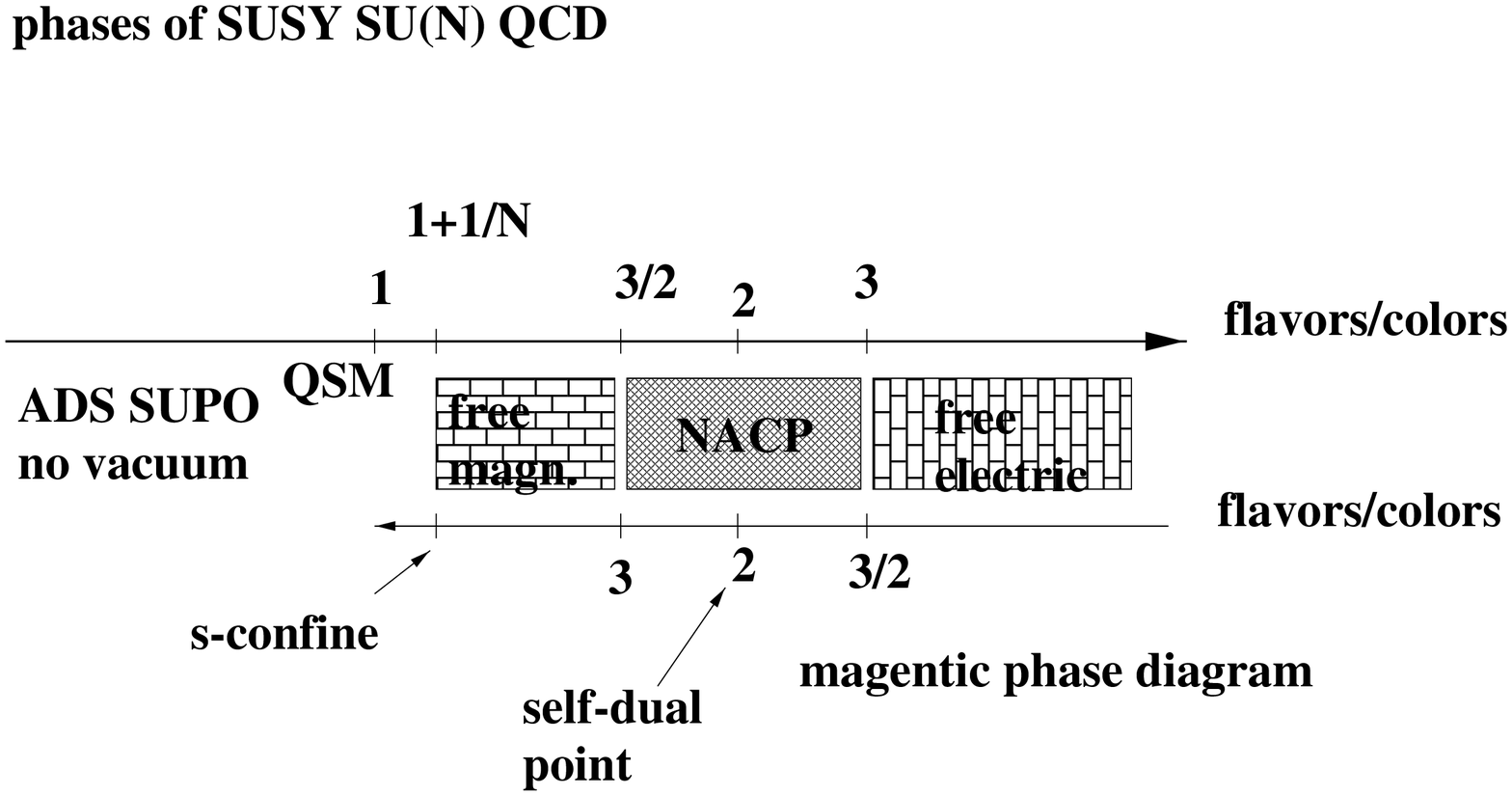}{10truecm}
\figlabel{\phases}
\vspace{1cm}

For small $r$ holomorphy and symmetries allow for a unique superpotential
whose minimum is at infinity in all the moduli (the ADS superpotential).
An instanton calculation at $N_f=N_c-1$ shows that it is indeed
generated for all $r<1$. For $r=1$ the quantum moduli space is
described by mesons and baryons with a unique
quantum constraint again fixed by symmetries and holomorphy.
For $N_f=N_c+1$ again baryons and mesons are the right degrees
of freedom. The classical moduli space stays uncorrected. This
behaviour is referred to as ``s-confining'' in the literature
\cite{sconfine}.
It is a special case of Seiberg duality with a trivial magnetic
gauge group.

For $r$ a little bit below 3
\footnote{To be more precise: $N_f$, $N_c \rightarrow
\infty$, $g_{YM} \rightarrow 0$ with $g^2_{YM} N_c$ fixed
and $\frac{N_f}{N_c}=3-\epsilon$.} one can establish the existence
of a non-trivial IR fixed point following the
arguments of Banks and Zaks
\cite{bankszaks}. The relation between
R-charge and conformal dimension contained in the superconformal
algebra tells us that this fixed point behaviour has to break
down at $r \leq 3/2$. Assuming that the non-trivial fixed point
theory, usually referred to as Non-abelian Coulomb phase,
really holds in the whole regime $3/2 < r< 3$, a beautiful
and consistent picture emerges. It is in this NACP that
the dual description comes into play. Using the same borders
in the dual group one can see that it is also in its NACP.
The duality then states that this is indeed the same fixed point.
There is plenty of evidence for this conjecture
\begin{itemize}
\item the `t Hooft anomaly matchings are satisfied
\item the moduli spaces match
\item under perturbations via superpotential
terms we flow to new consistent dual pairs
\item the ring of chiral operators matches
\end{itemize}

For $r \geq 3$ the electric theory loses asymptotic freedom and the
theory is free in the IR. For $r \leq 3/2$ the electric theory is
strongly coupled and intractable, however extending the duality
conjecture to this regime one finds a free magnetic phase.
For special values of $r$ one encounters self-dual pairs.
In the case of SUSY QCD this happens for $r=2$. As shown
in \cite{selfdual} in more general gauge theories self-duality
might show up at $r=1+1/n$ for some integer $n$ (in this
notation SUSY QCD realizes
selfduality at $n=1$). 

It was shown by \cite{egk} that the two ``dual''
theories can indeed be connected via a motion of branes in the 6
direction. One important thing one has to remember 
from our discussion in Chapter 2
is the Hanany-Witten effect \cite{HW} of brane creation
when an NS5 brane passes an D6 brane. 
Taking the NS5 brane all the way 
through the flavor branes and then ``$\epsilon$'' around the
NS5' brane one obtains a brane theory that is described
by the dual SYM.

\subsubsection{Mirror symmetry}

Another example of universality is mirror symmetry in 3 dimensions.
It was first discovered by Intriligator and
Seiberg in their study of 3 dimensional ${\cal N}=4$ theories (that
is 8 supercharges). In 3 dimensions the vector is dual
to a scalar, so that both the VM and the HM just
contain 4 scalars as bosonic degrees of freedom. Supersymmetry
requires the Coulomb and the Higgs branch both to be hyper-Kahler
manifolds. There might hence be
a symmetry mapping the Higgs branch of
one onto the Coulomb branch of another theory.
Since the scalar one obtains from dualizing the vector lives on a circle
of radius $g^2_{YM}$ while all the other scalars are in general
non-compact, one should expect that such
a symmetry can only exist at infinite coupling
(that is in the far IR since the coupling has mass dimension
1).

Taking into account quantum corrections the Higgs branch remains
untouched while the Coulomb branch is corrected at 1-loop.
Intriligator and Seiberg constructed theories
which are mirror to each other in the sense that their
quantum moduli spaces agree in the far IR with the
role of Coulomb and Higgs branch swapped.

The 
implementation of mirror symmetry
in string theory uses $SL(2,Z)$ duality of type IIB
and again the irrelevance of the motion in the 6 direction. This
analyzes was performed in the original HW paper \cite{HW} and
was one of the main motivations for introducing these brane setups.
Acting with S-duality on a 3d HW setup changes NS5 branes and D5 branes.
After rearranging branes in the 6 direction one
winds up with a system that has again a SYM interpretation.
If we start with a theory with a single gauge group and
a lot of matter, we will end up with a product of many
gauge groups and just two fundamental matter multiplets
from the two original NS5 branes turned into D5 branes.
I will present examples later on.

E.g. the mirror of $U(N)^k$ with bifundamentals and a single fundamental
in one of the gauge groups is $U(k)$ with an adjoint and $N$ flavors.
In the HW realization we put the theory on a circle. The electric
theory is obtained via $k$ NS5 branes and one D5 brane in one of the
gauge groups. The mirror has just one NS5 brane, which
yields $U(k)$ with an adjoint \footnote{Below I will give
a more detailed discussion of HW setups with a compact
$x^6$ direction, but it is straightforward to see that
a single NS5 on the circle gives us a single gauge group with an
adjoint from the ``bifundamental'' starting and ending in the same
group.}. The $N$ D5 branes add the $N$ matter multiplets.

Counting quaternionic dimensions 
of the branches is very simple. The dimension of the
Coulomb branch is just the rank of the gauge group, so it is
$N \cdot k$ in the original and $k$ in the mirror.
The Higgs dimension is obtained by counting the number of
HMs not eaten by the Higgs mechanism, so it is the
number of HMs minus the number of VMs, that is
$N k^2 +k - N k^2 =k$ for the original and
$N k + k^2 -k^2=N k$. So we find the expected agreement. Doing
a 1-loop calculation one can check that not only the dimension,
but really the full hyper-Kahler metric agrees \cite{yaron}.

Generalization to $USp$ groups needs some ``song and dance'' since
we now have to deal with the S-dual of the O5. This analysis was performed
in \cite{russe}. We can get mirror symmetry in this case also via 
combined S- and T-duality
in the brane setup: consider the setup
with a single NS5 and $N$ D5s. T-dualizing this yields
$k$ D2 branes probing a background of $N$ D6 branes.
In the S-dual picture we have just a single D5 and $N$ NS5s.
T-dualizing now yields D2 branes probing an $A_{N-1}$ singularity
\cite{dualNS}
with one extra matter multiplet coming
from the D5. We will see later that these
give indeed rise to the gauge groups presented above.
I think this is a very beautiful realization of
mirror symmetry, since it relates directly
the two most prominent backgrounds which were studied using the
brane probe technique. In this picture generalization to $USp$
gauge groups is straight forward and yields a correspondence
between D2 branes probing a $D_N$ singularity and
D2 branes probing an O6 $+N$ D6 system.

\subsubsection{Combining both: gauge theory with ${\cal N} = 2$ in d=3}

Mirror symmetry can be generalized to ${\cal N} = 2$ in d=3 \cite{many,
yaron2}. The mirror can be thought of as a theory of vortices.
In addition one can still perform the EGK brane move 
that yielded Seiberg duality in 4 dimensions, so one
might expect that these theories do exhibit two
different kinds of dualities, mirror symmetry and
Seiberg duality. Indeed it was argued in \cite{karch3d}
and \cite{ofer} from the field theory point of view, that
the Seiberg duals suggested by the brane picture
still hold in the 3d setup. This way
one can produce not just two but really very many gauge theories
that live in the same universality class.

\subsubsection{Irrelevance of the 6 position?}

I have presented several
examples of theories in the same universality class by assuming
that the 6 position in HW setups is irrelevant. This assumption was
based on the fact, that in order to identify the low-energy field
theory we effectively KK reduced on the interval thereby throwing
away all modes that could probe any structure along this
direction.

This argument could be spoiled if we encounter phase transitions
when moving the branes around. Especially dangerous are points,
where we move branes past each other. In the case of
a NS5 brane crossing a D5 brane we are saved by the brane-creation
mechanism and the s-rule. However there are situations when the IR physics
indeed does change with the relative 6 position of branes. 
This happens when NS5 branes meet parallel flavor branes, like it is
possible in HW setups with 4 supercharges (the NS5' is parallel to
the D6 in the conventions used so far). As examples I will
discuss the enhancement of the chiral global symmetry and the phenomenon
of flavor doubling, both related to D6 branes crossing NS5' branes.
Another example of a similar phase transition was discussed
in \cite{aharony} where it was shown that passing flavor branes
which are rotated with respect to each other past each other
changes the superpotential of the corresponding gauge theory.

As a first example let me discuss the appearance of enhanced
chiral symmetry as suggested in \cite{brodie,zaff1}.
Let us consider a 4d gauge theory.
The global symmetry of flavor rotations
is visible as the decoupled $SU(N_f)$ gauge theory on
the D6 worldvolume. With 8 supercharges that is all we expect
since the rotations of fundamental and antifundamental
${\cal N}=1$ chiral multiplets are linked via
the ${\cal N}=2$ superpotential $Q X \tilde{Q}$, where
$X$ is the adjoint scalar from the VM. For the same reason
$SO$ and $USp$ gauge groups will have $USp$ and $SO$ global
symmetry respectively (these
are the subgroups of the full $SU(N_f)$ flavor
rotations that leave the superpotential
invariant). However in the situation with 4 supercharges
no such superpotential is present. In the $SU$ case we expect
the full chiral $SU(N_F)_L \times SU(N_F)_R$ symmetry, while
the orthogonal and symplectic gauge groups will have a full
$SU(N_f)$ rotation group. Generically the branes
only see the ${\cal N}=2$ remnant. If we move all D6 branes
on top of the parallel NS5' brane they can split (this is just
a 6d HW setup). The half D6 branes will realize the full
global symmetry. For this to be possible we had to choose
a very particular 6 position of the D6 branes. 

This is just a special example of the more general phenomenon of
``flavor doubling'' \cite{chiralwithhan}
whenever a D6 brane meets an NS5'. Consider the
following cross configuration:
\vspace{1cm}
\fig{A ``cross'' configuration.
An intersection of a D4 brane, a D6 brane and an NS5$'$ brane.} 
{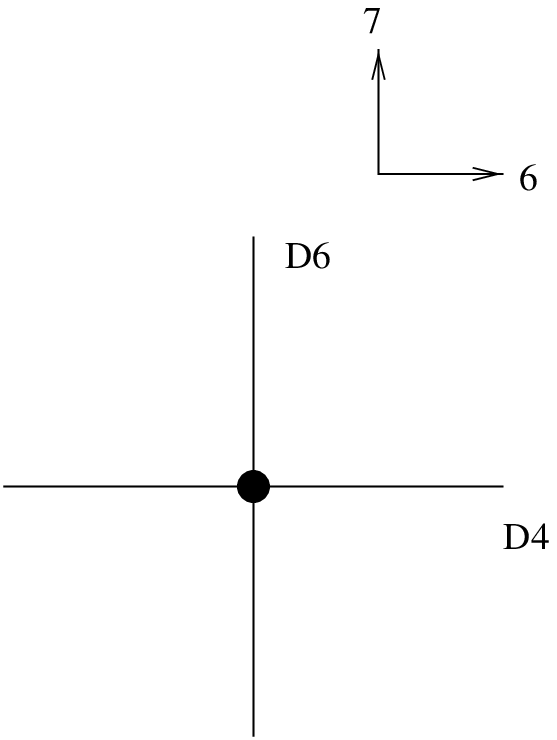}{5truecm}
\figlabel{\cross}
\vspace{1cm}
Let me identify the matter content corresponding to this configuration.
First let us recall that when a D6 brane meets a D4 brane there is
a massless hypermultiplet $Q$ which transforms under the $U(1)\times U(1)$
symmetry groups which sits 
on the D4 and D6 branes. The number of supersymmetries
for such a configuration is 8 and so there is a superpotential which is
restricted by the supersymmetry to be $(m-x)\tilde QQ$, where $m$ is the 45
position of the D6 brane, $x$ is the 45 position of the D4 brane.

We can slowly tune the
position of an NS5$'$ brane to touch the intersection of the
D4 and D6 branes. Locally the number of supersymmetries is now 4.
At this point, both the D4 and the D6 branes can break and
the gauge symmetry is enhanced to
$U(1)_u\times U(1)_r\times U(1)_d\times U(1)_l$.
The $u,d$ indices correspond to the two parts of the D6 branes and the $l,r$
indices correspond to the two parts of the D4 branes.

As usual for the transitions which lead to breaking of the
D branes, we should look for an interpretation as a Higgs mechanism.
Since,
the number of vector multiplets is increased by two, we need to look for two
more massless chiral fields.
By applying the same logic as in the case of the enhanced chiral
symmetry,
we see that we have four copies of chiral multiplets $\tilde Q, Q, \tilde R, R$.
They carry charges (1,-1,0,0), (0,1,-1,0), (0,0,1,-1), (-1,0,0,1), respectively
under the gauge groups.

In addition there are bi-fundamental fields for the intersection of the two new
D4 branes, $\tilde F, F$ with charges (0,1,0,-1) and (0,-1,0,1), respectively.
Two more bi-fundamental fields come from the two new D6 branes.
For the moment, we will ignore the bi-fundamentals for the D6 branes,
since they
have six dimensional kinetic terms and so are not dynamical for the four
dimensional system.

The system has 4 supercharges and we cannot exclude the possibility of a
superpotential.
Studying possible deformations and comparing with possible
branches of field theory the superpotential can be uniquely fixed to be
\begin{equation}
\tilde Q\tilde F R - \tilde R F Q.
\label{pupo}
\end{equation}
What happens if we rejoin the D6 brane and move it away from the NS5'?
Depending on whether we move it to the left or right, either
$\tilde Q Q$ or $\tilde R R$ will become massive, since the strings
stretching between the one D4 piece and the D6 have to stretch
a non-zero distance. If we embed this cross configuration in an HW
setup realizing product gauge groups we find that any D6 brane
basically contributes a fundamental hypermultiplet to all
product factors! All but one of them will have a finite mass.
At the special points when a D6 touches an NS5' and is allowed to
split, the fundamental hypermultiplets in the two neighbouring groups
will become massless simultaneously. We have created a situation in
which the matter content does depend on the 6 positions of the
branes involved.

\subsection{S-dual ${\cal N}=1$ pairs revisited}

After this long discussion about Seiberg duality and all its
cousins we can readress the question: what is the Seiberg
dual of a finite theory? Many of the examples of Seiberg duality
include theories that can be made finite upon adding an
appropriate superpotential term. Is in these cases the Seiberg
duality an S-duality?

Just checking through several examples one finds that in general
the Seiberg dual of a finite theory is not finite
\cite{leighstrassler,george1}, so
that a $g_{YM} \rightarrow 1/g_{YM}$ S-duality cannot be true.
However in all examples the dual does have a very special property:
it contains at least one marginal operator \cite{george2}.
The existence
of an  exactly marginal operator corresponds to having an arbitrary coupling
in the fixed point theory, parametrizing a whole fixed line.
On the fixed line of marginal couplings the theory is
superconformal, i.e. all $\beta$-functions are vanishing.
Following the work of Leigh and Strassler \cite{leighstrassler},
a simple criterion for the existence of exactly marginal
operators is given by 
analyzing the exact Shifman-Vainshtein \cite{SV} formula for the $\beta$
function. The gauge and Yukawa $\beta$ functions get a 1-loop
contribution and all higher loop and non-perturbative corrections
enter as linear functions of the anomalous dimensions of the matter
fields. In general setting the $r$ $\beta$ functions to zero yields
$r$ conditions on the $r$ couplings, leaving at most a fixed point.
If some of them are linearly dependent, we get a line of solutions
and hence a marginal operator.

Now we can make the comparison: finite models are in general
S-dual to a superconformal theory parametrized by a 
free coupling constant multiplying a marginal operator.
In the special case that the dual is also finite (like
in the well known ${\cal N}=4$ example) this dual free
coupling is just the gauge coupling, whereas in general
it is a combination of gauge and Yukawa couplings.
Several examples along these lines have been presented in
\cite{leighstrassler,george1,george2}. One of the examples
found in \cite{george1} actually seems to give a finite
dual of a finite theory. The electric model
is based on the gauge group $SO(10)$ with
matter fields $V$ 
in $N_f=8$ vector and $N_q=8$ fields $Q$ in the
spinor representations. The model
with this superfield content has vanishing one-loop gauge $\beta$ function.
Under addition of a superpotential
\begin{equation}
W=h \sum_{i=1}^8 Q_iQ_iV_i,
\end{equation}
the theory becomes finite. The conjectured S-dual is based on an $SU(9) \times
USp(14)$ gauge group with a symmetric tensor in the $SU(9)$ factor
and several bifundamentals and fundamentals. Again the
1-loop $\beta$ functions vanish and the superpotential
is such that according to \cite{leighstrassler} the whole
theory is actually finite. For more technical details see
\cite{george1}.

\section{Non-trivial RG fixed points}

\subsection{Appearance and applications}

\subsubsection{Interacting fixed point theories}

An interesting phenomenon very familiar in 4 and lower dimensions
is the appearance of non-trivial fixed points of the renormalization
group. Since the $\beta$ function by definition vanishes at
the FP the theory at the FP is invariant under scale transformations.
In the simplest cases the 
theory at the FP is free, that is the coupling vanishes.
This is for example the case for the UV fixed point of asymptotically
free field theories like QCD. Scale invariance of the free theory is somewhat
trivial. There are just no interactions present that could set any 
scale. Sometimes however one encounters an interacting fixed point.
In this case, at least in the supersymmetric version, the theory
is believed to not only exhibit scale invariance, but the full
conformal invariance (which includes scale invariance).
Due to the lack of any scale and hence any
mass in this (super)conformal theory, one has to deal with a continuum
of states, making any particle interpretation impossible. All information
about the theory is contained in its correlation functions. Such conformal
theories are needed to describe critical systems 
and are of substantial interest in both particle and statistical
physics.

One of the big surprises coming out of the SYM - string theory correspondence
was that such fixed points can also exist in 5 and 6 dimensions. This
was first noted in \cite{comments}
for maximally supersymmetric
($ {\cal N} =(2,0)$) 2-form 
``gauge'' theory in 6 dimensions arising from compactifying
IIB string theory on a K3. Similar fixed points where shown
to also exist in 6d SYM theories. From the field theory point of view such
fixed points were believed to be impossible. Since the dimension of a
gauge field is 1 in any spacetime dimension, simple dimensional analysis
tells us that the gauge coupling must be dimensionful in dimensions
other than 4. While in lower dimensions the coupling becomes stronger
at low energies, in 5 and higher dimensions it becomes weaker. Therefore
all gauge theories in 5 and 6 (and higher dimensions) were believed
to be infrared free, at the same time becoming ill defined in the UV, 
since for the same reasons the coupling blows up and the theory is
non-renormalizable just from naive power counting.

The caveat in this argument is that we tacitly assumed that at least at
some energy scale gauge theory is a valid description. Then it follows
automatically that at all lower scales gauge theory is also a good description
since the gauge coupling becomes even weaker and we hit the free fixed
point in the IR. The only way to avoid this is to have a theory which is
intrinsically strongly coupled, so that gauge theory is never 
a good description. A heuristic way to say this is that even though
the $F^2$ gauge kinetic operator is an irrelevant operator (turning it on 
doesn't move us away from the free theory in the IR), we might be able
to reach an interacting theory by taking the coupling parameter of this
irrelevant operator (the gauge coupling) formally to infinity, leaving
us with a strongly coupled gauge theory. From the field theory this only
teaches
us that gauge theory is not the right arena to discuss the appearance of
non-trivial fixed points in 5 and 6 dimensions. 
All we can do is try to write down consistency
conditions that have to be satisfied by the theory in order to have
a chance to have a well defined strong coupling limit (where we expect
the fixed point to be). These conditions where analyzed by Seiberg 
\cite{seiberg6d,seiberg5d}. In 6 dimensions one has to assure that
the theory is anomaly free, while in 5 dimensions the relevant
criterion is that once we make a small perturbation from the
fixed point, that is we turn on small $1/g^2$, the resulting 
(infrared free) theory 
be free of UV divergencies. Surprisingly
the branes then teach us that ALL theories satisfying these criteria
in fact do give rise to non-trivial strong coupling fixed points,
as I will partly show in the following for the 6 dimensional theories.

Another criterion is that there should exist a superconformal
algebra in the corresponding dimension with the right amount of
supersymmetry. These superconformal algebras where classified
by Nahm \cite{werner}. Again one finds that the branes realize all 
possible superconformal algebras.

\subsubsection{Applications}

Just establishing the existence of these higher dimensional fixed point
theories is certainly interesting on its own, since they were
not expected to exist. But they also have some interesting 
applications. For one we can basically learn about 4d field theories
from compactifying 6d FP theories on a torus. 
In a certain limit the physics of the compactified
theory reduces to pure Yang-Mills theory. 
Let me briefly discuss the (2,0) FP compactified on a torus
and how it reduces to pure (non-supersymmetric) QCD.
Compactifying the $A_{N-1}$ (2,0) theory (the theory on $N$
coinciding M5 branes) on a circle of radius $R_2$ one obtains
maximally supersymmetric $SU(N)$ Yang-Mills theory in 5d with coupling
$g^2_{YM}=R_2$ (the theory on the resulting D4 worldvolume).
In order to obtain 4 dimensional, non-supersymmetric QCD we
compactify on yet another circle of radius $R_1$ and include a
non-trivial twist by the R-symmetry, basically choosing anti-periodic
boundary conditions for the fermions. The resulting 
${\cal N}=0$ $SU(N)$ gauge theory has gauge coupling
\begin{equation}
\label{gQCD}                                
g^2_{QCD}=\frac{R_2}{R_1} 
\end{equation}
according to the standard KK ansatz. Besides the gauge
bosons of QCD this theory certainly contains many other states.
There are the KK modes with masses $1/R_2$ and $1/R_1$, the fermions
with masses of order $1/R_1$ and the scalars, which get masses at
one-loop from the fermions and hence have a mass of order
$$\frac{ g_{QCD} }{R_1}=\sqrt{ \frac{R_2}{R_1}}/R_1=\sqrt{ R_2/R_1^3}.$$
The question is whether we can find a limit
in which all these states become very massive while keeping the
QCD scale fixed.

This is indeed possible \cite{wittenba,ganorba}. Let us first read off
the QCD scale from the information we have so far. This is
we have to take into account the running of the
gauge coupling with the energy scale $\mu$,
$$g_{QCD}^{-2}(\mu)\sim \log{\frac{\mu}{\Lambda_{QCD}}}.$$
(\ref{gQCD})
gives us the gauge coupling at the compactification
scale, hence $$\frac{R_1}{R_2}=g^{-2}_{QCD} (1/R_1) \sim \log{\frac{1}{R_1 \Lambda_{QCD}}} .$$
From this we read off that 
\begin{equation}
\label{LQCD}
\Lambda_{QCD}  \sim e^{-\frac{R_1}{R_2}} / R_1
.
\end{equation}
Now consider the limit $R_1 \rightarrow 0$, $R_2 \rightarrow 0$,
with $R_1 >> R_2$. All
KK states, the fermions and the scalars become very massive in this
limit, the lightest ones being
the scalars with mass $\sqrt{\frac{R_2}{R_1}} \cdot 1/R_1$.
However due to the exponential suppression in
(\ref{LQCD}) for sufficiently small $R_2/R_1$ the QCD scale will
be much smaller than any of the other masses in the problem, leaving
us with pure QCD as advertised. Note that this limit corresponds
to very weak coupling in (\ref{gQCD}). This is not
surprising, since as discussed above, (\ref{gQCD}) determines
the coupling at the energy scale, where all the other fields
become important. We want this to happen far above the
QCD scale, that is at very weak coupling due to asymptotic
freedom of QCD.

A second very important application is to study the deformations
of the fixed points. As we will see in what follows this can be done
very easily using branes. 
A given brane setup represents a certain phase of string theory. This will
become more transparent once we have shown that
brane setups are actually equivalent to the language of geometric
compactifications. By tuning parameters of this compactification,
that is by moving around the branes, we encounter critical points as
certain branes collide, the non-trivial FP. Often at the FP we see
new deformations that allow us to perform a phase transition
into a topologically distinct vacuum of string theory.

Last but not least these FP theories play an important
role in the recent matrix conjecture \cite{bfss}. As explained in
Chapter 2 this
conjecture elevates the correspondence between gauge theory
and non-perturbative string theory to a principle, defining
all of M-theory in terms of the world-volume theory of certain
branes. For Matrix compactifications on a $T^4$ the relevant
brane theory is the strong coupling limit of the worldvolume
of $N$ coinciding D4 branes, that is the worldvolume
theory of $N$ M5 branes: the (2,0) FP theory \cite{brs}.

\subsection{Physics at non-trivial FP from branes}

\subsubsection{General idea}

From what we have learned so far it should be clear that 
branes are the essential tool to prove the existence of strongly
coupled fixed points in 5 and 6 dimensions. The principal idea is
as follows: one considers string theory in the background of certain
branes. Since this is a well defined theory, we can try to take the limit
in which gravity and the other bulk modes decouple. In
some cases this leaves us with an interacting theory on the brane.
The strong coupling limit then corresponds to having some
branes coincide, usually exhibiting an infinite tower
of states becoming massless as we would expect from a
conformal field theory. So proving the existence of an interacting
fixed point amounts to analyzing whether string theory allows for
a decoupling limit that leaves an interacting theory on the branes.
To demonstrate this procedure let us briefly consider the brane
realization of the (2,0) fixed point that was found by \cite{comments}
using a geometric picture.

I will also discuss the maximally supersymmetric cases in 5,7 and higher
dimensions as well as in 6d with (1,1) supersymmetry. In these cases
the analysis of \cite{werner} tells us that there are
no superconformal algebras. Indeed we will find in the brane picture that
in these cases decoupling of the bulk modes leads automatically
to a free theory on the brane. 

\subsubsection{The (2,0) theory from the M5 brane}

Consider a system of $N$ parallel M5 branes.
In the same spirit as in (\ref{decouple}) we want to
decouple bulk gravity by taking the limit $M_{pl} \rightarrow$
infinity. We want this to do in such a fashion that the theory
on the M5 branes stays interacting. Recall
that the theory on the M5 is that of $N$ ${\cal N}=(2,0)$
tensor multiplets. This theory does not have a coupling constant
which we could keep fixed. However we do know that these tensor
multiplets couple to the strings that describe the ends of
M2 branes. The tension of these strings is $\frac{D}{l_{pl}^3}$
where $D$ is the characteristic distance between two M5 branes.
These tensions correspond to the vevs of the scalars in 
the tensor multiplet, since they are given by the M5 positions
and have mass dimension 2 in the natural normalization. Therefor
the decoupling limit will be $D \rightarrow 0$, $l_{pl}
\rightarrow 0$, holding $u=\frac{D}{l_{pl}^3}$ fixed.
At the origin of the moduli space, that is if all the $u$
go to zero, we expect a superconformal fixed point. The
strings become tensionless and provide the continuum of
massless states of the conformal theory.

One might worry that this fixed point could be a free theory. One
way to see this cannot be the case is to consider the theory
on a large circle $R$. The resulting theory will
be 5d SYM with gauge coupling $g^2_{YM}=R$ \footnote{
One way to see this relation is to consider the compactification
in terms of branes. The M5 turns into a D4 with $g^2_{YM}= l_s g_s =R$.}.
We see that this is an interacting theory and the large
$R$ limit even corresponds to very strong coupling.

\vspace{1cm}
\fig{Brane realisation of the (2,0) theory.
}
{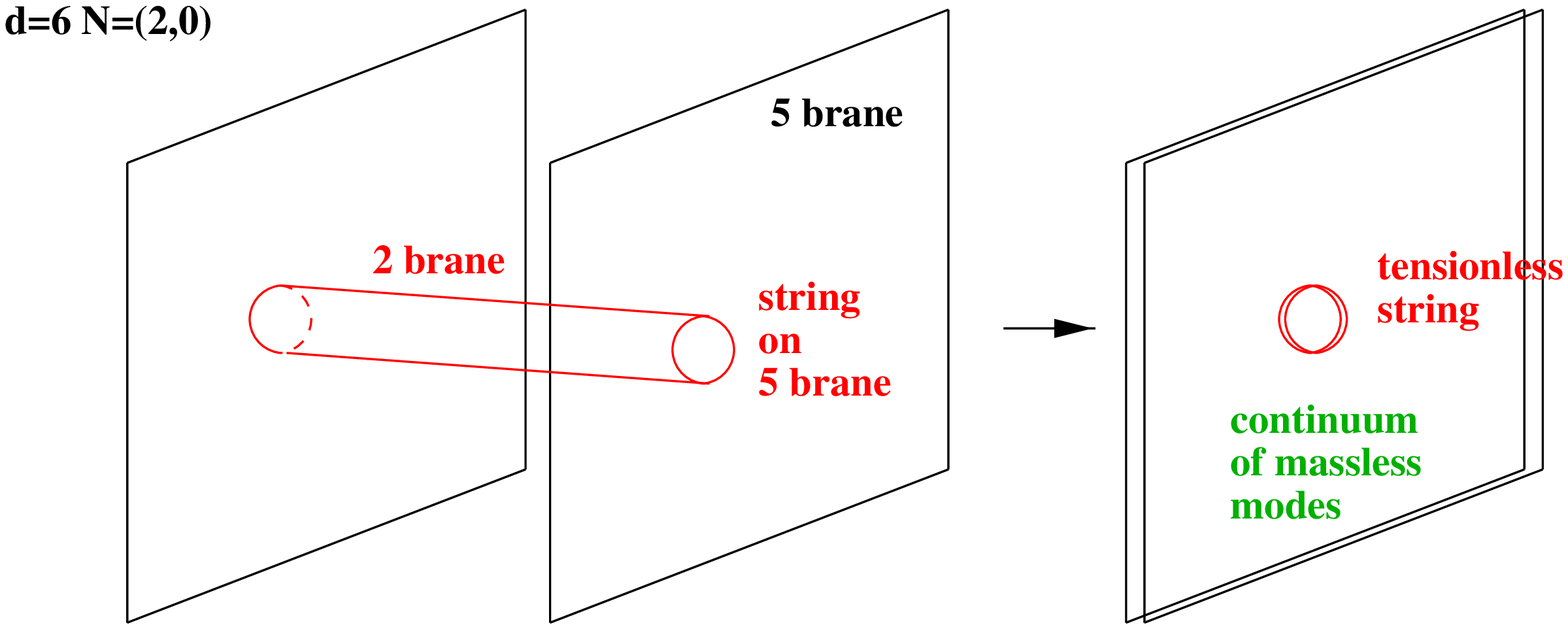}{11truecm}
\figlabel{\branefix}
\vspace{1cm}

\subsubsection{The other maximally supersymmetric theories}

One can try to do a similar construction for SYM theories in 5,6,7 and higher
dimensions, which are also maximally supersymmetric (where this time
in 6d we have non-chiral (1,1) supersymmetry). These
can be realized as the world volume theories of D4,5,6 and higher
branes. We want to
send $M_{pl}$ and $M_s$ to infinity in order to decouple all bulk modes
\footnote{The former is required to decouple gravity while the
latter decouples the higher order terms from the DBI action.}.
If we would again predict non-trivial FPs we would be in trouble since
the analysis of \cite{werner} shows that for these cases there 
are no superconformal
algebras. We expect a non-trivial theory once we put several
branes on top of each other.
$N$ colliding branes give rise to $U(N)$ gauge theory.
In order to obtain an interacting theory on the branes we need
to keep $g^2_{YM}$ on the branes finite. According to (\ref{gYM})
$$g^{-2}_{YM}= M_s^{d-3}/g_s$$
for the d+1 dimensional gauge theory on the worldvolume of a D$d$ brane.
Since
$$M_s^4 =g_s M_{pl}^4$$
We see that for the D7 brane $g^{-2}_{YM}=M_{pl}^4$ and hence $g_{YM}$ goes to
zero in the decoupling limit $M_{pl}$, $M_s$ $\rightarrow \infty$.
This still is true in higher dimensions. We are left with a free theory!

For the D6 $g^{-2}_{YM}=M_{pl}^3/g_s^{1/4}$ we see we can keep $g_{YM}$
finite if we simultaneously with $M_{pl}$ take $g_s$ to infinity. According
to \cite{various} strongly coupled IIA string theory is better thought of as
11d SUGRA. The duality tells us that the 11d Planck scale $M_{pl,11}$ is
given by
$$ M^3_{pl,11}=M_{pl}^3/g_s^{1/4}=g^{-2}_{YM}. $$
In order to decouple the 11d bulk we again have to stick to a free
theory on the worldvolume. For the D5 and D4 branes a similar story is
true. In both cases $g^2_{YM}$ can be kept finite only in the strong string 
coupling
limit. S-dualizing, the D5 brane turns into an NS5 brane at weak coupling
with gauge coupling $g^{-2}_{NS5}=M_s^2$ which is again free
in the decoupling limit.
One can slightly modify the decoupling limit by relaxing the condition
that $M_s$ is supposed to go to infinity. In this case still all
the bulk modes decouple. However we are no longer left with
just SYM on the brane. But whatever it is
we are left with, string theory
tells us that it exists. This way Seiberg proved \cite{newstrings} 
the existence
of a 6d string theory. It has a string scale $M_s$ and hence does
not correspond to a conformal theory.

For the D4 brane the strong string coupling limit once more decompactifies
the 11th dimension. The D4 brane becomes an M5. Hence the 5d SYM
at strong coupling grows an extra dimension. We do obtain
a non-trivial strong coupling fixed point, but it is again the
6 dimensional (2,0) theory which we encountered before.

\subsection{6d Hanany-Witten setups}

\subsubsection{Motivation}

In order to study less supersymmetric theories we need more involved
brane setups. In what follows I will present a Hanany-Witten setup
describing fixed points in $d=6$ with ${\cal N}=1$.
This setup was presented in \cite{andreas1}
and further analyzed in \cite{andreas3,zaff1, zaff2}.
The same fixed points where analyzed from the branes
as probes point of view
by \cite{intblum,int} and were geometrically
engineered from F-theory in \cite{engineer6dFP}. The motivation for analyzing
these theories instead of their more supersymmetric cousins is threefold.
\begin{itemize}
\item As described above we can learn about 4d gauge theory by
putting these fixed points on a torus. Since the dynamics of 
maximally supersymmetric SYM in 4d is rather constrained it is definitely
interesting to do this with less supersymmetric theories. If we
again
want to study non-supersymmetric QCD by imposing anti-periodic
boundary conditions on the fermions, we have this time the
choice to put different boundary conditions on fermions coming
from vector and hypermultiplets, opening up the possibility
to realize QCD with matter.
\item The $(2,0)$ fixed point has only one possible deformation,
which corresponds to moving the M5 branes apart. To really
study phase transitions we need to study the more elaborate
fixed points with only 8 supercharges, which in general do
allow several different perturbations, leading to
a transition between distinct phases.
\item When used as matrix models these fixed points should be
used to describe DLCQ string theories with 16 supercharges, like the
heterotic string on a $T^4$ or type II string on a K3. These
theories have a much richer dynamics and are hence more interesting
than their more supersymmetric cousins.
\end{itemize}

\subsubsection{The consistency requirement: Anomaly cancellation}

As discussed above, even though field theory is not
the right arena to discuss the appearance of non-trivial FPs,
it nevertheless imposes several constraints on the existence
of consistent strong coupling limits. In 6 dimensions this
criterion is anomaly cancellation. Let me briefly review
the relevant mechanism from the field theory point of view.
In what follows we will see that the branes actually know
about this mechanism and only yield anomaly free
theories.

The anomaly in
six dimensions can be characterized by an anomaly eight form.
For a theory
with a product gauge group $G= \prod_{\alpha} G_{\alpha}$
with matter transforming in the representation $R$,
the anomaly polynomial reads
\begin{equation}
I = \sum_{\alpha} \lb \Tr F_{\alpha}^4 - \sum_{R} n_R \tr F^4 \rb
- 6 \sum_{\alpha \alpha'\, R R'}
n_{R, R'} \tr_R F_{\alpha}^2 \tr_{R'} F_{\alpha'}^2.
\end{equation}
Here, $\Tr$ denotes the trace in the adjoint and $\tr_R$ is the
trace in Representation $R$.
The symbol $\tr$ is reserved for
the trace in the fundamental representation. $n_R$ denotes the number
of matter multiplets transforming in a particular representation
$R$ and $n_{R R'}$ is the number of multiplets transforming
in the representation $R\times R'$ of a product group
$G_{\alpha} \times G_{\beta}$.
For a consistent theory, the anomaly should be cancelled in some way:
Either the anomaly polynomial vanishes or we cancel the anomaly
by a Green--Schwarz mechanism \cite{gs}. So let us look at the
anomaly in more detail. First, we want to rewrite the
anomaly polynomial using only traces in the fundamental
representation. Formally, the polynomial then looks like
\begin{equation}
\label{anomalie}
I = \sum_{\alpha} a_{\alpha} \tr F_{\alpha}^4
  + \sum_{\alpha \alpha'} c_{\alpha \, \alpha'} \lb \tr F_{\alpha}^2\rb
\lb \tr F_{\alpha'}^2 \rb
\end{equation}

For concreteness let me discuss the case of a single gauge
group factor. The anomaly reduces to
\begin{equation} 
I =  a \tr F^4
  +  c (F^2)^2
\end{equation}
For $a=c=0$ the anomaly cancels completely. If $a \neq 0$ the
anomaly can't be cancelled and the theory is sick.
However if $a=0$ we can cancel the anomaly by the Green-Schwarz
mechanism by coupling it to a two-index tensor field.
In 6 dimensions the tensor splits into an anti-selfdual and an
selfdual part. While the former sits in the gravity multiplet,
the latter is part of the tensor multiplet. Depending
on the sign of $c$ we have to use one or the other. 
If $c>0$ we can cure the anomaly by coupling to a tensor multiplet,
for $c<0$ the theory is sick without coupling to gravity. 
If realized in terms of branes, in the latter case 
there can't be a consistent decoupling
of the bulk modes. For a very detailed discussion
of the anomaly polynomials for product gauge groups see e.g.
\cite{ilkathesis}.

In the case of $SU(N)$ gauge theory using the group theoretical
identity \cite{erler}
\begin{equation} \label{erlergl}
\Tr F^4_{SU(N)} = 2N \tr F^4_{SU(N)} + 6 \lb\tr F^2_{SU(N)} \rb^2 
\end{equation}
we find that 
\begin{equation}
I = (2 N - N_f) \tr F^4 + 6 \lb \tr F^2 \rb^2.
\end{equation}
Therefore, the deadly $F^4$ term cancels precisely
for $N_f = 2 N$.
The prefactor $c$ of the $ \lb F^2 \rb^2$
term is always bigger than zero, so that we can cancel
the anomaly by coupling to a single tensor multiplet.

Adding the Green-Schwarz counterterm, the action contains
a coupling of the scalar $\phi$
in the tensor multiplet to
the gauge fields, so that the gauge kinetic term becomes \cite{seiberg6d}
$$
\frac{1}{g^2} {\rm tr} F_{\mu \nu}^{2} + \sqrt{c} \phi {\rm tr} F_{\mu \nu}^2.
$$
We see that one can absorb the bare gauge coupling into the
expectation
value of $\phi$ by just a shift of the origin. We
obtain an effective coupling
$$
\frac{1}{g_{eff}^2} = \sqrt{c} \phi.
$$
At $\phi =0 $ we expect the possibility
of a strong coupling fixed point \cite{seiberg6d}.

\subsubsection{The brane configuration}

In order to study Hanany-Witten setups in 6 dimensions,
we need as fundamental ingredients D6 branes, on whose
worldvolume we realize the gauge theory. In addition
one has the obligatory NS5 branes, between which the
6 branes are suspended. As usual this compact interval
gets rid of one of the 7 worldvolume dimensions of the D6
brane via KK reduction. In addition the boundary conditions
for the D6 branes ending on the NS5 branes projects
out some of the fields. In total we break $3/4$ of the
supercharges ($1/2$ by the D6 and another $1/2$ by the NS5)
leaving ${\cal N}=1$ in d=6. 

As usual there are two ways to introduce flavors in the setup. One is
to include the additional flavor branes, which in our case will be D8 branes.
I will explain how to deal with them later. For now I will include flavors
via semi-infinite D6 branes to the right and left. With this the players
in our game are: 

\begin{center}
\vspace{.6cm}
\begin{tabular}{|c||c|c|c|c|c|c|c|c|c|c|}
\hline
&$x^0$&$x^1$&$x^2$&$x^3$&$x^4$&$x^5$&$x^6$&$x^7$&$x^8$&$x^9$\\
\hline
NS 5&x&x&x&x&x&x&o&o&o&o\\
\hline
D 6&x&x&x&x&x&x&x&o&o&o\\
\hline
D 8&x&x&x&x&x&x&o&x&x&x\\
\hline
\end{tabular}
\vspace{.6cm}
\end{center}

The presence of these branes breaks the $SO(9,1)$ Lorentz symmetry
of type IIA down to $SO(5,1) \times SO(3)$ corresponding to
rotations in the 789 space and Lorentz transformations in 012345 space.
The $SO(3)$ can be identified with the $SU(2)$ R-symmetry of the
d=6 ${\cal N}=1$ SUSY algebra. To understand the basic issues that
are different in 6d from the ``usual'' Hanany-Witten setups
discussed above let me first focus on the simplest setup.

\vspace{1cm}
\fig{The brane configuration under consideration, giving rise
to a 6 dimensional field theory. Horizontal lines represent
D6 branes, the crosses represent NS5 branes.}
{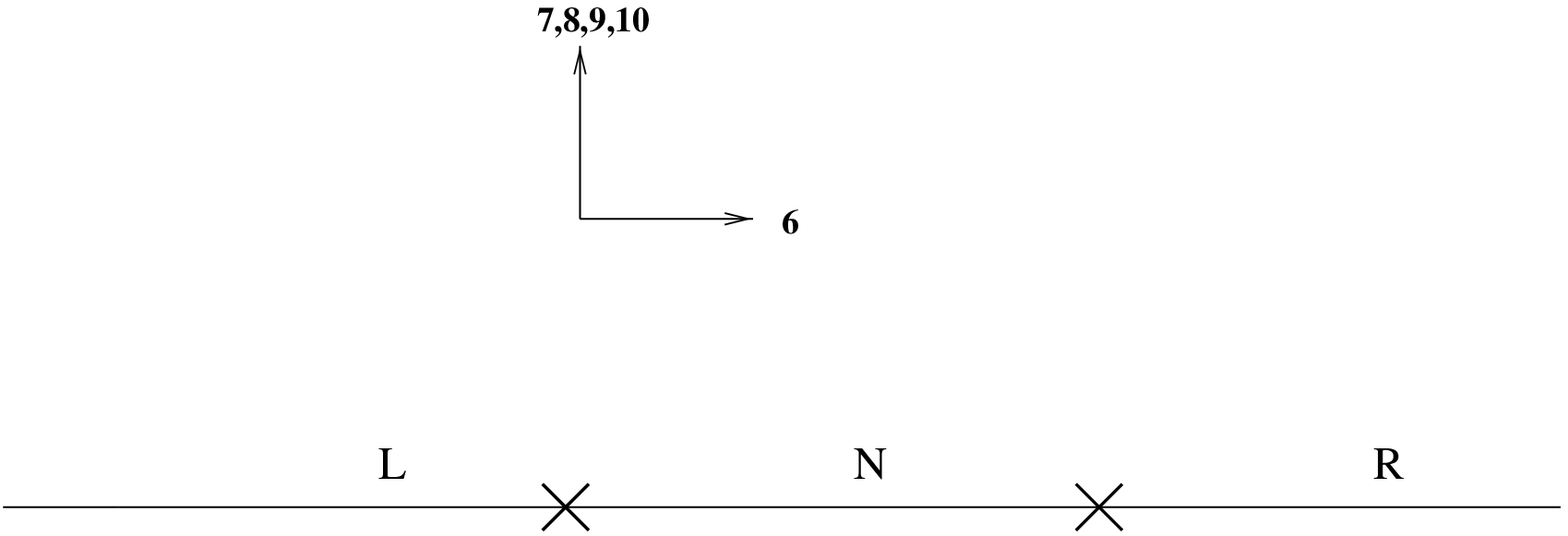}{10truecm}
\figlabel{\basic}
\vspace{1cm}
Figure \basic shows a configuration of $N$ D6 branes suspended between
two NS5 branes. There are $L$ ($R$) semi-infinite D6 branes to
the left (right). Let us first discuss the matter content corresponding
to this setup. As usual we get an $SU(N)$ vector multiplet from the
color branes and $L+R$ fundamental hypermultiplets from the
semi-infinite flavor branes. But these are not all the matter
multiplets we get. According to our standard philosophy we
keep all the light modes coming from
the lowest dimensional branes. In all other dimensions
this meant decoupling the worldvolume fields of the NS5 branes. But here
the NS5 branes live inside the D6 brane, they also have a 6d worldvolume
as the finite D6 brane pieces we are considering. Hence we should
also include the matter from the NS5 branes.

This can also be discussed in a more quantitative manner, looking
at the precise decoupling limit. As discussed in our introduction
to Hanany-Witten setups we want to take $M_{pl}$ and $M_s$ to 
infinity in order to decouple the bulk, sent the length $L$ of
the interval to zero in order to decouple the KK modes and do this
all in such a way to keep the gauge coupling of the 6d theory fixed.
According to (\ref{gYM}) we get $g^2_{YM}=\frac{g_s}{M_s^3 L}$.
The theory on the NS5 branes is a theory
of tensor multiplets, so there is no gauge coupling. To see
what is going on let us translate into 11d units via (\ref{11d}).
We see that $g^2_{YM}=\frac{l_{pl,11}^3}{L}$. But this is
precisely the tension of the M2 branes stretching between the
5 branes. When discussing the decoupling of the (2,0) theory
we found that this quantity basically governs the interaction
strength of the theory on the 5 brane.
Keeping it fixed means that we should
indeed keep the interacting modes on the 5 branes as well as those
on the worldvolume of the finite D6 brane. 

So now what is this additional matter?
 The theory on a IIA NS5 brane is the theory of a (2,0)-- tensor
multiplet. This multiplet consists of a tensor and 5 scalars (and
fermions). Because of the presence of the D6 branes, one half of
the SUSY is broken and we are left with a (1,0) theory. The tensor
multiplet decomposes into a (1,0) tensor, which only contains one
scalar, and a hypermultiplet, which contains 4 scalars. The
hypermultiplet is projected out from the massless spectrum because
the position of the semi-infinite D6 branes fixes the position
of the NS5 branes. Moving the NS5 brane out of the D6 brane 
(remember that we have the 5 brane embedded in the D6 brane)
corresponds to turning 
on the 3 FI terms and a theta angle.
The scalar in the tensor multiplet corresponds
to motions of the 5 branes in the $x_6$ direction. This is going
to be a modulus of our theory. We have two
NS5 branes and therefore two tensor multiplets, but effectively
we keep only one of them because one of the scalars can be taken to
describe the center of mass motion of the system. The vev of the
other scalar gives us the distance between the NS5 branes.
On the other hand, we know that the distance between the NS5 branes
is related to the inverse Yang-Mills coupling of the six-dimensional
gauge theory according to the usual KK philosophy. The vev
of the scalar hence plays the role
of a gauge coupling, as expected from field theory.

Now let us turn to the quantum picture. As in the other theories
with 8 supercharges, the 1-loop information will be encoded
in the bending of the NS5 branes and there won't be any further
corrections. We in particular expect the bending to provide
us with the information about the anomalies, since those are
a 1-loop effect. Indeed the analysis of the bending is really
easy in 6d, as discussed before. Since the NS5 branes are embedded in the D6
branes there is no transverse worldvolume. The NS5 brane can't absorb
any flux coming from the charged ends of a D6 brane. The only
way we can have a consistent brane setup is if the net charge cancels.
The net charge is given by the
number of D6 branes ending from one side minus the number of D6 branes
ending from the other side.
Thus, we only get a consistent picture
if:
$$
N = L = R,
$$
where $L$ ($R$) denotes the number of D6 ending from the left (right).
The total number of flavors is
\begin{equation}
\label{rrcancel}
N_f = L + R = 2 N.
\end{equation}
Together with the tensor multiplet from the NS5 branes this is indeed the
(only) right \footnote{For $SU(2)$ and $SU(3)$ there
are also some other possibilities since
they do not have an independent fourth order Casimir
and hence $a$ vanishes automatically. Global anomalies \cite{vafa}
restrict us to $N_f=4,10$ for $SU(2)$ and $N_f=0,6,12$ for
$SU(3)$. The 10 flavor case can be achieved by
realizing $SU(2)$ as $USp(2)$ with an orientifold.
The other $SU(3)$ theories do not have an HW interpretation.} 
matter content to cancel the anomaly.

\subsubsection{Realization of the fixed point}

Our theory has a Coulomb branch parametrized by the vev of the
scalar in the tensor multiplet. At the origin, the effective
gauge coupling becomes infinite and we find the strong coupling
fixed point. Since the tensor multiplet corresponds to the distance
between the NS5 branes this happens precisely when they coincide.
The strings from D2 branes stretching between the NS5 branes
become tensionless. Building product groups is straight forward.
Again we find FPs whenever the NS5 branes coincide.

An interesting generalization is to put the
theory on a circle. In this case we will see an $SU(N)^k$ gauge
group, where $k$ denotes the number of NS5 branes.
Tuning the moduli in such a way that we obtain
the fixed point theory amounts to moving all the 5 branes on top
of each other. Since the distance between the 5 branes is the
effective coupling constant of the corresponding gauge theory
we see that there is always one gauge group who's color branes
stretch around the entire circle, no matter where we choose to
bring our 5 branes together. This
means that we will always have
one gauge factor with
a finite gauge coupling that becomes free in the IR and hence
becomes a global symmetry.
From the field theory point of view this statement can also be seen to
be true. Defining the effective gauge couplings by absorbing
the classic gauge couplings in the vev of the scalars in the 
tensor multiplets, one finds that there is always one
subgroup for which $g_{eff}$ becomes negative at large $\phi$
\cite{int}, so that we should start of with a finite value
for this coupling. 

The value of $\phi$ at which this sign change happens is
given by $g^{-2}_{dec. fact.}=\frac{R_6}{l_s^3 g_s}=\frac{R_6}{l_{pl,11}^3}$ 
(remember that $\phi$ has
mass dimension 2). That is in the decoupling limit,
where we send $l_{pl,11} \rightarrow 0$ keeping
$\frac{L}{l_s^3 g^s}$ fixed, this critical value goes off to infinity,
so we do not see any effect of the finite radius. If
however we choose to also send $R_6$ to zero as well,
keeping $\frac{R_6}{l_{pl,11}^3}$ fixed, we can engineer
a theory with the following properties:
\begin{itemize}
\item it decouples from all bulk modes, hence it is 6d
\item gauge theory breaks down at energy scales $\frac{R_6}{l_{pl,11}^3}$,
new degrees of freedom become important
\item the theory contains string like excitations with
tension $\frac{R_6}{l_{pl,11}^3}$
\end{itemize}
One usually refers to those theories as little string theories. The
way we constructed them is the T-dual of the original construction
of \cite{intlittle} as a generalization of Seiberg's work 
\cite{newstrings} on the theory with 16 supercharges.
 
\subsubsection{Including 8 branes}

So far we have seen how to realize fixed points in branes. Now I will
move on to more complicated setups in order to classify all fixed points
realizable by HW setups. In \cite{sweden} all anomaly
free matter contents in 6d have been classified. Strictly speaking
this list should be seen as the list of all theories that could possibly
have a consistent strong coupling fixed point. The nice
stringy result is that indeed all of them do.
HW setups will not be able
to generate the exceptional gauge groups, but will be very efficient at 
classifying those FP allowed with classical gauge groups and products
thereof. The few missing cases can be realized via geometry \cite{
engineer6dFP,smallinonsin}.

In order to build more general fixed point theories, let us
now look at the second possibility to include flavors, via
D8 branes living in 012345789,
as discussed in \cite{zaff1}.
D8 branes introduce additional complication, since
D8 branes are not a solution of standard IIA theory but require
massive IIA \cite{romans}, that is the inclusion of a cosmological
constant $m$. The D8 branes then act as domain walls between regions
of space with different values of $m$.
$m$ is quantized and hence
in appropriate normalization can be chosen to be an integer.

The presence of $m$ makes itself known to the brane configuration via
a coupling
\begin{equation}
\label{m}
-m\int dx^{10} B\wedge
*F^{(2)},
\end{equation}
where $B$ is the 2 form NS gauge field under which the NS5 brane is charged
and F is the 8 form field strength for the 7 form gauge field under
which the D6 is charged.
From the Bianchi identity for the 2 form field strength dual
to $F^{(8)}$ in the presence of a
D6 brane ending on a 5 brane, $dF^{(2)}=d*F^{(8)}= \theta
(x^6)\delta^{(789)} - mH$ (where $H=dB$), one gets $mdH=\delta^{(6789)}$.
The $\delta$ function comes from the charge of the D6 brane-end.
From here we get back the old result, that for $m=0$
D6 branes always have to end on a given NS5 brane in pairs
of opposite charge (that is from opposite sides). However for arbitrary
$m$ this relation shows that RR charge conservation requires that
we have a difference of $m$ between the number of branes ending from
the left and from the right
\footnote{Throughout this work I will use the following conventions
to fix the signs: by passing through an D8 from left to right
$m$ increases by 1 unit. In a background of a given $m$ the number
of D6 branes ending on a given NS5 from the left is by $m$ bigger than
the number of D6 branes ending from the right.}.

\vspace{1cm}
\fig{Basic Hanany Zaffaroni Setup}
{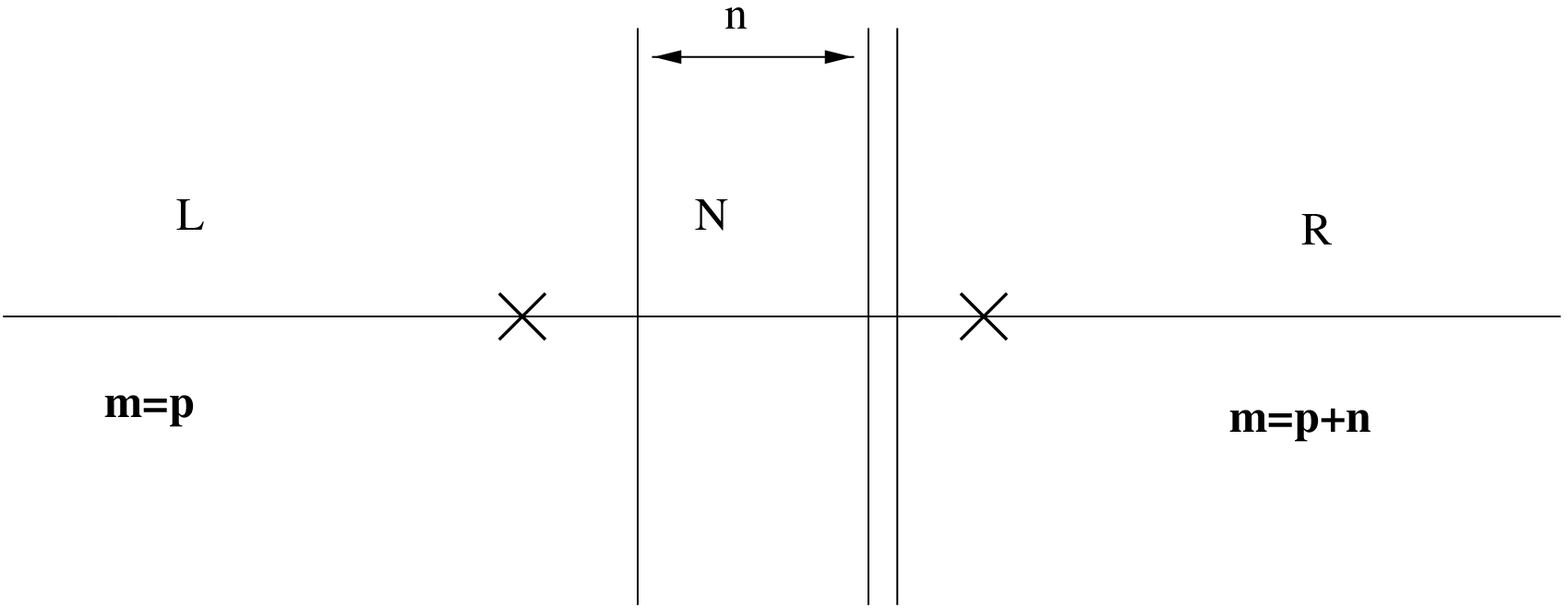}{10truecm}
\figlabel{\han}
\vspace{1cm}

Figure \han shows the basic brane configuration involving D8 branes.
The $n$ D8 branes in the middle give rise to $n$ flavors for the
$SU(N)$ gauge group. In addition they raise the cosmological constant
$m$ from $p$ to $p+n$. In the background of these values of $m$
the modified RR charge conservation tells us
\begin{eqnarray*}
N&=& L- p \\
R&=& N -(p+n) 
\end{eqnarray*}
With the $R+L$ flavors from the semi-infinite D6 branes the total number
of flavors is $L+R+n=2N$ still in 
agreement with the gauge anomaly considerations
on the 6 brane 
for every possible value of $p$.
So far we have not succeeded in 
obtaining any new FPs, since SUSY QCD with $N_f=2N_c$
is just what we were already able to realize with semi-infinite D6 branes.
However as we move on we will need the D8 branes in more elaborate setups.

\subsubsection{Orientifolds}

There is one more element we can incorporate in the HW brane setup,
the orientifold. Since an orientifold breaks the same supersymmetries
as the corresponding D-brane, we can introduce O8, O6 or both.
Each of them comes with two possible signs.
Let us first consider the
O8. We have to distinguish two possibilities: O8 planes with negative or
positive D8 brane charge (that is -16 or +16) \cite{polchinski}.
The former are the T-dual of the O9 projecting
IIB to type I. On the D8 worldvolume they project the symmetry group to
a (global) $SO$ group, while on the D6 we get a local $USp$ group.
The positively charged O8 projects onto global $USp$ and local $SO$.
If we want to have vanishing total D8 charge, we should restrict
ourselves
to either 2 negatively charged O8s with 32 D8 branes or one O8 of each type.

There is indeed a physics reason that we should restrict ourselves to
satisfy this requirement. What becomes important is that
not only the cosmological constant jumps when we cross a D8 brane,
but also the dilaton. If the dilaton is constant to the
right of a given D8 brane, $e^{-\Phi}$ (the inverse string coupling) will 
rise linearly on the left. For symmetry reasons this means that
the dilaton will run with slope -8 into an O8 from the left
and leave with slope
+8 on the other side. In order to be back to slope -8 again before hitting
the next O8, one has to put precisely 16 physical D8 branes in between.
One interesting application of this behaviour is that this
way the coupling may diverge at the orientifold planes, while
it is finite everywhere else and even constant in between the 8th and
the 9th physical D8.
Usually one refers to 
this whole setup as type I' string theory.
Of course it is no problem to include just one O8 and fewer D8 branes
by just keeping far away from the others.

\vspace{1cm}
\fig{Various possibilities to introduce O8 planes}
{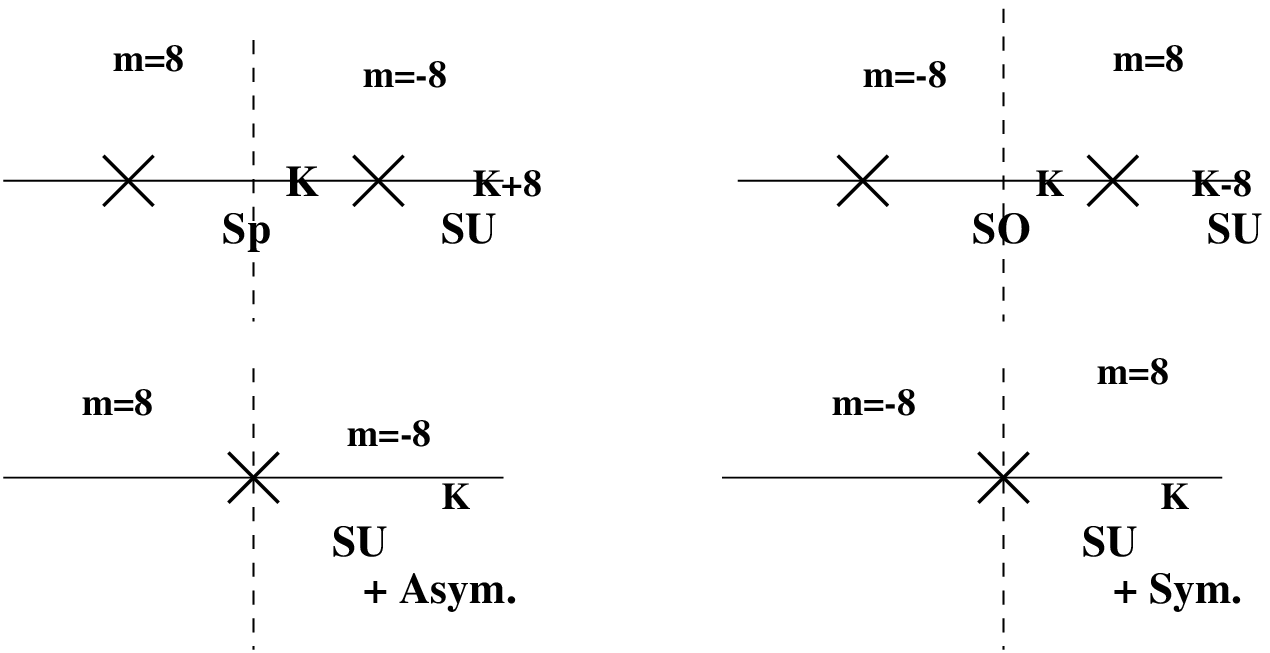}{10truecm}
\figlabel{\orient}
\vspace{1cm}

The gauge groups corresponding to brane setups in Figure \orient have been
analyzed in the equivalent setup for 4d in \cite{LL}.
If the O8 is in between
two 5 branes the `center'
gauge group is projected to $USp(K)$ or $SO(K)$ respectively.
All other gauge groups stay $SU$, however the $SU$ groups to the right
are identified with those to the left and one effectively projects out
half of the $SU$ groups. In addition to this, in 6d we get the special
situation that the O8 also changes the cosmological constant. For symmetry
reasons we have to choose $m=\pm 8$ on the two sides of the orientifold.
Therefore we obtain in total
\begin{eqnarray*}
 USp(k) &\times& \prod_i SU(k+8 i) \; \; \; \mbox{  or  } \\
 SO(k) &\times& \prod_i SU(k - 8 i) 
\end{eqnarray*}
respectively. From strings stretching in between neighbouring
gauge groups we get bifundamentals $(\fund,\fund)$.
We have taken all the D8 branes that are required to cancel
the O8 plane charge to be far away. Of course including them
we get some more fundamental matter fields and new contributions to
the cosmological constant. 

The other possible situation considered in \cite{LL}
is that one of the 5 branes is stuck to the O8. In this case
all the groups stay $SU$, again with the left and the right ones identified,
leading to effectively half the number of gauge groups. In addition
the middle gauge group has a matter multiplet in the antisymmetric
or symmetric representation for positive/negative charge O8 planes.
In 6d we again have the effect of the cosmological constant
and as a result get
$$ \prod_i SU(k \pm 8i)$$
with antisymmetric/symmetric tensor matter in the first gauge group factor
in addition to the bifundamentals.

The other realization of $SO$ and $USp$ groups is to introduce
an O6 along the D6 branes. In discussing O6 planes we have to be
careful to take into account an effect first discussed in
\cite{shapere}: whenever an O6 passes through an NS5 brane, it changes
its sign. Originally this was argued for O4 and NS5 based on
consistency of the resulting 4d physics. Here we could basically
do the same: without including this effect on would construct
anomalous gauge theories. The sign flip miracously removes all
anomalies, providing strong arguments in favor of this assumption.
There also exist some worldsheet arguments, that further support
the existence of the flip \cite{chiralis}.

Due to this sign flip the O6 also contributes to the
RR charge cancellation. Since its charge is +4 on one side
and -4 on the other, the number of D6 branes ending to
the left and right also has to jump by 8. We therefor
are left with product gauge groups forming
an $$ \prod_i SO(N+8) \times USp(N)$$
chain with bifundamentals. Bringing the 3 ingredients - D8, O8 and
O6 - together, we can clearly cook up very complicated models.
I will write down some hopefully illustrative examples later on.

All these gauge theories I constructed with the help of orientifolds
are anomaly free \cite{ilkathesis,sweden,intblum}
upon coupling to the tensor multiplets associated
to the independent motions of the 5 branes. Note that in the case of the
O8, already a single NS5 brane gives rise to a tensor multiplet,
parametrising the distance to the fixed plane. There is no decoupled center 
of mass dynamics since presence of the orientifolds
breaks translation invariance along the 6 direction.

Again it is straight forward to put the theory on a circle.
I will restrict myself to the discussion of O8s.
The O6 case is pretty obvious. The only thing one has
to take care of is that only an even number of NS5 branes
is allowed, so that the ``global''
groups from semi-infinite D6 branes
are either both $USp$ or both $SO$
and we are able to close the circle.
We should either
take 2 negatively charged O8s with
32 D8 branes (that is the 16 physical D8 branes
and their 16 mirrors) or a pair of oppositely charged O8s.

\vspace{1cm}
\fig{Brane configuration yielding a product of two $USp$ and several
$SU$ gauge groups with 
bifundamentals and an antisymmetric tensor.}
{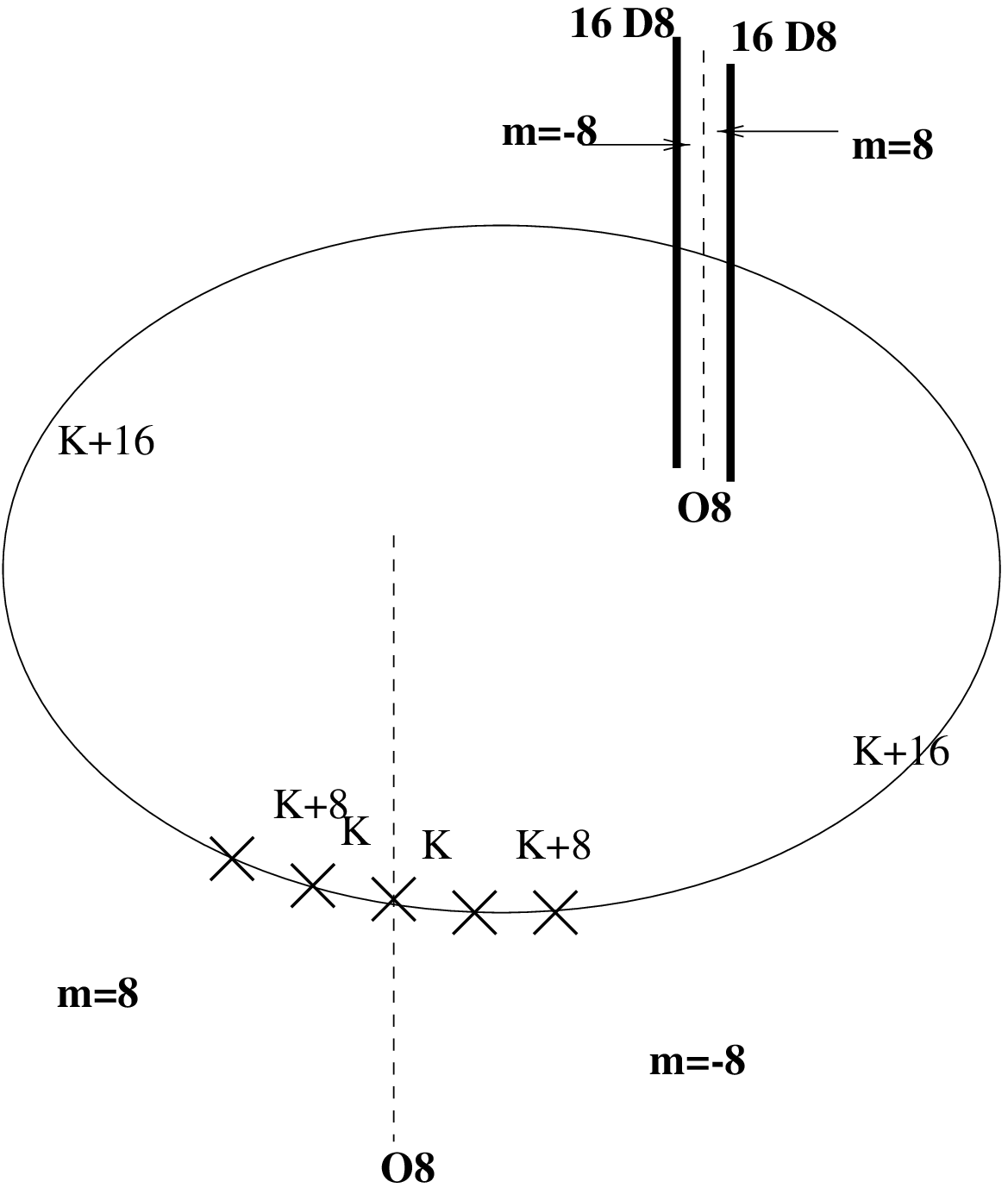}{7truecm}
\figlabel{\sp_cp}
\vspace{1cm}

Encoding the position of the D8 branes in a vector
$w_{\mu}$, where the entries in $w$ denote the number
of D8 branes in the gauge group between NS5 brane number $\mu$
and number $\mu+1$, one can write down formulas for the resulting
gauge groups. Clearly $\sum_{\mu} w_{\mu}= 32$. In addition
we introduce a quantity $D_{\mu}$ that encodes in a similar
way the information about the orientifold: the entry
corresponding to
the gauge group that includes an orientifold is 16,
since an O8 carries 16 times the charge of a D8.
All other entries are zero.

Let me only discuss the case of 2 negatively charged O8s with
32 D8 branes. This will have a dual description in
terms of $SO(32)$ small instantons. The case of
oppositely charged orientifolds describes
a disconnected part of the moduli space. It is dual to string theory
compactifications with reduced rank of the gauge theory, like the CHL string
\cite{wittenwithout}.

We have to distinguish three cases
\begin{itemize}
\item the number of NS5 branes is odd, so one of them has to be stuck
at one of the O8s and the others live in mirror pairs on the circle
\item the number of NS5 branes is even, one is stuck at each
of the orientifolds and the others appear in mirror pairs
\item the number of NS5 branes is even and all of them appear
in mirror pairs on the circle and are free to move
\end{itemize}

Let me discuss in some detail the case where $k+1$, the number
of NS5 branes, is odd. The other two cases can be worked out
the same way \cite{andreas3,intblum}.
As in Figure \sp_cp we have two orientifolds on the circle, both with
negative charge. An example also explaining the notation
is illustrated in the following figure:
\vspace{1cm}
\fig{Notation in the example of 7 NS5 branes on the circle}
{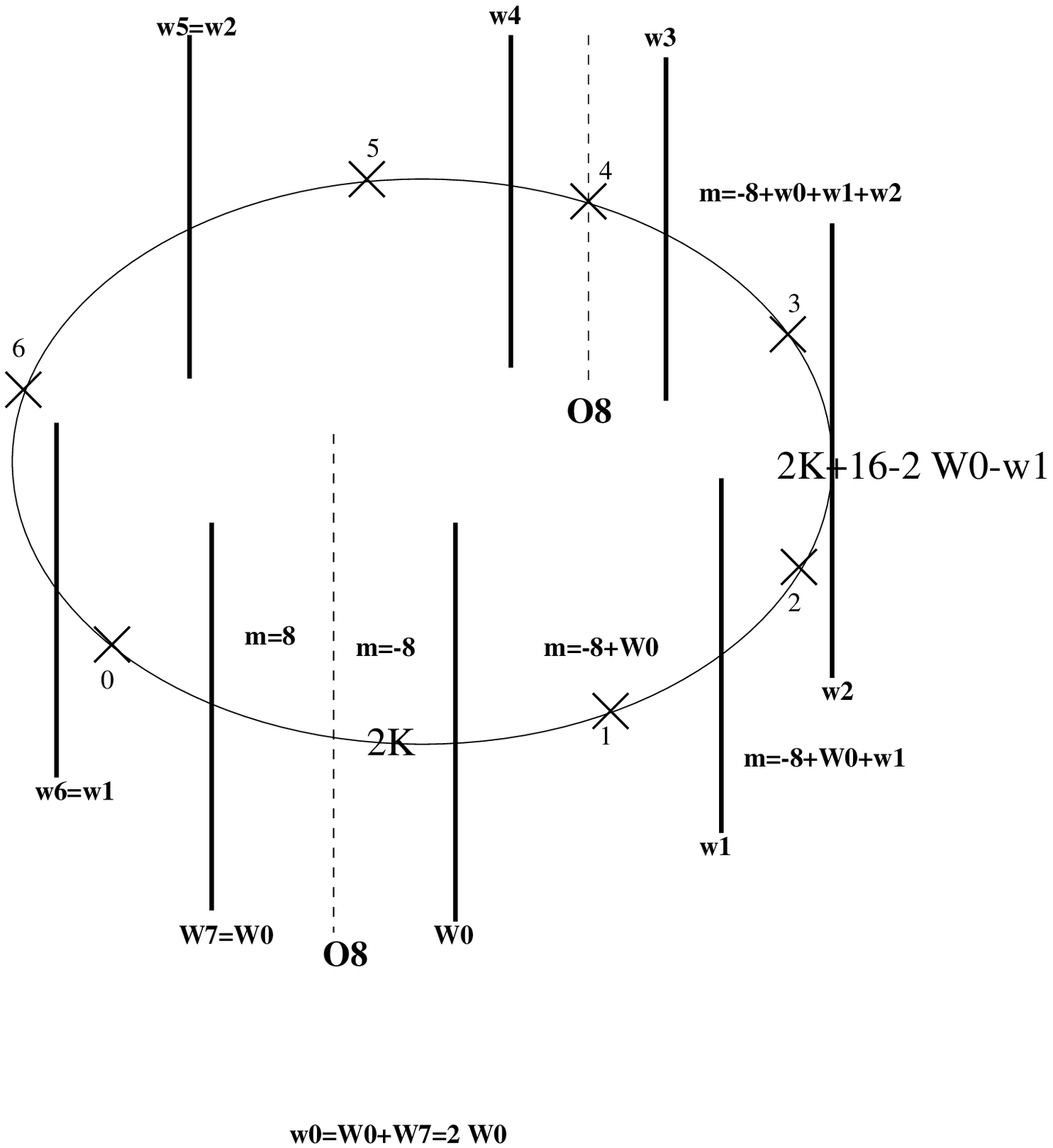}{9truecm}
\figlabel{\su3}
\vspace{1cm}

One of the 5 branes is stuck on the one orientifold.
We label the 5 branes with $\mu$ running from 0 to $k$ as in Figure \su3
\footnote{Note that in this numbering the stuck 5 brane is number $(k+2)/2$.}.
In addition there are 32 D8 branes. Let  again $w_{\mu}$ denote the number of
8 branes in between the $\mu$th and $(\mu +1)$th NS5 brane. Obviously
$\sum_{\mu} w_{\mu}=32$. Also symmetry with respect
to the orientifold plane requires $w_{\mu}=w_{k+1-\mu}$.
It is useful to slightly modify the definition
of the vector counting the D8 branes by introducing $W_{\mu}$
in order to take care of the effect
that for every gauge group whose color branes stretch over the
orientifold half of the $w_{\mu}$ branes have to be located on either
side. Therefore in these cases we define $W_{\mu} = \half w_{\mu}$ while
in all other cases $W_{\mu}= w_{\mu}$.

According
to the rules of how to introduce 8 branes and orientifolds
\footnote{The D8 branes have two effects: first they introduce a fundamental
hypermultiplet in the gauge group they are sitting in, second they decrease
the number of colors in every following gauge group
to their right by one per 5 brane in between,
since the value of the cosmological constant
$m$ is changed.} , we find that the gauge group
is
\begin{equation}
\label{odd2}
USp(V_{0}) \times \prod_{\mu=1}^{k/2} SU(V_{\mu})
\end{equation}
with
$$ V_{\mu} = 2K - \sum_{\nu=0}^{\mu-1} (\mu-\nu) W_{\nu} + 8 \mu $$

We have
$\half w_0 \fund_0, \oplus_{\mu=1}^{k/2} w_{\mu} \fund_{\mu},
 \oplus_{\mu=1}^{k/2}
(\fund_{\mu-1}, \fund_{\mu})$ and
$\asym_{k/2}$ matter multiplets (subscripts label
the gauge group) and $k/2$ tensor multiplets.

Using the symmetry property $W_{\mu}=w_{\mu}=w_{k+1-\mu}=W_{k+1-\mu}$,
$W_{0} = \half w_0$ the requirement of having a total
of 32 D8 branes $\sum_{\nu=0}^{k} w_{\nu} =32$ can be rewritten
as $$\sum_{\nu=0}^{k/2} W_{\nu} = 16 $$
which just says that 16 of the 32 D8 branes have to be on either
side of the orientifold.
Therefore $$\sum_{\nu=0}^{\mu-1} \mu W_{\nu} = 16 \mu -
 \mu \sum_{\nu=\mu}^{k/2} W_{\nu}$$
and (\ref{odd2}) becomes
$$V_{\mu} = 2K - 8 \mu + \sum_{\nu=0}^{\mu-1} \nu W_{\nu} +
\sum_{\nu=\mu}^{k/2} \mu W_{\nu} $$

It should come as no surprise that all these theories
are anomaly free. This was explicitely checked in \cite{intblum}.

\section{Extracting information about the FP theory}
\subsection{Global Symmetries}

So far I have discussed how to realize FP theories in a given
brane setup in the limit that all the bulk modes decouple.
In the spirit of Seiberg \cite{seiberg5d,seiberg6d}, 
identifying a decoupling limit
in which interacting physics survives on the brane is basically
an existence proof for these FP theories. As I discussed earlier, this
is a result, that could not be obtained from pure field theory
reasoning. Field theory gave us consistency requirements, branes
show us the existence.

But now we can move on and ask ourselves whether we can actually learn
something about the FP theories from their brane realization and
our (partial) knowledge of the ``embedding'' string theory.
One of the facts we can learn about are enhanced global symmetries.
Some information about the global symmetries can be gotten already from
field theory: once we perturb the FP by turning on the relevant
perturbation $1/g^2 F^2$, as discussed in length above, we will flow
to the free IR fixed point. Dealing with a free theory, it is
straight forward to identify the global symmetries. In our case
there will be for example a global flavor symmetry $SU(N_f)$. It will
be realized in the branes as the gauge group on the higher dimensional
branes, e.g. those branes whose worldvolume fields decouple together
with the bulk modes. In our case the $SU(N_f)$ global flavor symmetry
is realized on the worldvolume of the $N_f$ D8 branes. 

Once we go to the fixed point (that is turn off the $1/g^2$ perturbation
and go to infinite coupling) these symmetries are still present. But
sometimes it happens that we gain some new or enhanced global symmetries at
the fixed point. This can occur if some symmetry breaking operators
become irrelevant at the fixed point. This phenomenon is well known from
field theory where it goes under the name accidental symmetries:
some symmetries which appear to be valid in the IR may be broken by
higher dimensional operators. So while in the full theory they are
no good symmetries, they become better and better approximate symmetries
when we go to the IR and at the fixed point they are exact.

Enhanced global symmetries in the brane setup are again
realized as worldvolume gauge symmetries on the
decoupled heavy branes. Consider an easy example:
Take an HW setup, realizing pure $U(1)$ gauge theory with ${\cal N}=4$
in d=3. That is we just suspend a single D3 brane between two type IIB
NS5 branes. The coupling is given by the separation between the NS5 branes.
This theory is known to lead to a nontrivial IR fixed point, as can
be seen
following arguments like those in
 \cite{mirrorintsei}.
Going to the IR fixed point is equivalent to going to infinite
coupling, since in 3d the coupling has dimension of mass and therefor
sets the scale. Taking the coupling to infinity amounts to looking at
energies way below the intrinsic scale, that is the far IR of our theory.
The strong coupling fixed point corresponds to taking
the two NS5 branes to coincide. Since the worldvolume theory of IIB
NS5 branes is just SYM, this leads to an enhanced global $SU(2)$
symmetry at the fixed point. This same mechanism works in almost
all other HW-like setups.

However in 6d the situation is special and we won't get away with this
trick: as shown above in 6d the dynamics of the NS5 branes does not
decouple from the D6 brane gauge theory. In fact we needed to include the
tensors from the NS5 branes in the dynamics in order to cancel anomalies.
So coinciding NS5 branes do not correspond to global symmetries.
On the other hand the global $SU(N_f)$ symmetries on the
D8 branes are already visible at finite coupling, that is in the free
field theory. It therefore seems that in the 6d setup we do not see
enhanced global symmetries. There is however a way to obtain
such enhanced global symmetries. Roughly speaking one can go to
the fixed point by taking the ``surrounding'' IIA string coupling to infinity
instead of letting the NS5 branes coincide \cite{zaff2}.

It is well known that in the strong coupling limit
the $SO(2N)$ global symmetries on $N$ D8 branes coinciding on top
of an O8 plane (with $N=1,\ldots,7$) is enhanced to $E_{N+1}$
\cite{notesond}.
The missing gauge bosons are realized by D0 branes stuck on the
D8 brane, whose mass goes like $\frac{1}{g_s l_s}$ and hence they
become massless in the strong coupling limit \cite{shamiteva}.
This leads to fixed points with exceptional enhanced global
symmetries. The best known example is the one with 
no D6 branes and just one NS5 brane and 7 D8 branes all sitting 
on top of the O8. At strong coupling this becomes Horava-Witten
M-theory with one M5 brane sitting on the ``end of the world brane''. This
is the well known realization of the small $E_8$ instanton \cite{smalle8}.

\subsection{Deformations and Phase Transitions}

It is very easy to see deformations of the fixed point theories
in the brane picture. For example we can move
one of the NS5 branes away from the fixed point. If we do this
along the 6 direction, we go back on the Coulomb branch of
our theory, if we move it off along 789 we turn on some FI
terms. For the FI terms it doesn't matter at all whether
we turn them on at the fixed point or at any point on the Coulomb branch.

More interesting are deformations that are not possible
away from the FP, that is if the Coulomb branch and a Higgs
branch meet and the FP sits at the common origin. In this case
we can actually perform a phase transition. That is we start
with branes realizing the theory on its Coulomb branch as one
phase of string theory. Tuning the moduli we can go to the fixed point.
At the fixed point we see a new deformation that was not possible before,
taking us out to the Higgs branch. Tuning the Higgs moduli away
from the fixed point, we have performed a transition to a topologically
distinct phase of string theory. 

Let me show one example of such a transition. In hindsight of
the relation to geometry which I will exhibit in the next Chapter,
I will call this a small instanton transition. It was first
discussed in the context of 6d brane setups in \cite{zaff2}.

\vspace{1cm}
\fig{The small instanton transition trading 1 tensor for 29 hypers.
}
{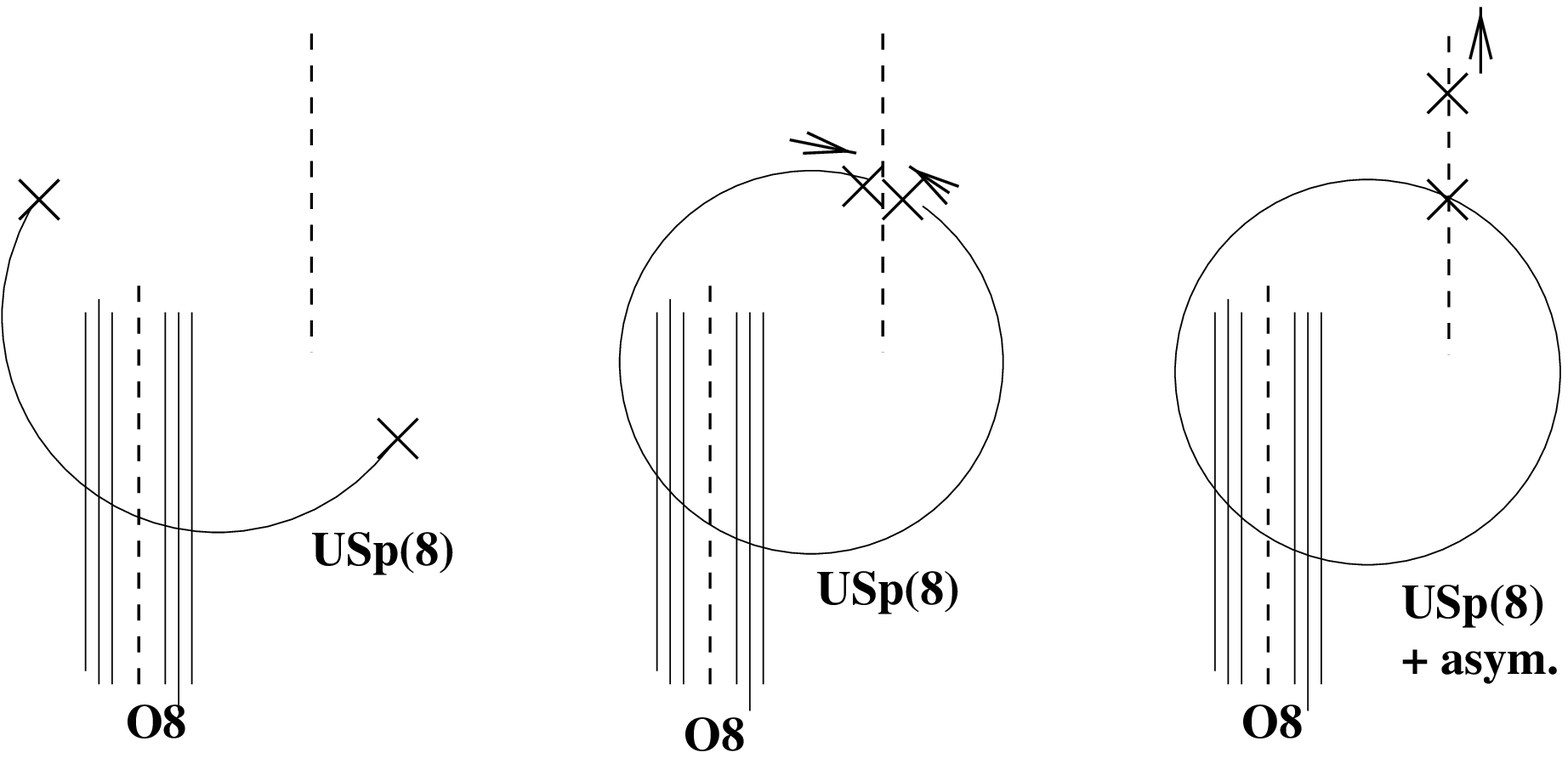}{10truecm}
\figlabel{\who}
\vspace{1cm}

We start with a special case of a theory with 2 O8s and 32 D8 branes
on a circle. All of the D8s are on top of a single O8, so
we will see an $SO(32)$ global symmetry. In addition
we will study a single pair of NS5 branes. The cosmological
constant on each side will be $\pm 8$ respectively, so
that we have to put at least 8 D6 branes on this interval. They
will all terminate on the NS5 branes. The corresponding
gauge group can easily be read off to be $USp(8)$. There are
going to be 16 fundamental hypermultiplets from the D8 branes
and one tensor multiplet describing the position of the NS5 and its
mirror on the circle. Tuning the scalar in the tensor multiplet
we can reach a fixed point by letting the NS5 collide with its mirror
at the far orientifold. At this point we see a new deformation:
one of the NS5 branes can be moved of in 789 space. 
According to our rules this will still be a $USp(8)$ gauge
theory. However one obtains an additional hypermultiplet
in the antisymmetric tensor representation from
the far orientifold. This gives us $27+1=28$ new hypermultiplets.
In addition one of the NS5 branes moves freely in 789 space,
corresponding to a 29th hypermultiplet. But now
the second NS5 brane is stuck at the orientifold, so
there is no tensor multiplet left \footnote
{The gauge theory has $a=c=0$ and hence doesn't need a
Green Schwarz mechanism to cancel the anomaly.}.
All in all this describes a phase transition trading 1 tensor
for 29 hypers. These numbers are required for any theory that
can be coupled consistently to gravity: the contribution
of a single tensor multiplet to the gravitational anomaly
in 6d is the same as that of 29 hypers. So if the gauge theory
can be coupled consistently to gravity in one phase 
(and since our setup is a limit of string theory we should
definitely be able to do so), then the other phase
which lost one tensor and has the same number of vectors has
to have 29 more hypermultiplets in order to have a consistent
coupling to gravity, too.

\subsection{Correlators from branes}

So far we have seen that quite a lot can be learned about 6d
fixed points using the brane realization I have presented in this work.
But the things we are really interested in are the correlation
functions defining the superconformal field theory. So far they
have escaped our control.

Recently a tool has emerged that might finally help us to tackle
this goal. According to Maldacena there is a duality that
relates the theory on the worldvolume of a given brane setup to
gravity in the background of the branes. At large
$N$ the gravity calculation reduces to a classical calculation.
It will be of great interest to apply these techniques to
the theories I have been discussing. Some progress
in this direction has been made in \cite{ferrara} where
the dimensions of the chiral operators have been found.
In this way the brane setups I have presented 
serve as the natural starting point for any more elaborate
analysis of the physics of these 6 dimensional theories.

\subsection{A chiral / non-chiral transition}

By the very nature of Hanany-Witten setups there is a close
connection between physics in $d+2$ and $d$ dimensions. As I
explained, if we realize our desired $d$ dimensional gauge theory on
the worldvolume of $Dd$ branes on an interval, matter multiplets
can easily be incorporated by orthogonally intersecting $D(d+2)$
branes. Now instead of just using $D(d+2)$ branes we can use
a full Hanany-Witten setup, corresponding to a gauge theory
in two more dimensions with twice the amount of supersymmetry.
Applying this idea let me use our 6d brane setups to learn
something about 4d physics. We will follow what happens
to our setup under the same brane motions we have performed above,
again giving 
rise to phase transitions, this time for ${\cal N}=1$
supersymmetric 4 dimensional vacua. As a highlight we will present
a transition between a chiral and a non-chiral phase \cite{chiralwithhan}
\footnote{The brane realization of this chiral theory
was also discussed in \cite{chiralis,chiralLL}.}.

Let me start with the standard ingredients realizing
an ${\cal N}=1$ theory in 4d, that is we will have to deal with
NS5 branes, D6 branes and D4 branes. We will allow NS5 and D6
to be rotated from 45 into 89 space by an arbitrary angle.
As usual we will refer to NS5 branes if they live along 012345
and NS5' if the angle is 90 degrees, that is they live
in 012378 space. When using more general branes the
we will call them A,B or C NS5 branes and
the following conventions will be used for the angles
\fig{
Notation for the angles in the product group setup.}
{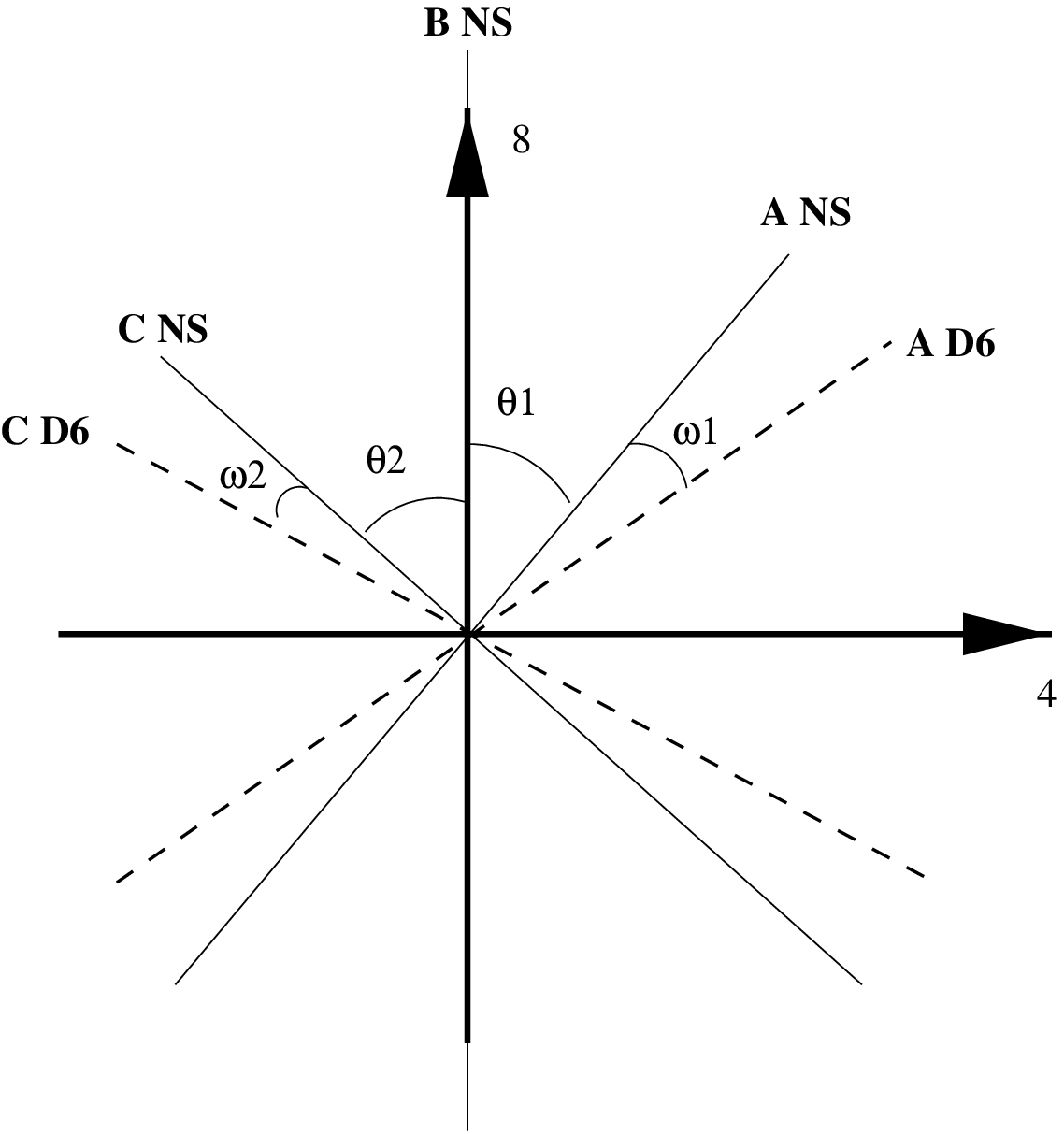}{6.5truecm}
\figlabel{\notation}
\vspace{.5cm}
In what follows I will at some point replace a D6 by a full
6d HW setup, that is by D6 branes with NS5' branes embedded in them.
In our 6d analysis we already realized that in order to get
interesting phase transitions we should really include an
orientifold. This way a single NS5' brane will be stuck, while
bringing it together with another one we will realize
a non-trivial fixed point which has a new branch opening up.

Hence the easiest setups we are going to study have at least
2 group factors and 3 NS5 branes: the NS5' stuck at the orientifold
and then another NS5 and its mirror. In order to study the resulting theory
it is very helpful to first discuss the matter arising without
the orientifold.
That is we first study the following setup for
a product gauge group analyzed in\cite{brodie}:

\vspace{.5cm}
\fig{The product gauge group considered in \cite{brodie}. The thick lines are
NS branes at arbitrary angle in 45-89 direction. The D6 branes are parallel to
the A and C NS5 branes.}
{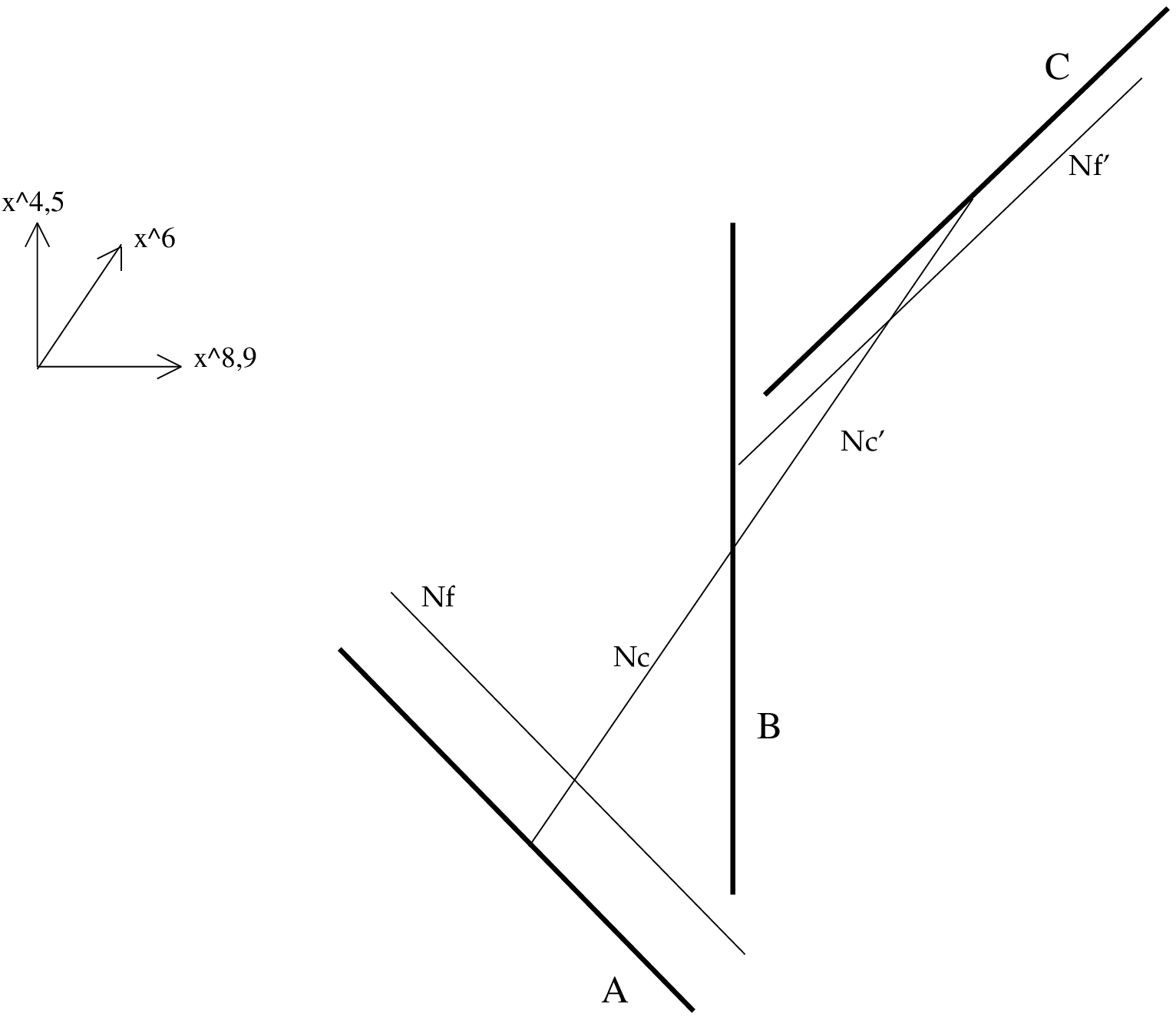}{6.5truecm}
\figlabel{\ils}
\vspace{.5cm}

Let me choose the B branes to be still NS5' branes (just fixing
the overall orientation). 
Let $(k,k',k'')$ denote the number of A, B and C type NS5 branes respectively.
The resulting SYM will have $SU(N_c) \times SU(N_c')$
gauge group with bifundamental matter $F$ and
$\tilde{F}$ and fundamental chiral multiplets $Q$, $\tilde{Q}$ and $Q'$,
$\tilde{Q}'$ from the D6 branes in the first and second group
factor respectively. In addition there will be adjoints
$X_1$ and $X_2$. For $k=k'=k''=1$ they will be massive and
can be integrated out.

According to \cite{brodie} the $(k,k,k)$ case then leads to a superpotential
\begin{equation}
\label{SUPO}
W=m_1 X_1^{k+1} + m_2 X_2^{k+1} + X_1 \tilde{F} F + X_2 \tilde{F} F +
\lambda_1 Q X_1 \tilde{Q} + \lambda_2 Q' X_2 \tilde{Q}'
\end{equation}
while the $(k,1,k)$ case leads to
\begin{equation}
\label{SUPO2}
W=-\frac{1}{2} \left ( \frac{1}{m_1} + \frac{1} {m_2} \right )
( F \tilde{F})^{k+1}+
\lambda_1 Q X_1 \tilde{Q} + \lambda_2 Q' X_2 \tilde{Q}'.
\end{equation}
In the following, I will identify these superpotentials by their triple 
number, $(\cdot,\cdot,\cdot)$.
The coefficients in the superpotential are determined in terms of
the angles as
\begin{eqnarray}
\lambda_1 &=& \sin \omega_1 \\
\lambda_2 &=& \sin \omega_2 \\
m_1 &=& \tan \theta_1 \\
m_2 &=& \tan \theta_2.
\end{eqnarray}
As usual all these statements can be verified by studying
allowed brane motions and comparing with the classical
moduli spaces of the gauge groups in the presence
of these superpotential terms.

Now we are in a position to introduce the orientifold.
I will put an O6 on top of the middle NS5' brane,
so that the NS5' is embedded inside the O6 and we
are basically dealing with a 6d HW setup.
In principle we can also include an O6'. However this won't
be related to any 6d theory. In order to be symmetric
with respect to the orientifold we have to restrict ourselves
to
\begin{itemize}
\item $N_c=N'_c$
\item $N_f=N'_f$
\item $\theta_1=-\theta_2$ and $\omega_1 = -\omega_2$
(the A branes have to be mirrors of the C branes)
\item equal number of A branes and C branes,
that is we consider only $(k,k',k)$ configurations
\item one can only include an O6 or an O6$'$ (the B NS5$'$ brane, carrying one
unit of NS charge,  has to be
self-mirror)
\end{itemize}
Since one possible deformation will be to move the NS5' brane
along the orientifold (corresponding to a baryonic
branch of the gauge theory) I will also have to study
setups with $k'=0$. This will again allow
me to uniquely fix the matter content and the interactions
by comparing to the flat directions expected from classical
field theory.
So we want to consider the following possible setups:
\vspace{.5cm}
\fig{Theories with orientifolds and their deformations}
{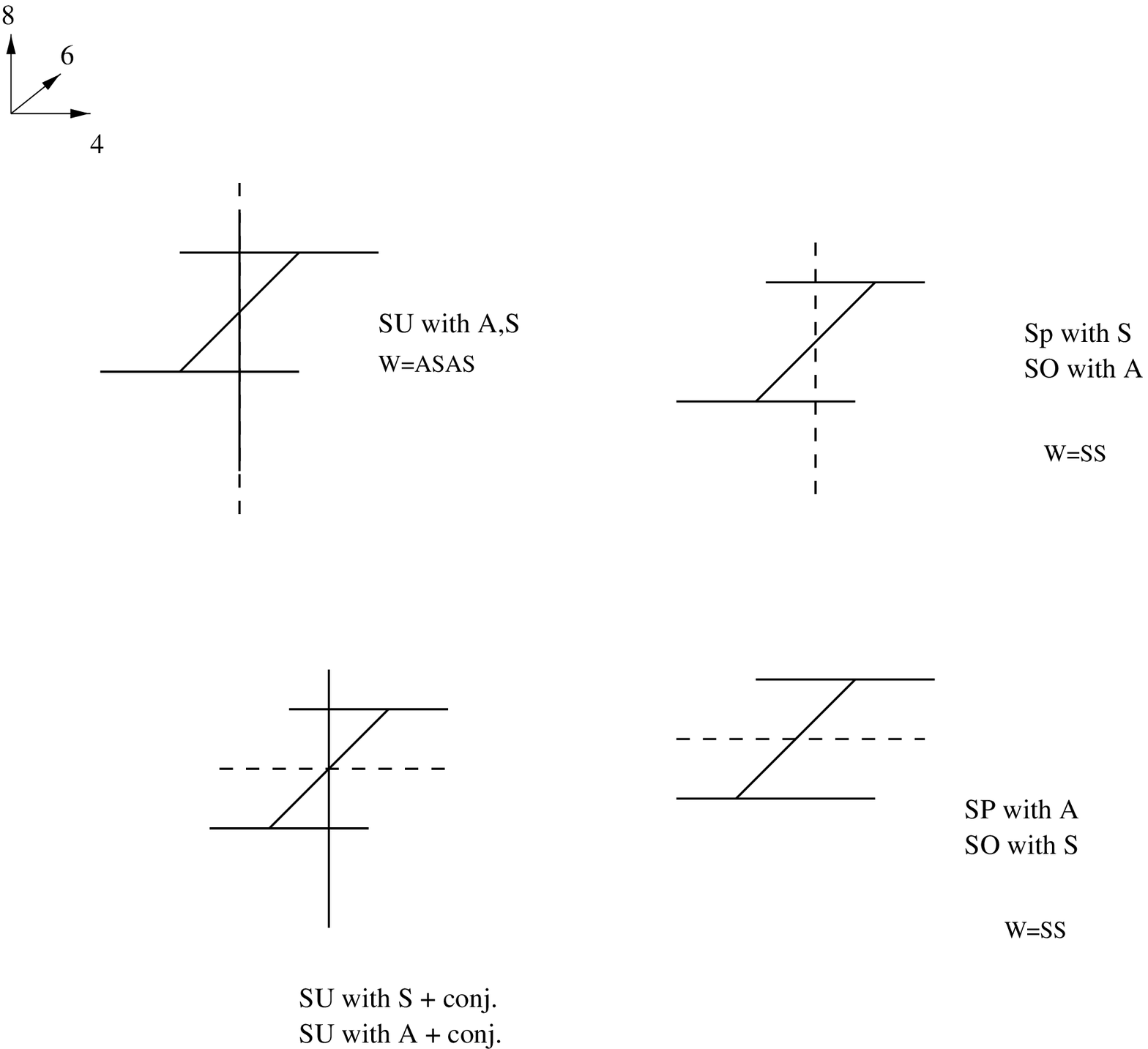}{11truecm}
\figlabel{\seat}
\vspace{.5cm}

On the left in figure \seat we find the two possible projections
of the product gauge group setup, the right hand side displays
two possible setups involving just 2 NS5 branes and an orientifold, which
arise as possible deformations of the theories on the left, as we will see
in the following. Let us briefly summarize the resulting gauge groups
and matter contents. A more detailed discussion of this identification
will be presented in the following sections.

The configuration in the upper right corner of figure \seat
is just the ${\cal N}=2$ setup analyzed in \cite{LL}. The corresponding
gauge group is $SO$ ($USp$) depending on the sign of the
orientifold projection. Rotating the NS5 branes breaks
${\cal N}=2$ to ${\cal N}=1$ by giving a mass to the adjoint chiral 
multiplet
in the ${\cal N}=2$ vector multiplet. We are left with an $SO$ ($USp$).
Note that this mass is
already infinite at $\theta=\frac{\pi}{4}$,
since this time it is given by the angle
between
the outer NS5 branes which is twice the angle $\theta$
between outer NS5 and NS5$'$ that
determined the mass before. Rotating further I claim that instead of
the adjoint tensor that became infinitely heavy a new tensor
with the opposite symmetry properties is coming down from infinite mass.
As usual the way to prove this is to compare the possible
brane moves 
with results from classical field theory.
Note that if we were dealing with D branes instead of the NS5 branes
the rotation we performed would precisely change their worldvolume
gauge theory from $SO$ to $Sp$. It's therefore reasonable to
assume that something similar happens to the NS5 branes, too.
This way we identify the gauge group corresponding to the brane
configuration in the lower right corner of figure \seat to
be an $SO$ ($USp$) gauge theory with a symmetric (antisymmetric) tensor.

In the two pictures on the left the two $SU$ factors are identified
under the orientifold projection. One adjoint field is present, whose
mass is given by the angle $\theta$ between NS5 and NS5$'$ brane. In
addition there are degrees of freedom that gave rise to
bifundamentals in the product gauge groups.
According to the analysis
of \cite{chiralwithhan} one finds that in the orientifolded theories
the O6' will give rise to a full flavor
\footnote{that is the tensor and its conjugate} of symmetric 
or antisymmetric tensors for the
O6$'$, depending on the sign of the projection.
The really interesting case is the one with the O6.
This now is part of a 6d HW setup. The O6 changes sign when
passing through the NS5'. This means that in order to conserve
RR charge it has to have 8 half D6 branes embedded in it
on one side. The corresponding matter content will
be an antisymmetric tensor, a conjugate symmetric
tensor and 8 fundamentals.
Analyzing the field theory one indeed finds two baryonic branches
corresponding to moving the NS5' to the $SO$ or $USp$ side of
the O6 and leaving one of the two theories on the right side
of figure \seat respectively. Note that this is a chiral theory
with anomaly free matter content. The 8 fundamentals are precisely
what we need to cancel the contribution from the tensors.

Now we have all the ingredients we need in order to perform the transition.
We perform the ``small instanton transition'' very similar
to how it was performed in the 6d case and follow through what
happens to our 4d physics.

\vspace{1cm}
\fig{A small instanton transition which lead to chirality change in the
spectrum. An NS5$'$ 
brane comes from infinity in the $x^7$ direction and combines
with the stuck NS5$'$ at the origin. At this point they can both leave the
origin in the $x^6$ direction. The resulting four dimensional theory is no
longer chiral.}
{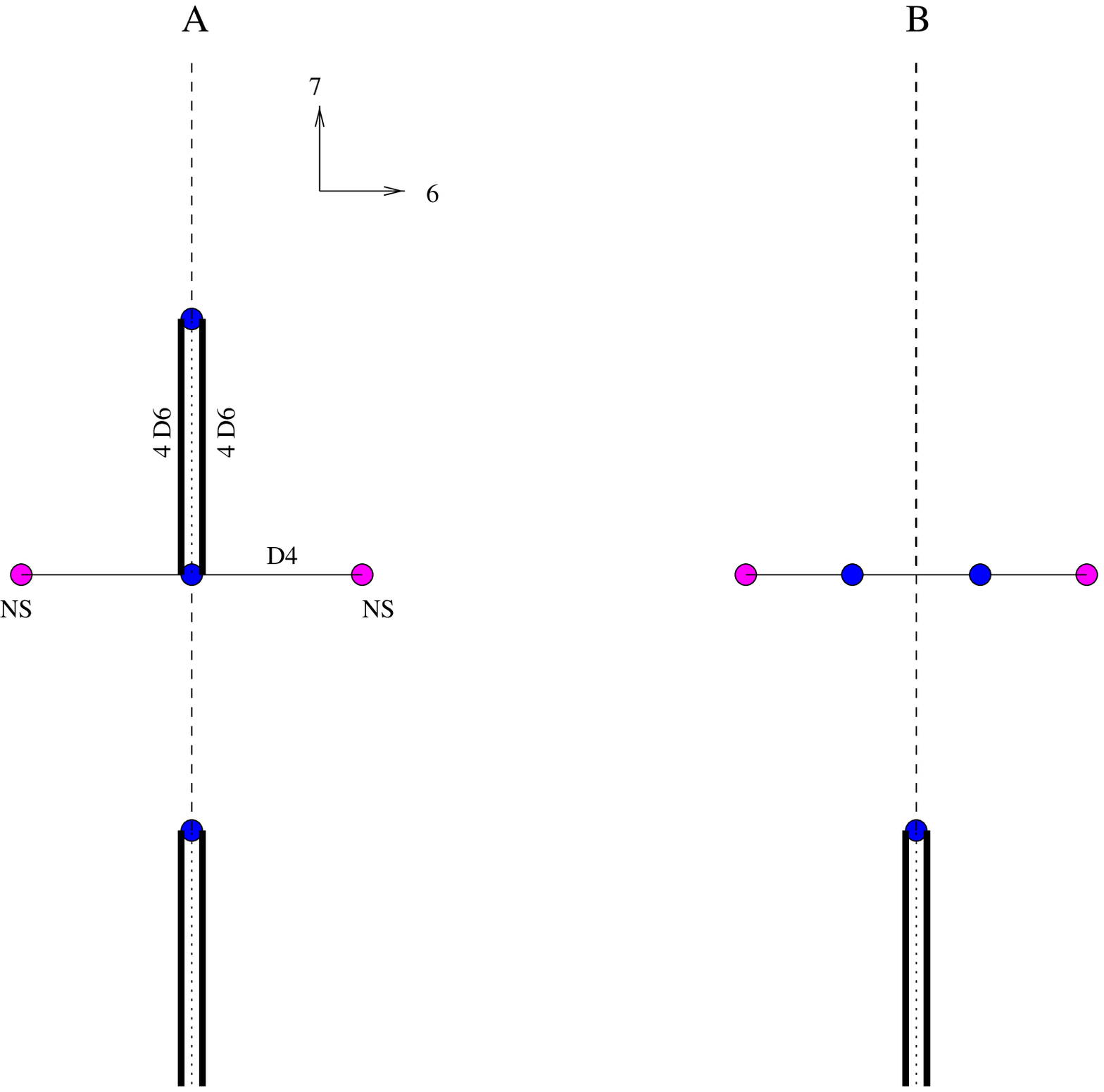}{10truecm}
\figlabel{\smallin}
\vspace{1cm}

Consider the configuration depicted in figure \smallin A.
It realizes the chiral theory I just described.
I included two other NS5' branes far away so that
they have no effect on the dynamics. We would like to
bring them in from infinity in order to perform the transition.
We will call them upper and
lower branes, respectively. The original NS5$'$ brane, to which the four
dimensional system is attached, will be called central.

The NS5$'$ branes are not allowed to move in the 456
directions, since they can only move off the orientifold
as a mirror pair. On the other hand, the relative
motion of the NS5$'$ branes along the $x^7$ direction corresponds to a change in
the two tensor multiplets in the 6d theory. 
There is, however, an option for a pair of NS5$'$ branes to move in the 456
directions. We can move two NS5$'$ branes to touch in the $x^7$ direction.
At this point, the orientifold planes from above and below the pair of NS5$'$
branes are identical, as is clear from figure \smallin.
 From a six dimensional point of view such a motion corresponds to taking one of
the gauge couplings to infinity and thereby obtaining
 strings with vanishing tension - a non
trivial fixed point.
The pair of NS5$'$ branes can then move in the 456 directions, as in figure
\smallin B.
The resulting six dimensional gauge group is now completely broken.

Let us go back to reinterpret this transition in our four dimensional system.
When the NS5$'$ branes are far, we have our chiral theory.
When the NS5$'$ branes move in the 456 directions, the theory is different.
We can read it off from figure \smallin B. We have $SO(N_c)\times SU(N_c)$
with
a symmetric tensor for the $SO$ group and a
pair (chiral and its conjugate) of bi-fundamental fields. Note
that this is a non-chiral theory as advertised.

\chapter{The equivalence of the various approaches}

As mentioned in the beginning there are several ways discussed
in the literature to embed gauge theories in string theory.
So far I have only discussed in detail the Hanany-Witten approach.
The advantage of this approach is that the setups and especially
their moduli and deformations have a very intuitive meaning.
One just has to move around flat branes. In this section I'd like
to show that indeed all the approaches are equivalent, in the sense
that the resulting string theories are dual to each other. This
way I will show that the phase transitions I studied in the previous
chapter by moving branes together and apart again in a different fashion
can indeed be interpreted on the dual side as a transition from one
geometrical compactification to a topologically distinct vacuum. 
This analysis
will hopefully provide one further step towards an understanding of
the dynamical processes relating the various string vacua, finally
leading to a determination of the true vacuum of string theory.

\section{The duality between orbifold and NS5 brane}

The basic relation for our discussion will be the observation
of \cite{dualNS} that string theory in the background of
an NS5 brane is T-dual to string theory on an A-type
ALE space.
This is already an example of an equivalence between branes
as probes and geometrical engineering in the restricted case
of theories with 16 supercharges. Embedding a given 6d field theory
in string theory via probing flat IIA/B with an NS5 brane or
by engineering it via IIB/A on an ALE space is guaranteed to
give the same results, since these two setups are actually equivalent
as string backgrounds.

Let me discuss this duality in some detail. I will mostly follow
the reasoning of \cite{9708086,wittenpq}. In order to apply T-duality
we should look for a $U(1)$ isometry of our solution. $C^2/Z_{k+1}$
does have some $U(1)$ isometries, 
but the corresponding radius grows as we move away from the orbifold
point, so that at infinity it becomes infinite. The would
be dual should then be some geometry with a vanishing circle at infinity
and hence would be hard to interpret.

In order to make some progress we consider a slight modification of
the original geometry to something that looks asymptotically like
an $S^1$ fiber over $R^3$ and still has the A$_k$ singularity at the origin 
\footnote{
The decoupled physics in 6d won't
depend on the radius of the circle, as one could see by recalling
once more the decoupling limit. It is only encoded in the local singularity
and not in the global structure.}. These are precisely
the properties of a $k+1$ centered KK monopole in the limit that
all the centers coincide. Now we just T-dualize
the $U(1)$ isometry we introduced by hand. One finds that
the dual metric is the NS5 brane metric as advertised. A short-cut
argument is that T-duality exchanges the two $U(1)$ fields
in 9d generated by the component of the metric and the B-field
around the circle respectively. Since the KK-solution
is the monopole of the former while the NS5 brane is the monopole
of the latter, the two solutions get interchanged under T-duality, too.

\section{Hanany Witten versus branes at orbifolds}
\subsection{The orbifold construction}
Now we want to move on to the more interesting case of 8 supercharges.
Let me first demonstrate that Hanany-Witten setups
are equivalent to branes at orbifolds, a particular realization
of the branes as probes idea. Here I will follow
the discussion in \cite{zaff2,andreas3}. 
We consider D5 branes moving on top of a $C^2/\Gamma$ orbifold space.

At this point it is necessary to actually work out
the field theory of $K$ D$p$
\footnote{Of course we have to restrict
ourselves to $p \leq 5$ so that
the transverse space is big enough to carry an 4d ALE space. This
restriction again just reflects the classical fact that theories
in above 6 dimensions have at least 16 supercharges.}
branes moving on the orbifold. So let me briefly
review the relevant analysis due to Douglas and Moore
\cite{douglasmoore}.

Even though the orbifolded space is singular, string theory physics on the
orbifolded space is smooth. String theory automatically provides
additional modes in the twisted sectors that resolve the singularity.
In the case of an orbifold these twisted sectors are taken care
of by including also the $k$ $Z_{k+1}$ mirror images of each original D$p$
brane.
At the orbifold point the original D$p$ coincides with all its mirrors,
so that we start off with maximally supersymmetric SYM in $p+1$
dimensions and $U((k+1) K)$ gauge group. In this
theory we impose the projection conditions by throwing away
everything that is not invariant under $Z_{k+1}$. Since
we chose our $Z_{k+1}$ action in such a way that it is a
subgroup of $SU(2)$ (remember that this is what we mean
by the orbifold limit of an ALE space), we are guaranteed that
the remaining spectrum will be that of a theory with 8 supercharges,
so it is enough to study only the bosons and the fermions just follow
by SUSY. The $Z_{k+1}$ acts
\begin{itemize}
\item on the $SO(4)$ subgroup of the R-symmetry, acting on the scalars
and fermions given in the obvious way by embedding of the geometric action
in internal spacetime (the scalars are the spacetime Xs!)
\item on the $U((k+1) K)$ Chan Paton indices according to
some representation of $Z_{k+1}$.
\end{itemize}
If we want to describe D3 branes which are free to move away from
the orbifold fixed point we should restrict ourselves to embed
the orbifold group into the Chan Paton factors via
the regular representation $R$ of $\Gamma$, that is the $| \Gamma |$ dimensional
representation that accounts for every mirror once. 
Note that this representation is reducible and decomposes
in terms of the irreducible representations $r_i$ as $R=\oplus_i dim(r_i) r_i$.
Doing so we take into account the D3 branes and all its mirrors, as
is required if we want to have branes that are free to move away from the
fixed point.
Other representations can be considered as well (at least at a classical
level) and lead to fractional branes \cite{douglasmoore,Enhanced,
Diaconescu,C3orbi}. We will have more to say about these in what follows.
For the moment let's restrict ourselves to the regular representation.

We get the following actions on the fields ($i,j$
labelling the columns, of length $K$, of vector
indices transforming under the same irreducible representation
$r_i$ of $\Gamma$ and $a=1,\ldots,4$ a vector
index of the $SO(4)$ R-symmetry.

{\bf Vectors:}
$$A^i_j \rightarrow r_i \times \bar{r}_j A^i_j$$
leaving a $\prod_i U(dim(r_i) N)$ gauge group, since
$r_i \times \bar{r}_j$ contains the identity only for $i=j$.
For our case of $\Gamma=Z_{k+1}$ all the $r_i$ are one dimensional,
since $Z_{k+1}$ is abelian. Together with the gauginos
and eventually the scalars corresponding to motions transverse to
the ALE space and transverse to the brane for $p <5$
\footnote{e.g. a D3 brane in 0123 with the ALE space in 6789
will have a complex scalar corresponding to 45 motions that
does not transform under the spacetime action of
the orbifold and is hence projected as the vectors.},
they will combine into full VMs.

{\bf Scalars:}
$$\Phi^{i a}_j \rightarrow a^4_{ik} r_k \times \bar{r}_j \Phi^{i a}_j$$
where $a^4_{ik}$ denotes the Clebsch Gordan coefficient in the decomposition
of $r_i \times {\bf 4} = \oplus_k a^4_{ik} r_k$, where
where {\bf 4} denotes the 4 dimensional representation of the scalars under
the R symmetry. We hence obtain $a^4_{ik}$ scalars transforming as
bifundamentals under the $i$th and $k$th gauge group. For $i=k$
this is interpreted as an
adjoint. 
For $Z_{k+1}$, ${\bf 4} \rightarrow 2 r_1 + 2 r_{-1}$ and
we obtain bifundamentals in neighbouring gauge groups. Together
with the fermions these scalars will form HMs.

\subsection{The classical branches}

Having constructed the field theory corresponding to D5 branes on
top of the singularity we can move on and study its classical branches.
The Higgs branch corresponds to moving the branes away from the singularity,
its metric will be the metric of the ALE space (the resolved orbifold) itself.
Non-trivial fixed points will occur when the branes actually sit
on top of the singularity. For $p<5$ we will also have a Coulomb branch
from the scalars in the VM. This will have to be interpreted in terms
of ``fractional branes''. For $p=5$, that is in 6d, there is
no such Coulomb branch associated to the VM. However we already found
before that in 6d we need to couple the gauge theory to tensor multiplets
in order to cancel the anomalies. These bring in
new scalars which parametrize a branch that
is now referred to as the Coulomb branch (even though
it has quite a different interpretation from
in the other cases).
In the HW setup these tensors were
automatically provided by the NS5 brane motions.
In the orbifold construction they correspond to degrees
of freedom coming from reducing the 10d two-form on
the vanishing two cycles. Of course these degrees of freedom
are also there in fewer than 6 dimensions. But 
as in the HW case, taking the limit in which the bulk fields decouple
we find in $d<6$ that these extra matter multiplets decouple as well,
whereas in $d=6$ we have to keep them since their interaction strength is
of the same magnitude as the 6d SYM coupling.

\subsection{Including D9 branes and orientifolds}

Let me for a moment just discuss the classical theory, so that
we do not have to worry about anomalies and charge cancellation.
We are hence free to study $K$ D5 branes in the background of
$N$ D9 branes and an ADE singularity. Without the D9 branes
the Higgs branch still describes the ALE space itself. Without the
ADE singularity, the Higgs branch corresponds to the moduli
space of $K$ $U(N)$ instantons, since
a D$p$ brane is just the instanton within the D$p+4$ brane
\cite{BranesWithin}. This can be put together
naturally to the statement that the theory with ALE space
and the D9 branes together yield a theory whose Higgs branch
is the moduli space of $K$ $U(N)$ instantons on the ALE
\cite{douglasmoore,intblum}. 
Kronheimer and Nakajima \cite{kronheimer} have introduced these
theories as a hyper-Kahler quotient construction previously,
exactly with the goal of describing these moduli spaces. It
is beautiful to see that they naturally appear
in the classical analysis of brane theory.
Of course the same Higgs branch arises
in any other D$p$ D$p+4$ system with the D$p+4$ brane wrapping
the ALE space.

Besides the obvious data $N$ and $K$ we need one more piece
of information: the Wilson lines at infinity. 
Since $\pi_1$ at infinity is the discrete ADE
group $\Gamma$ such
non-trivial Wilson lines can exist. They are characterized
by a matrix $\rho_{\infty}$ representing $\Gamma$
in the gauge group. That is $\rho_{\infty}$ is given
by
$\rho _\infty =\oplus _\mu w_\mu R_\mu$, where $R_{\mu}$
are the irreducible representations of $\Gamma$
and $w_{\mu}$ some integers, so that $\rho _\infty$ really
is a representation of $\Gamma$. In addition for
$\rho _\infty$ to be an element of the $U(N)$ gauge
group it better be an $N \times N$ matrix, so
we have to demand that $\sum_{\mu} n_{\mu} w_{\mu} =N$
where the $n_{\mu}$ are the dimensions of the
corresponding representation. To keep the notation
readable I will from now on focus on A-type (that is abelian $Z_{k+1}$) orbifold
groups only, so that all $n_{\mu}=1$. The formulas for the general
case can be found in \cite{intblum}. In addition
one could also include non-trivial first Chern classes. They
would correspond to turning on B fluxes in the type IIB background.
For simplicity I will also neglect those. From here on it
is a tedious but straightforward analysis to find the resulting
gauge group. It is $\prod_{\mu=0}^k  U(V_{\mu})$ where
\footnote{ $C^{-1}$ denotes the inverse Cartan metric of $\Gamma$,
that is for $Z_{k+1}$ $C^{-1}_{i<j}=i (k+1-j)/(k+1)$.}
$$V_0=2K, \; \; \; V_{i \neq 0} = 2K+\sum_{j=1}^k C^{-1}_{ij} w_j .$$
As matter we have bifundamentals under neighboring gauge groups
and $w_{\mu}$ fundamentals in the $\mu$th gauge group.
This indeed coincides with the hyper-Kahler construction.
Now let us discuss the quantum theory. According to
\cite{int} anomaly freedom of the above gauge theory demands
$N=0$. This is no big surprise: we are not allowed to put
any D9 branes in type IIB.

Similarly one can discuss $K$ $SO(N)$ instantons. 
In
the brane picture this is done by introducing an additional
O9 plane. 
Here
anomaly freedom will demand $N=32$, since
we are now dealing with type I. However the classical
construction can still be carried out for any $N$, yielding
again a description of instanton moduli spaces in terms
of the classical Higgs branch of a given gauge theory. 
The resulting gauge group is modified slightly:

\begin{itemize}
\item For $k+1$ even there is an additional discrete choice
for $\rho_{\infty}$ due to the fact that the space-time
gauge group is really $Spin(32)/Z_2$ and not $SO(32)$. The
$Z_2$ is generated by the element $w$ in the center
of $Spin(32)$ which acts as $-1$ on the vector, $-1$ on the spinor of
negative chirality, and $+1$ on the spinor of positive
chirality. Because only representations with $w=1$ are in the
$Spin(32)/\IZ _2$ string theory, the identity element $e\in \Gamma _G$
can be mapped to either the element $1$ or $w$ in $Spin(32)$.
\item The orientifold projection will of course change the gauge
group. This is again a straight forward exercise in applying projection
operators and one finds for example for $k+1$ odd:
\begin{equation}
\label{odd}
Sp(V_{0}) \times \prod_{\mu=1}^{k/2} SU(V_{\mu}).
\end{equation}
The matter content consists of hypermultiplets transforming according to
$\half w_0 \fund_0$, $\oplus_{\mu=1}^{k/2} w_{\mu} \fund_{\mu}$, $
 \oplus_{\mu=1}^{k/2}
(\fund_{\mu-1}, \fund_{\mu})$ and
$\asym_{k/2}$
\item
the $V_{\mu}$s pick up an additional contribution from a
vector $D_{\mu}$ basically reflecting the orientifold charge
so that we obtain
$$V_0=2K, \; \; \; V_{i \neq 0} = 2K+\sum_{j=1}^k C^{-1}_{ij} (w_j + D_j)$$
\end{itemize}

Choosing $D_{\mu}=- \delta_{\mu,0}$ one reproduces the right classical
theory whose Higgs branch is the moduli space of instantons,
while $D_{\mu} = -16 \delta_{\mu,0} -8 \delta_{\mu,
k/2} - 8 \delta_{\mu, (k+2)/2}$ yields an anomaly free
gauge theory once we couple to the additional tensor
multiplets from the 2-form fluxes. Choosing the anomaly
free theory one finds that one misses some degrees
of freedom that are necessary to describe the Higgs branch.
All in all 29 hypermultiplet
degrees of freedom a missing for every tensor present
on the Coulomb branch. We again
interpret this as a small instanton transition. The Higgs branch
describes the D5 dissolved as an instanton inside the D9. If
the instanton shrinks down to a point we reach a non-trivial fixed point.
From there the Coulomb branch opens up, described by the anomaly
free gauge theory living on the worldvolume of the D5.
The new branch is parametrized by the vev of the scalars
in the tensor multiplets, which as I mentioned arise
in this setup from reducing the 10d two form on the
vanishing cycles.

More intuitive is the same process for the $E_8$
small instanton. Here the Higgs branch
still corresponds to an M5 dissolved
inside an M-theory end of the world brane carrying
an $E_8$ gauge theory \cite{howiI,howiII}. However
the Coulomb branch this time has a real geometric
interpretation, moving of the 5 brane into the bulk.
As we have seen, both these cases are also beautifully realized in
the HW setup. Here the $SO(32)$ as well as the $E_8 \times
E_8$ small instanton's Coulomb branch are visible as motions
of the 5 brane.

\subsection{Applying T-duality}

Of course it is no coincidence that we find the same small
instanton transition
in both realizations. In fact the two approaches are related
by a simple T-duality. Considering an HW setup on a circle and
T-dualizing this 6-direction, the NS5 branes turn into
the orbifold singularity as above \cite{dualNS},
the D6 becomes the the D5, the small $SO(32)$ instanton.
The O8s and D8s turn into O9s and D9s. We reproduce precisely
the same low-energy theories if we identify the $w_{\mu}$ and
$D_{\mu}$ I defined in the HW setup as the number of D8 branes
in the $\mu$th gauge group and the contribution of
the orientifold charge to the $\mu$th gauge group,
with the $w_{\mu}$ and $D_{\mu}$ defined above,
basically carrying the information about the non-trivial
Wilson lines at infinity. The $Z_2$ choice for even $k+1$
(number of NS5 branes in the HW setup) corresponds
to the choice of having all NS5 branes free to move or one
stuck at each of the orientifold planes. These identifications
reflect the well known fact \cite{polchinski}, that
under T-duality the Wilson lines translate into brane positions.

Note that this way a small $SO(32)$ instanton transforms
into a D6 brane in the HW setup with the NS5 branes playing the
role of the singularity. On the other hand taking the strong
coupling limit of the HW the O8 turns into an M-theory
end of the world brane. So the strong coupling limit of
HW perfectly captures the small $E_8 \times E_8$ small instanton
as well, but this time the NS5 branes are the instantons
(after all they become M5 branes in 11d) while the D6 branes
in 11d turn into the singularity. This way one can reproduce
the theories of \cite{smallinonsin} describing
small $E_8$ instantons colliding on top of an A or D type
singularity by HW setups.

\subsection{Adiabatically expanding to d=4 ${\cal N}=1$}

As I explained in the last chapter, a HW setup
in d dimensions is closely related to a HW setup in d+2 dimension
with twice the amount of supersymmetry, by replacing
the flavor system in d dimensions by a whole HW setup in d+2.
This was how we realized the chiral / non-chiral transition
in 4d by performing the 6d small instanton transition.
This correspondence of course has an equivalent in the geometric
language: having analyzed string theory on ALE spaces, one
can next consider theories on an ALE fibration
over a $P^1$. These will also lead to theories
with half the amount of supersymmetry in 2 less
dimensions. In the limit of large base size one can apply
the 6d results fibrewise. This is usually referred to
as the adiabatic argument.

\section{Branes and geometry:  \\obtaining the Seiberg-Witten curve}
\subsection{The 3 approaches}
As we have seen, the branes as probes and the HW approach to field theories
are actually equivalent. In both cases we cook up a certain string
background whose low energy field theory realizes the gauge setup we want.
But then it turns out that these two string backgrounds are dual to each
other. This duality allowed us to analyze certain aspects in one or the other
frame: e.g. deformations were obvious in the HW setup,
the identification of the Higgs branch as the moduli space of $SO(32)$ 
instantons on an ALE space can easily be motivated from the orbifold
point of view. Let me finally show in the example of
Seiberg-Witten theory that this kind of relation also extends to the
third approach, geometrical engineering. This will complete our
dictionary from phase transitions enforced by brane motions to the
purely geometrical language of string compactifications
\footnote{A similar connection has been established in 5d in
\cite{vafaand5d}.}.

The three approaches to engineer gauge theories have already
been mentioned in general 
in the introduction. Let me show in more detail how they
work in the special case of ${\cal N}=2$ theories in 4d. I will
especially focus on the quantum aspects, that is I will identify
the quantum contributions and show how they can sometimes be solved
for. 

\subsubsection{Geometrical Engineering}

In the introduction I already discussed how geometric engineering
works in principle and how to engineer 6d theories with 16
supercharges. Now we want to go down to ${\cal N}=2$ in d=4.
So we have to compactify two more dimensions and break half
of the supersymmetry. This is done via the adiabatic expansion.
That is we look at non-compact Calabi-Yau manifolds, which locally
look like an ALE fibration over a 2d base. The simplest base is
just a $P^1$, but we will also need bases built out of several
$P^1$s with non-trivial intersection patterns.

Let me first discuss the case of a $P^1$ base. If we choose the
ALE fiber to be of  A$_k$ type, we will start with
an $SU(k+1)$ gauge theory in 6 dimensions. The massless
states arise from D2 branes wrapping the vanishing
cycles of the ALE in the orbifold limit. Taking into account
the fibration structure only the monodromy invariant states
survive. It turns out that if one deals with
a genus $g$ base with ADE singularity one expects to have ADE gauge
symmetry with $g$ adjoint hyper multiplets \cite{BVS,KMP,KM}.
That is for our case of genus zero we only keep a pure ${\cal N}=2$
SYM as desired. For $g=1$ one obtains the finite ${\cal N}=4$
spectrum and for higher $g$ we would loose asymptotic freedom.

Decoupling gravity is again achieved by effectively decompactifying
the Calabi-Yau manifold and focusing on the
local singularity structure. However this time there are
two scales involved, the size of the base $P^1$ and the characteristic
size of the fibers, so that some care is required.

As an illustrative example \cite{EngineerII} consider the geometric engineering
of a pure $SU(2)$ gauge symmetry which is related to an $A_1$ singularity
in $K3$. Locally, one needs a vanishing 2-sphere $P^1$,
around which the D-branes, being the $W^\pm$ bosons, are wrapped.
This $P^1_f$ has
to be fibered over the base $P^1_b$ in order to have ${\cal N}=2$ supersymmetry
in four dimensions. The different ways to perform this fibration are encoded
by an integer $n$, and the corresponding fiber bundles are the
Hirzebruch surfaces $F_n$.
The mass (in string units)
of the $W^\pm$ bosons corresponds to the area of the fiber
$P^1_f$, whereas the area of the base $P^1_b$ is proportional to
$1/g^2$ ($g^2$ is the four-dimensional gauge coupling at the string scale).
Now let us perform the field theory limit which means that we send the
string scale to infinity. Asymptotic freedom implies that
${g^2}$ should go to zero in this limit; thus the K\"ahler class
of the base $P^1_b$ must go to zero: $t_b\rightarrow \infty$. Second,
in the field theory limit the gauge boson masses should go to zero,
i.e. $t_f\rightarrow 0$. In fact these two limits are related by
the running coupling constant,
\begin{equation}
{1\over g^2}\sim\log {M_W\over\Lambda},
\end{equation}
and the local geometry is derived from the following double scaling
limit:
\begin{equation}
t_b\sim - \log t_f\rightarrow\infty .\label{doublesc}
\end{equation}

Clearly, this picture can be easily generalized to engineer higher rank ADE
gauge groups. In this case there is not only one shrinking $P_f$ in the
fiber, but several such that the fiber acquires a local ADE singularity.
For the close comparison with the Hanany-Witten set up, it is also
interesting how product gauge groups can be geometrically
engineered \cite{KMV}. For concreteness consider the group $SU(n)\times
SU(m)$ with a hypermultiplet in the bifundamental representation.
To realize this we have an $A_{n-1}$ singularity over $P^1_b$ and an
$A_{m-1}$ singularity over another  $P^1_b$. The two $P^1_b$'s intersect
at a point where the singularity jumps to $A_{m+n-1}$. This can
be seen in 6 dimensions as symmetry breaking of $SU(n+m)$
to $SU(n)\times SU(m)\times U(1)$ by the vevs of some scalars in the
Cartan subalgebra of $SU(n+m)$. It is straightforward to generalize
this procedure to an arbitrary product of $SU$ groups with matter in
bi-fundamentals. One associates to each gauge group a base $P^1_b$ over
which there is the corresponding $SU$ singularity;
to each pair of gauge groups connected by a bi-fundamental representation
one associates an intersection of the base $P^1_b$'s, where over
the intersection point the singularity is enhanced to $SU(n+m)$.

Now it is interesting to see how the quantum effects are incorporated. 
Basically we took care of the 1-loop effects by the double scaling limit,
the non-perturbative effects are due to worldsheet instantons,
that is non-trivial embeddings of the string worldsheet in the target space.
Actually the way this appears in string theory is slightly more involved.
From the field theory point of view all the information we want
is encoded in the so called prepotential
\begin{equation}
{\cal F}(A)=
{1\over 2} \tau_0A^2+{i\over \pi}A^2\log\biggl({A\over\Lambda}\biggr)^2
+{1\over 2\pi i}A^2\sum_{l=1}^\infty c_l\biggl({\Lambda\over A}\biggr)^{4l}.
\label{fieldprep}
\end{equation}
Here $\tau_0={1\over g^2}$ is the classical gauge coupling, the logarithmic
term describes the one-loop correction due to the running of
the gauge coupling and the last term collects all non-perturbative
contributions from the instantons with instanton numbers $c_l$.
This is the function that is solved for by the SW solution.

The way this prepotential appears in IIA string theory is by
a classical computation (that is no $g_s$ corrections). We only
get corrections from worldsheet instantons. This is basically due
to the fact that the dilaton, which controls the $g_s$ corrections,
sits in a hypermultiplet and therefore doesn't talk to
the prepotential which controls the vectormultiplets.
For a Calabi-Yau it has the following structure \cite{cand}:
\begin{eqnarray}
 {\cal F}^{\rm II} = -{1\over 6}C_{ABC}t_At_Bt_C
- \frac{\chi \zeta(3)}{2 (2 \pi^{3})}+
\frac{1}{(2 \pi)^3} \sum_{d_1,...,d_h}
n_{d_1,...,d_h} Li_3({\bf exp} [i\sum_Ad_At_A]),
\label{ftype2}
\end{eqnarray}
where we work inside the K\"ahler cone ${\rm Re}(t_A)\geq 0$.
The polynomial part of the type-IIA
prepotential is given in terms of the classical intersection
numbers $C_{ABC}$ and the Euler number $\chi$,
whereas the coefficients $n_{d_1,...,d_h}$
of the exponential terms denote the rational instanton numbers of
genus 0 and multi degree $d_A$.
For our non-compact version of the CY there are just two
moduli, namely $t_b$ and $t_f$ and the
corresponding instanton numbers $n_{d_b,d_f}$ denote
instantons wrapping the base $d_b$ times and the fiber $d_f$ times.

Now we want to see that in the double scaling limit
(\ref{doublesc}) the string prepotential gives
us back the field theory prepotential with the right
logarithmic one-loop behaviour.
First note that in this limit the $n_{0,d_f}$ yield
precisely the logarithmic terms from the
1-loop corrections (well, after all that's how we chose our limit.
). The worldsheet instantons wrapping non-trivially around the
base will yield the non-perturbative contributions.

But how can we sum them up?
What comes to our rescue is mirror symmetry. Type IIA on a given
Calabi Yau is dual to IIB on the mirror Calabi Yau.
The two hodge numbers counting ``shape'' and ``size'', $h^{2,1}$ and
$h^{1,1}$ get interchanged. It is interesting to see where
the quantum corrections come from. As I already mentioned,
the $g_s$ corrections will only affect the HM moduli space.
In addition worldsheet instantons will correct moduli associated
with $h^{1,1}$, since they are associated with the non-trivial
two cycles. But this means that the vector multiplet moduli
space on the IIB side is exact!
\begin{center}
\begin{tabular}{c|cc c c|cc}
IIA& $\alpha'$ &$g_s$& $\rule{2cm}{0cm}$& IIB& $\alpha'$ &$g_s$ \\
\cline{1-3}  \cline{5-7}
$\rule{0cm}{.6cm}$ VM$_{h^{1,1}}$ & x & 0 & & VM$_{h^{2,1}}$&0&0\\
$\rule{0cm}{.6cm}$ HM$_{h^{2,1}}$ & 0 & x & & HM$_{h^{1,1}}$&x&x\\
\end{tabular}
\end{center}
All we have to do now is find the mirror of our non-compact
Calabi Yau. There is a well defined description of how to
perform this so called ``local mirror map'' \cite{EngineerII}.
This way we can solve for the SW curve by a classical
computation in string theory. To give the SW curve a geometrical
interpretation we once more perform a T-duality,
taking the singularity of the IIB fiber into an NS5 brane \cite{Klemm}.
This way the beautiful final answer is obtained:
The SW curve appears as the physical object 
a IIA NS5 brane wraps on.

\vspace{1cm}
\fig{Solving in Geometric Engineering via Local Mirror Symmetry}
{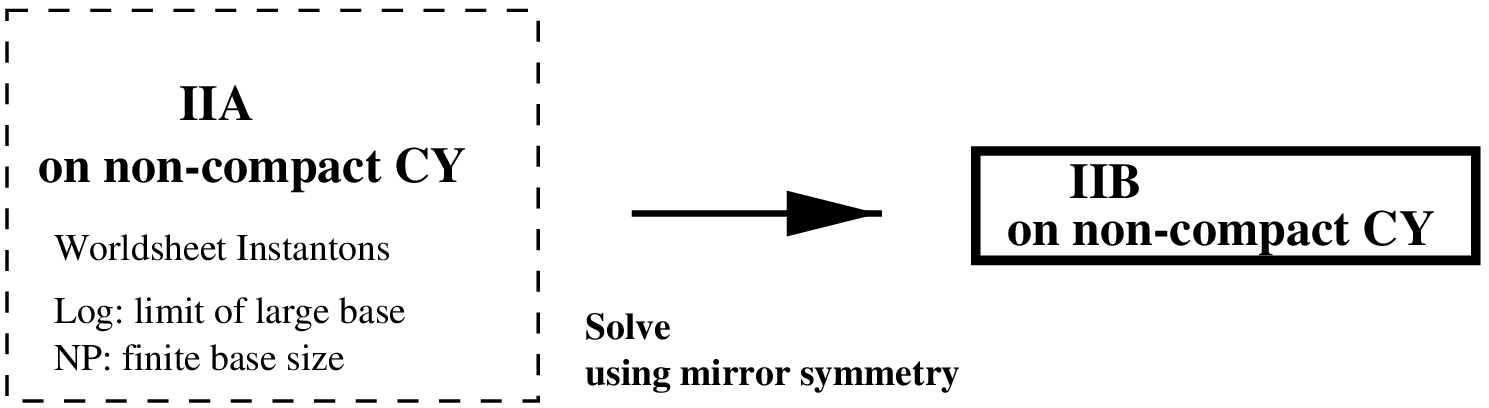}{10truecm}
\figlabel{\eng}

\subsubsection{Fractional Branes}

Above I have shown how to construct the field theory of $K$ D$p$
branes on top of an orbifold singularity.
Their Higgs Branch
was identified as moving the branes away from the singularity,
breaking the $U(K)^{k+1}$ gauge group to its diagonal $U(N)$ subgroup.
But, as I mentioned, for $p<5$ we will also see a Coulomb branch,
corresponding to the scalars in the VM, along which the gauge
group is generically broken to $U(1)^{K (k+1)}$. How do
we see this Coulomb branch in the brane at orbifolds description? 
I will explain this in the case of D3 branes relevant for the
discussion of ${\cal N}=2$ in 4d. Generalizations to other
D$p$ branes should be obvious.

The answer was given in \cite{Enhanced} (see also \cite{douglasmoore}):
on the Coulomb branch, a single D3 brane splits into $|\Gamma|$ fractional
branes of equal mass $\frac{m_{D3}}{|\Gamma|}$. These fractional
branes can be interpreted as D5 branes wrapping the vanishing
cycles of the orbifold. Therefore they are stuck to the singularity
in the orbifolded part of spacetime, but are free to move in the
transverse space, giving rise to the Coulomb branch.
The example considered in \cite{Enhanced} was D0 branes moving
on an ALE space. In the orbifold limit these new degrees
of freedom supported on the Coulomb branch are precisely those that
are necessary for the Matrix description of enhanced gauge
symmetry for M-theory on the ALE space.

Looking just at a single fractional brane is achieved by
embedding the orbifold group into the Chan Paton factors via
a representation other than the regular one. This is consistent
with the fact, that only by choosing the regular representation
can we describe objects that are free to move away from the orbifold.
This identification can be proven by an explicit world-sheet calculation
\cite{douglasmoore} showing that the fractional branes
carry charge under the RR fields coupling to D5 branes. This
charge vanishes if and only if one chooses the regular representation.
So instead of just obtaining fractional branes by splitting D3 branes
to move on to the Coulomb branch, one can just add
these fractional branes by hand. This is the approach chosen
in \cite{ALE1,ALE2} to explain the wrapped membranes in Matrix
theory. We will give a more detailed analysis of the
branches, parameters and the quantum behaviour of these theories as we
proceed.

\subsubsection{Hanany-Witten}

The 4d Hanany-Witten setup and its solution via the M-theory lift has
already been discussed in great detail in Chapter 2. Let me summarize
Witten's solution I reviewed there 
in the form of a diagram like in the other case,
so that we are able to compare:

\vspace{1cm}
\fig{Solving HW setup via M-theory}
{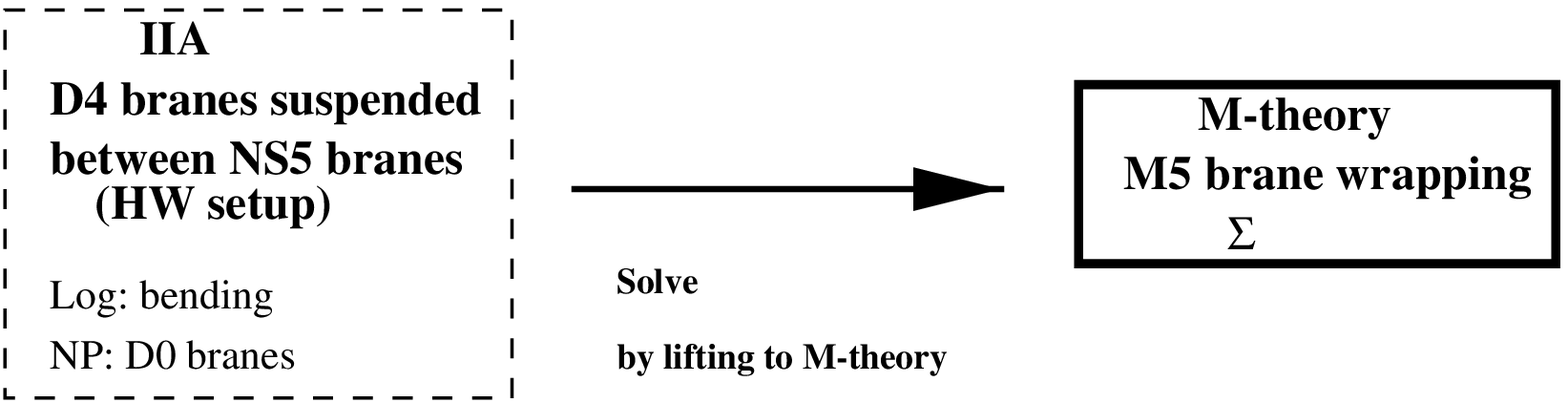}{9truecm}
\figlabel{\hw}

\subsection{A T-duality for bent branes and  \\ branes ending on branes}

From our discussion of the 6d case we are by now used to the idea that
branes as probes are dual to HW setups. The T-duality as
in the 6d setup will work the same for D4 branes suspended
between NS5 branes and D3 branes at ADE singularities, as long
as in the HW setup the brane connects all the way around the circle.
In 6d we were forced to only consider those setups from anomaly
reasons. Here they only correspond to the very special case
of finite theories at the
origin of the Coulomb branch. To understand the more general cases we will
need a generalization of the notion of T-duality 
to the case of branes ending on branes and bent branes.
This will be the purpose of this section. To summarize what we will
find: the dual of a brane living on a interval bounded by two other branes
will turn out to be the fractional brane discussed in the context of
branes at orbifolds. Or again phrased in our diagrammatic language:

\vspace{1cm}
\fig{Duality relation between HW setup and fractional branes}
{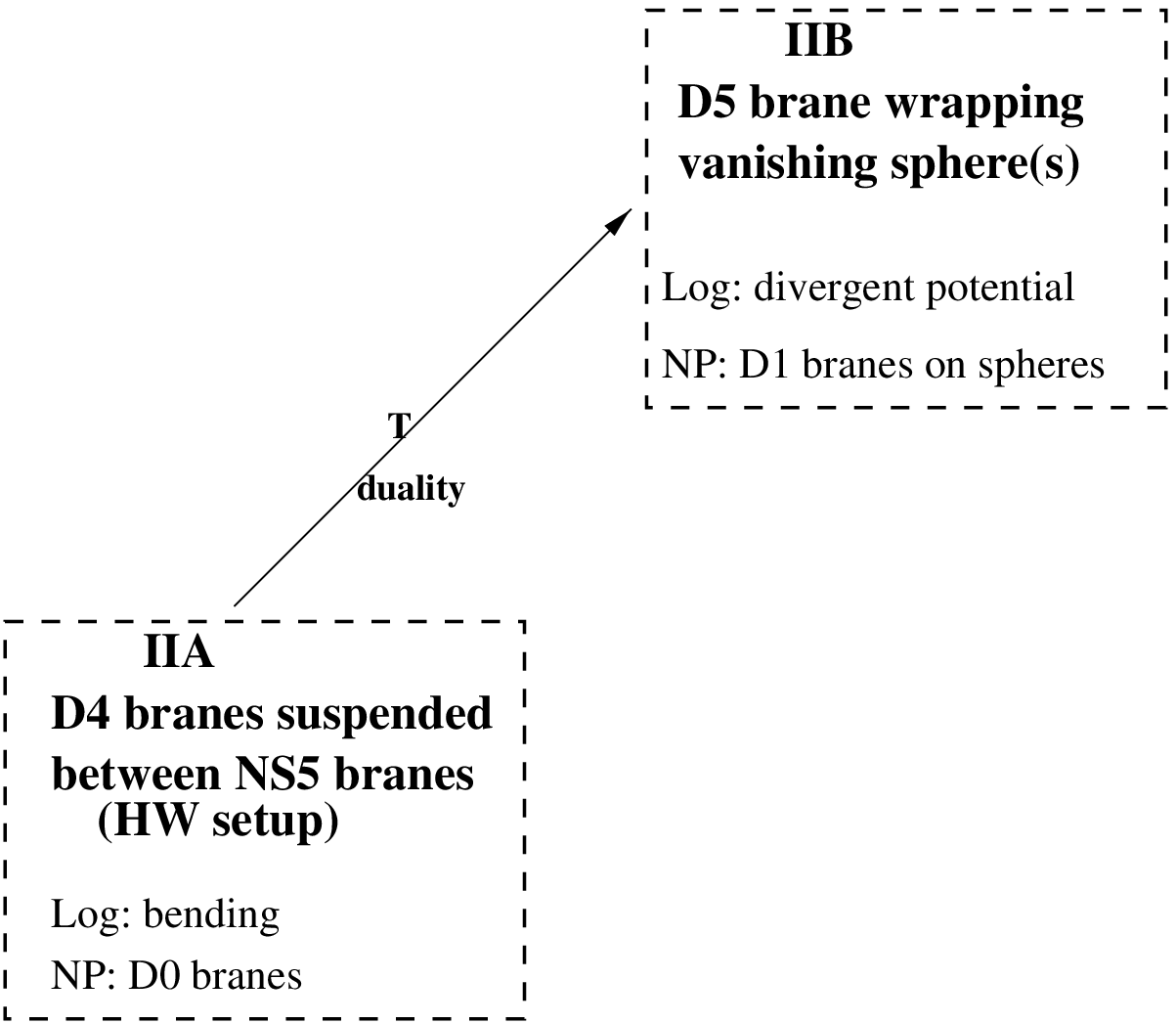}{7truecm}
\figlabel{\tdual}

Consider again $K$ D3 branes at a transverse $A_{k}$ singularity, that
is we take the D3 branes to live in the 0123 space and put these
on top of an $A_{k}$ singularity.
Since the D3 branes can move away from the singularity we should
choose to embed the orbifold group in the Chan Paton factors
via the regular representation, that is include the D3 brane and all its
$k$ images.
The corresponding gauge group is $U(K)^{k+1}$.
Upon T-duality in the 6 direction this can be mapped into an
elliptic Hanany-Witten setup, that is $k+1$ NS5 branes living
in the 012345 directions, with $K$ D4 branes living in 01236 space
suspended between every pair of consecutive NS5 branes on the circle.
Let us label
the NS5 branes with $i=0,\ldots,k$.
This is just the T-duality between branes at orbifolds and HW setups
I discussed above in detail for the 6d case.

But now consider a setup where we have just $K$ D4 branes
stretching between the $i$th and
the $(i+1)$th NS5 brane and no other D4 branes.
By just comparing the resulting gauge theory (a single $U(K)$)
it is natural to assume that the same T-duality
in the 6 direction as we used above now translates this into
$N$ fractional branes in the sense of \cite{Enhanced}, that
is $N$ D5 branes wrapping the vanishing homology two-cycle
$\sigma_i$ \footnote{where $\sigma_{k+1}= -\sum_i \sigma_i$ \cite{Enhanced}}
of the $A_{k}$ orbifold in the dual picture. Or again in the language of
\cite{Enhanced}, we consider the same orbifold group, but this
time choose to embed the orbifold group into the Chan Paton factors
via the 1d irreducible representation
$r_i$
associated with the $i$th node of the extended Dynkin diagram.
Or more generally, if we have $K_i$ D4 branes connecting the $i$th
NS5 brane with the $(i+1)$th NS5 brane the dual setup will again be given
by a $Z_{k+1}$ orbifold, this time with the orbifold group embedded in
the Chan Paton indices via a general representation $R$ given by
\begin{equation}
R= \oplus_i N_i r_i.
\end{equation}

As a first check note that even in this
more general case the resulting gauge groups and matter contents
agree: both constructions yield a $\prod_i U(K_i)$ gauge
group with hypermultiplets transforming as bi-fundamentals
under neighboring gauge groups.
Similarly we can compare the classical parameters and moduli.
First consider the case of all $K_i=K$, where the duality
is well established. This theory has two branches, a Coulomb
branch and a Higgs branch. As free parameters we can add FI terms $\xi_i$,
for each gauge group factor
which are triplets under the $SU(2)_R$ symmetry
and totally lift the Coulomb branch.
Since we have $k+1$ gauge group factors we have in addition
k gauge coupling constants $g_i^2$, which at least in the classical theory
are additional free parameters. Together with
the $\theta$-angles they form the complex coupling constants
$\tau_i=\frac{\theta_i}{2 \pi} + \frac{4 \pi i}{g_i^2}$

On the orbifold side the Higgs branch just corresponds to moving
the D3 branes away from the orbifold point in the ALE space.
Therefore the Higgs branch metric coincides
with the metric of the transverse ALE space.
Since all the $N_i$ are equal in the HW setup all the D4 brane
pieces can connect to form $N$ full D4 branes which can then move
away from the NS5 branes in the 789 direction. As usual, the 4th real
dimension of this branch is given by the $A_6$ component of the
gauge field on the D4 branes.

When the D3 branes meet the NS5 branes in 789 space, that is at the origin
of the Higgs branch, they can separate into the pieces which are then
free to move around 45 space along the NS5 branes giving rise to the
Coulomb branch. In the orbifold picture the same process is
described by the $K$ D3 branes splitting up into $(k+1) K$ fractional
branes. These are now localized at the singularity in 6789 space,
but are also free to move in the 45 space.
Now consider turning on the FI terms $\xi_i$. For the
orbifold this
corresponds to resolving the singularity by blowing up the $i$th vanishing
sphere $\sigma_i$. The mass of the states on the Coulomb branch
(which are after all D5 branes wrapping these spheres) become
of order $\xi_i^2$. The Coulomb branch is removed as a supersymmetric
vacuum. In the dual picture the FI terms correspond to motion of the
NS5 branes in 789 space. This resolves the singularity
since the 5 branes no longer coincide and again the mass of states on
the Coulomb branch (which in this case are D4 branes stretching
between the displaced NS5 branes) is of order $\xi_i^2$.

Last but not least we have to discuss the coupling constants. On the HW side
they are simply given in terms of the distances $\Delta_i$
between the NS5 branes in the 6 direction, or to be more precise
\begin{equation}
g_i^{-2}=\Delta_i / g^A_s l_s
\end{equation}
where $g^A_s$ and $l_s$ denote the string coupling constant and string length
of the underlying type IIA string theory
respectively. Since we are dealing with a theory with compact 6 direction
we have
\begin{equation}
\sum_i \Delta_i = R_6
\end{equation}
where $R_6$ is the radius of the 6 direction.
The easiest case to consider is if all $\Delta_i=\Delta$, that is
the NS5 branes are equally spaced around the circle.
In this case we get $g_i^{-2}=g^{-2}=R_6 / g^A_s k l_s$. Applying T-duality
to type IIB this
means that $g_i^{-2}=1 / g^B_s k$ since $g^B_s=g^A_s l_s/R_6$, in accordance
with the orbifold analysis \cite{douglasmoore,Lawrence}.
The same configuration can be
also viewed as $N_i$ fractional branes, that is $N_i$ D5 branes
wrapping the $i$th vanishing cycle. According to \cite{Lawrence}
in this case the coupling constant is given in terms of fluxes
of two-form charge on the vanishing sphere
\begin{equation}
\tau_i=\int_{\sigma_i} B^{RR} + i \int_{\sigma_i} F -B^{NS}
\end{equation}
At the orbifold
the values of the B-fields are fixed to yield again $g_i^{-2}=1/g^B_sk$
\cite{Enhanced}. Next consider the case where all the NS5 branes coincide.
In 6d this will correspond to an enhanced gauge symmetry on the NS5 brane.
As shown in \cite{9507012} this only happens if the B-flux
is zero on the dual side, again in accordance with our formulas.
\footnote{The one gauge factor with the finite gauge coupling
is asymptotically non-free. In the IR it is just a global symmetry.
At higher energies new 6 dimensional 
degrees of freedom come in from the self-intersection of the NS5 branes
- it is impossible to put bent branes on a circle without
any selfintersection.}
Under
T-duality these $\int_{\sigma_i} F -B^{NS}$ 
Wilson lines translate into the positions
of the NS5 branes on the circle, as expected.

It is easy to see that this discussion generalizes without any problems
to arbitrary values of $K_i$. In addition we can introduce D6
branes in the HW setup living in 0123789 space to give extra matter. These
simply T-dualize into D7 branes wrapping the ALE space in the dual picture.

One may wonder whether this is consistent as a quantum
theory. Since, as we discussed above, the fractional
branes are charged under the RR gauge fields we need `enough' non-compact
space in the transverse dimensions for this to be consistent. 
Since we are dealing with ${\cal N}=2$
SUSY there are only two kinds of corrections, one-loop and
non-perturbative corrections. All higher loop contributions
vanish exactly. The one loop beta function gives rise to
the usual logarithmic running of the coupling constant. The
instanton contributions have been summed up by the Seiberg-Witten solution.

On the HW side it is well known how to incorporate quantum corrections.
In the full string theory the D4 brane ending on the NS5 brane
actually has a back-reaction on the NS5 brane leading to a logarithmic
bending of the NS5 brane. Since the coupling constant is given
in terms of distance between the NS5 branes this fact just
reflects the 1-loop correction, the running of the gauge coupling.

The non-perturbative
effects in this case are given by Euclidean D0 branes stretching
along the 6 direction \cite{Barbon1,Barbon2,BrodieD0}
using the well known relation that
SYM instantons inside a D$p$ brane are represented by D$p-4$ branes
\cite{BranesWithin}.
Like the D4 branes themselves these Euclidean D0 branes can of
course split on the NS5 branes and become fractional D0 branes
\cite{BrodieD0}. According to the same T duality in
the 6 direction we applied to the D4 branes they should become
fractional D(-1) branes, that is Euclidean D1 branes wrapping
the vanishing cycles. This fits nicely with the picture
that this way the instanton corrections on the orbifold
side are again given by D$p-4$ branes living inside the
D$p$ brane.

But how does the 1-loop effect arise in the orbifold picture?
Note that the fractional branes carry charge under the appropriate
RR fields, as can be born out by a world-sheet computation
of the corresponding tadpoles
\cite{douglasmoore}.
They represent a charge sitting in the space transverse to
the D3 brane and transverse to the orbifold. In our example
this is the 2 dimensional 45 space. In two dimensions the Green's functions
are given by logarithms. Therefore we would expect a theory of
a charged object in 2 dimensions to be plagued by divergences.
To be more precise we have the following trouble: we want
to consider effectively a 3 brane sitting in 6d spacetime.
Since the transverse space is only 2d, the corresponding
4-form vector potential $C^{(4)}$ grows logarithmically.

From the D5 brane point of view the 3 brane vector potential
couples to the world-volume theory via
\begin{equation}
 \int dx^6 {\cal F} \wedge
C^{(4)}.
\end{equation}
After compactification on $\sigma_i$ this leads to a term in the
3 brane action of the form
\begin{equation}
\int_{\sigma_i} {\cal F} \cdot \int dx^4 C^{(4)}.
\end{equation}
But recall that we identified $\int_{\sigma_i} {\cal F}$ as the
gauge coupling $\frac{4 \pi}{g_i^2}$.
The growing of $C^{(4)}$ can be absorbed by introducing an effective
running coupling constant.
The divergences we encounter are just due to the 1-loop running
of the gauge coupling! Charge neutrality,
which is obtained if we choose the regular representation,
corresponds
to finiteness \cite{douglasmoore,Lawrence}. In this way we can see that
the correct dependence of the beta function on $N_c$ and $N_f$ is reproduced.
This can most easily be seen by observing that the charge clearly depends
linearly on the number of fractional branes and is zero when $N_f = 2N_c$,
corresponding to the case where the fractional branes can all combine to
form a full brane. The relative contributions originate from the
(self-)intersection numbers of the spheres $\sigma_i$ given by the
(extended) Cartan matrix (here for $A_{k-1}$).

In this way the bending of the
branes directly translates into the fall-off of the RR field in transverse
space in the orbifold picture. The correspondence also
holds in other dimensions (for D5  branes
the transverse space is 0 dimensional and we need neutrality
reflecting anomaly freedom of the underlying gauge theory, in 5,4,3,2,...
dimensions we get linear, logarithmic, $1/r$, $1/r^2$, ... fall off
for the field strength). This reflects the appropriate running of the gauge
coupling just as in the HW picture.

\subsection{Unifying the different approaches}

Now we are in a position to put all the three approaches together into
an unifying
picture. To visualize what we are about to do, let me
first show a diagram that summarizes the connection. I will
then explain the dualities involved step by step.

\vspace{1cm}
\fig{Connection between the various approaches. The non-compact
Calabi Yau spaces are ALE spaces fibered over a sphere (or several
intersecting spheres). These can be T-dualized via T-duality
on the fiber. The last T-duality, connecting to
the HW picture is the one discussed in this work.
}
{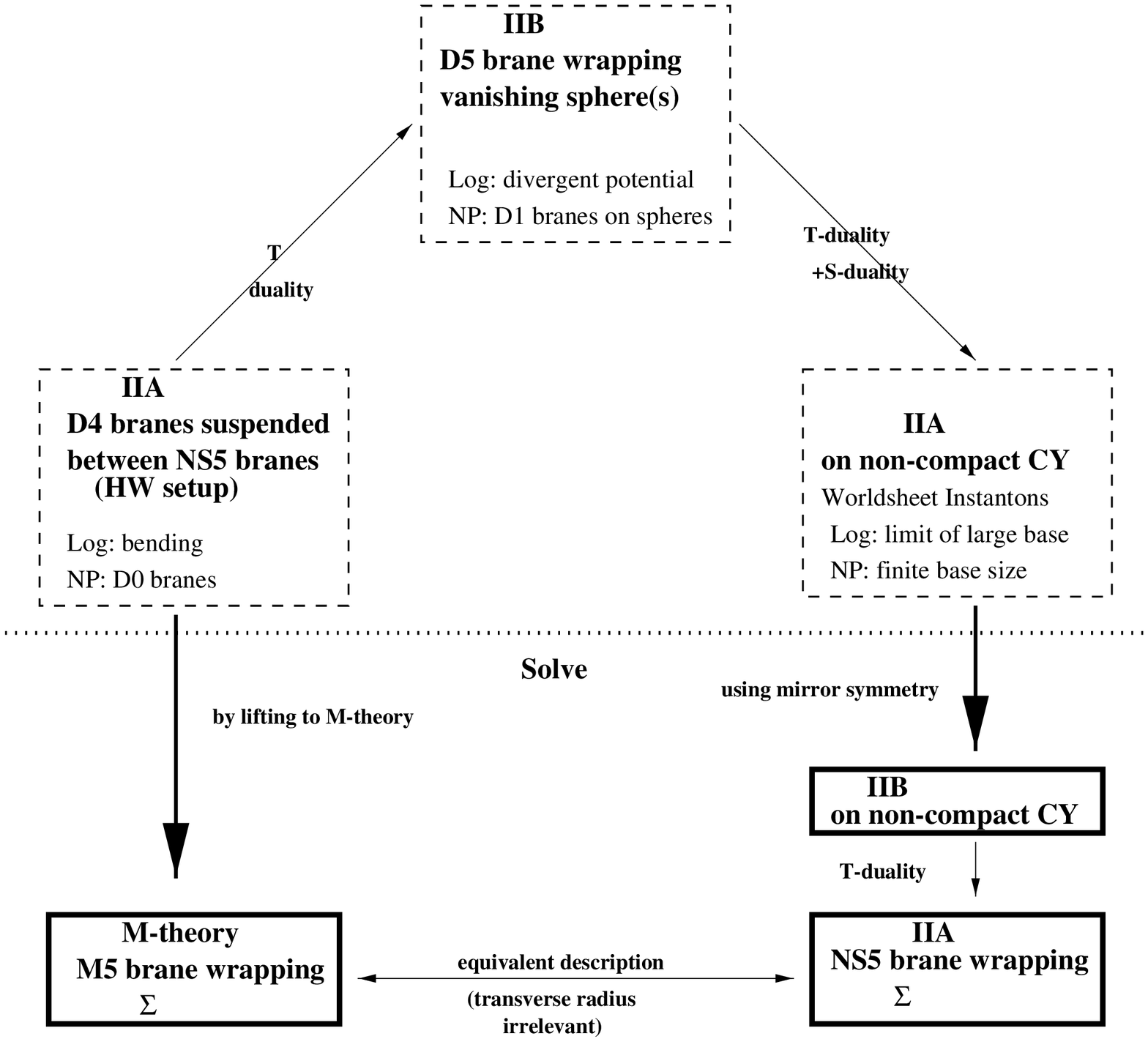}{14truecm}
\figlabel{\dia}
\vspace{1cm}

In Figure (\dia) the connections are displayed diagrammatically, putting
together the pieces of Figures (\eng), (\hw) and (\tdual).
For the
geometric engineering
approach of \cite{EngineerII} the starting point is IIA
string theory on a non-compact Calabi-Yau which is an ALE space
fibered over some base (a non-compact version of a K3 fibration).
Quantum corrections can be solved by using mirror symmetry
to a IIB picture. A T-duality on the ALE
fiber can be used to map this to a IIA NS5 brane wrapping
the Seiberg-Witten curve \cite{Klemm, EngineerII}. While in this approach
the solution is obtained via mirror symmetry, this last
T-duality leads to the very nice interpretation
of the SW curve: it appears as the physical object the
5 brane is wrapping.

Applying the same T-duality, which we have just applied to the IIB solution,
directly on the ALE fiber in the
IIA setup (which is the starting point for the geometric engineering approach),
followed by an S-duality in the resulting IIB string theory,
it is straight-forward to connect
this to a setup of D5 branes wrapping the base of the fibration.

Using the T-duality between fractional branes and branes
ending on branes proposed above one can translate this
into the very intuitive language of a HW setup \cite{HW},
which finally can be solved for by lifting to M-theory \cite{wittenn2}.
This way we not only obtain the same solution (the SW curve), but
also the same physical interpretation: the SW curve appears directly
as the space an M5 brane wraps on
\footnote{In the spirit of \cite{seisethi} the difference between
IIA and M5 brane is irrelevant, since the radius of the transverse
circle does not affect the low-energy SYM on the brane.}.

In the following we will discuss the steps in the above chain of dualities
in more detail. There are two important points one should account
for. For one thing there are the quantum corrections. Since we are
dealing with ${\cal N}=2$ there are only two types of these corrections:
the 1-loop contribution which gives rise to the logarithmic
running of the gauge coupling and the instanton corrections.
It is interesting to see how these quantum effects are incorporated
in the various pictures and how they are finally solved for.

The other thing one should treat with some care is how the singularity
type affects the gauge group.
In all the examples we are considering there are always two
pieces of information, which we will refer to as the singularity
types determining the ``gauge group'' and the ``product structure''.
This is easiest to understand in a particular example.
Consider $N$ D3 branes at an $A_{k-1}$ singularity. This gives rise
to an $SU(N)^{k}$ gauge group. In this example $N$ determines
the ``gauge group'' (the size and type of the single gauge group factors)
and $k$ the ``product structure'' (we have $k$ $SU(N)$ factors). Since
in the various duality transformations we are using we repeatedly
map branes into singularities and vice versa it is quite important
to distinguish which ingredient in the picture determines
gauge group and which determines product structure.
In the cases where the gauge group is determined by a geometric
singularity, generalization to D or E type singularities leads
to $SO$ or $E_{6,7,8}$ gauge groups, while in the
cases where the product structure is determined by a geometric
singularity the generalization to D or E type just leads to a more
involved products of $SU(N)$ groups determined by
the corresponding extended Dynkin diagram. For example in the D3
brane case $N$ D3 branes at an $E_6$ singularity lead to
an $SU(N)^3 \times SU(2N)^3 \times SU(3N)$ gauge group.

{\bf HW $\rightarrow$ fractional branes: }
This is the duality between branes ending on branes
and fractional branes I have discussed above. A system
of $k$ NS5 branes with $N_i$ D4 branes suspended between the
$i$th and the $(i+1)$th NS5 brane is dual to an
$A_{k-1}$ orbifold singularity with $N_i$ fractional branes associated
to the $i$th shrinking cycle. While in the HW setup the gauge
group is determined by the D4 branes and the NS5 branes determine the product
structure, the corresponding roles are played by D3 branes and the orbifold
type in the dual picture. That is, the vanishing 2-cycles are given by
spheres intersecting according to the
extended Dynkin diagram. The gauge group can be generalized
to D type by including orientifold planes on top of the D-branes.
It is not known how to achieve E type gauge groups this way. Using an E type
orbifold in the fractional branes only affects the product structure.

The 1-loop quantum effects correspond to bending of the NS5 branes in HW
language and get mapped to the logarithmic Green's functions in the
orbifold picture, as discussed in the previous section. Non-perturbative
effects are due to Euclidean D0 branes, which are mapped by the same T-duality
into Euclidean D1 branes wrapping the vanishing 2-cycles (fractional
D-instantons).

{\bf Fractional branes $\rightarrow$ IIA on non-compact CY: }
We can now apply a different T-duality on the setup described by
the fractional branes. First we S-dualize, taking the $N_i$ D5
branes into NS5 branes
and the Euclidean D1 branes into fundamental strings (that is world-sheet
instantons). Performing a T-duality in the overall transverse space
(that is 45 space) the NS5 branes turn into an $A_{N_i}$ singularity
according to the duality of \cite{dualNS} which we have already used
several times. The vanishing 2 cycle, which is still
the space built out of the $k$ spheres intersecting according
to the extended Dynkin diagram, stays unchanged.
The fact that the NS5 branes only wrap parts of this base ($N_i$ NS5 branes
on the $i$th sphere $\sigma_i$) translates into the fact that
the type of the ALE fibers over the base changes from one sphere to the other.
From the IIA point of view this looks like T-duality
acting on the fibers as described in
\cite{Klemm}. Since now everything is geometric, generalizations to E type
are straight forward: the product structure is determined by
the intersection pattern of the $k$ spheres, the ADE type of the
fiber determines the gauge group. This way we can even
engineer products of exceptional gauge groups.

The non-perturbative effects are now due to world-sheet instantons,
as expected. The log-corrections coming from one loop are incorporated
in the particular limit one has to choose to decouple gravity
\cite{DecouplingLimit}. The system can be solved via local mirror
symmetry.

\chapter{Open problems and directions for further research}

\section{Brane Boxes}

As I mentioned in the beginning there is a natural generalization of
HW setups: the brane box \cite{HanZaff}. In order to branegineer generic
models with 4 supercharges, one should consider living on
a rectangle bounded by two kinds of NS5 branes. Only recently
it has become clear \cite{leighrozali} that these models are indeed consistent
also at the quantum level. This calculation is done using a
generalization of the HW - branes as probes T-duality I have presented here.
Brane boxes are T-dual to D3 branes at a $C^3/\Gamma$ orbifold,
where this time $\Gamma \in SU(3)$, so that we are left with ${\cal N}=1$
instead of ${\cal N}=2$. Again a brane on a single box is T-dual
to fractional branes characterized by a given irreducible
representation of the orbifold group \cite{neueami}, just
as in the case I was discussing.

A tadpole associated to a generic orbifold element without
any fixed plane corresponds to a charge
in a 0d space. So it has to vanish. Indeed it was shown that vanishing
of these generic tadpoles is equivalent to anomaly freedom.
However there are also tadpoles corresponding to a twists that
leave a 2d fixed plane. For the same reasons as I discussed
in the previous chapter one should interpret the resulting
logarithmic divergence as the running of the gauge coupling.
Vanishing of these tadpoles therefore is not necessary for
consistency but implies finiteness. This condition once
more corresponds to a no-bending condition for the brane
boxes.

It would definitely be desirable to perform the ``lift to M-theory''
for the brane boxes. For one this should explain the anomaly
in terms of bending and hence explore some yet unknown aspects
of brane physics. It is still the wrong limit to actually solve
the theory, but again it should be possible to
solve for all the holomorphic quantities.
Some intrinsic ${\cal N}=1$ phenomena such as dynamical supersymmetry
breaking can be studied this way and find their natural place in string theory.
It is obvious that such a lift has to be performed via a SUSY
3-cycle instead of the SUSY 2-cycle. The conditions for
SUSY 3-cycles have also been worked out in \cite{becker,becker2}.
These equations are technically much more involved, but hopefully
the lift can be performed \footnote{After this work
has been finished, we performed this lift in \cite{andre}
in a very special class of models.}.

\section{Maldacena Conjecture}

Recently a really remarkable proposal has been made by
Maldacena \cite{Maldacena} generalizing the aspects of the brane / SYM
correspondence I have been discussing so far. It states
that the worldvolume theory of a brane is really dual to string
theory in the background of the brane. For macroscopic
systems (that is large number of coinciding branes) the curvature
of the brane solution becomes small, so that supergravity is
a good approximation of string theory. Taking
the limit $M_s \rightarrow \infty$ reduces
the worldvolume theory to SYM, while we focus
in to the near horizon region of the soliton,
leading to the statement
that large $N$ ${\cal N}=4$ $SU(N)$ SYM is dual
to supergravity in an $AdS_5 \times S^5$ background.

This conjecture already led to a variety of beautiful results.
Again it basically helps to understand aspects 
string theory as well as of field theory. As I explained
in Chapter 2 it is the most promising approach for actually
solving some of aspects of field theory that
so far have escaped our control. 
On the other hand we have learned several new aspects about quantum gravity.
As I mentioned it is the first realization of the concept of
holography, which is supposed to be a genuine property of
quantum gravity.
So it seems that there are still many aspects of
brane physics that need to be explored. Hopefully
in the end we will find a true understanding of
the fundamental problems we set out to solve.

\chapter*{Summary}
\addcontentsline{toc}{chapter}{\numberline{}Summary}

In this work I discussed several applications
of the connection of non-perturbative string theory and
SYM theory. In Chapter 1 I reviewed the physics of D-branes
as one example of a non-perturbative effect in string theory.
Their dynamics is dominated by gauge theory. 
This fact can be used to engineer certain string backgrounds
which yield interacting SYM theories as their low-energy
description. 

In Chapter 2 I then introduced one of the approaches in detail, the
HW setup. I gave a summary of the identification of the
classical gauge theory, showed how quantum effects manifest
themselves in the brane picture and how to solve them.

This way of embedding gauge theories into string theories has
several interesting applications. These were the topic
of Chapter 3. First I discussed dualities in field theory
and showed how they arise as a natural consequence of
string duality. As a second application I used branes to prove the
existence of non-trivial fixed point theories in 6 dimensions and
to study their properties. Some of these fixed points describe
phase transitions between two different brane configurations.
From a 4d point of view these 6d transitions can induce
a chiral non-chiral transition.

In Chapter 4 I discussed the relation of the HW setup
with the other approaches of
embedding gauge theory into string theory, 
especially the branes as probes approach.
The different ways of embedding gauge theories in string theory
are shown to be actually T-dual as string backgrounds. For
one this allowed us to explore several new aspects of T-duality,
like T-duality for bended branes and branes ending on branes.
In addition this relation can be used to show that
the transitions found in the brane picture can as well be understood
as transitions between topologically distinct compactifications
of string theory.

Some open problems and directions for further research were mentioned
in Chapter 5.

\chapter*{Acknowledgements}
\addcontentsline{toc}{chapter}{\numberline{}Acknowledgements}

I'd like to thank
H.~Dorn for advising me during this thesis
and D~L\"ust for his help and collaboration.
Large parts of my work have been done together with
Ilka Brunner. I am especially grateful to her.
I also thank Amihay Hanany, Andre Miemiec, Douglas Smith and George Zoupanos
for the good collaboration.
In addition I thank
M.~Aganagic, B.~Andreas,  G.~Curio,
J.~de Boer, J.~Distler,
M.~Faux, S.~Kachru,
A.~Krause, 
S.~Mahapatra, 
T.~Mohaupt, H.-J.~Otto, Y.~Oz,
C.~Preitschopf,
R.~Reinbacher, 
W.~Sabra, I.~Schnakenburg and E.~Silverstein for useful discussions.

\addcontentsline{toc}{chapter}{\numberline{}Bibliography}
\bibliography{arbeit}
\bibliographystyle{utphys}
\end{document}